\documentclass[twocolumn]{aastex63}

\shorttitle{A Detailed View of AFGL\,4176}
\shortauthors{Johnston et al.}
\begin{document}

\title{A Detailed View of the Circumstellar Environment and Disk\\ of the Forming O-star AFGL\,4176 }

\correspondingauthor{Katharine G. Johnston}
\email{k.g.johnston@leeds.ac.uk}

\author[0000-0003-4509-1180]{Katharine G. Johnston}
\affiliation{School of Physics \& Astronomy, E.C. Stoner Building, The University of Leeds, Leeds, LS2 9JT, UK}

\author{Melvin G. Hoare}
\affiliation{School of Physics \& Astronomy, E.C. Stoner Building, The University of Leeds, Leeds, LS2 9JT, UK}

\author{Henrik Beuther}
\affiliation{Max Planck Institute for Astronomy, K\"onigstuhl 17, D-69117 Heidelberg, Germany}

\author{Hendrik Linz}
\affiliation{Max Planck Institute for Astronomy, K\"onigstuhl 17, D-69117 Heidelberg, Germany}

\author{Paul Boley}
\affiliation{Moscow Institute of Physics and Technology, 9 Institutskiy per., Dolgoprudny 141701, Russia}
\affiliation{Ural Federal University, 19 Mira St., Ekaterinburg 620075, Russia}

\author{Rolf Kuiper}
\affiliation{Institute of Astronomy and Astrophysics, Eberhard Karls University T\"ubingen, Auf der Morgenstelle 10, D-72076 T\"ubingen, Germany}

\author{Nathaniel Dylan Kee}
\affiliation{Institute of Astronomy, KU Leuven, Celestijnenlaan 200D, B-3001 Leuven, Belgium}

\author{Thomas P. Robitaille}
\affiliation{Aperio Software, Headingley Enterprise and Arts Centre, Bennett Road, Headingley, Leeds, LS6 3HN, UK}

\received{\small 7 July 2019}
\accepted{\small 18 April 2020}

\begin{abstract}
% 250 word limit
We present a detailed analysis of the disk and circumstellar environment of the forming O-type star AFGL\,4176 mm1, placing results from the Atacama Large Millimeter/submillimeter Array (ALMA) into context with multiwavelength data. With ALMA, we detect seventeen 1.2\,mm continuum sources within 5$''$ (21,000\,au) of AFGL\,4176 mm1. We find that mm1 has a spectral index of 3.4$\pm$0.2 across the ALMA band, with $>$87\% of its 1.2\,mm continuum emission from dust. The source mm2, projected 4200\,au from mm1, may be a companion or a blueshifted knot in a jet. We also explore the morphological differences between the molecular lines detected with ALMA, finding 203 lines from 25 molecules, which we categorize into several morphological types. Our results show that AFGL\,4176 mm1 provides an example of a forming O-star with a large and chemically complex disk, which is mainly traced by nitrogen-bearing molecules. Lines that show strong emission on the blueshifted side of the disk are predominantly oxygen-bearing, which we suggest are tracing a disk accretion shock. The molecules C$^{34}$S, H$_2$CS and CH$_{3}$CCN trace a slow wide-angle wind or dense structures in the outflow cavity walls. With the Australia Telescope Compact Array (ATCA), we detect a compact continuum source ($<$2000 $\times$ 760\,au) at 1.2\,cm, associated with mm1, of which $>$96\% is from ionized gas. The ATCA NH$_3$(1,1) and (2,2) emission traces a large-scale (r$\sim$0.5\,pc) rotating toroid with the disk source mm1 in the blueshifted part of this structure offset to the northwest.

\end{abstract}




%\tableofcontents

\section{Introduction} \label{sec:intro}
Long before they were directly observed, disks were predicted to exist around forming stars of all masses, based simply on the fact that a collapsing cloud of gas will flatten perpendicular to the cloud's axis of rotation due to conservation of angular momentum. However, it was not until the 1980s that disks around young low-mass ($<$2\,M$_{\odot}$) stars began to be observed \citep[e.g.][]{beckwith84a, sargent87a}. Led by advances made by the Hubble Space Telescope, which revealed protoplanetary disks in silhouette or scattered light at optical and near-infrared wavelengths \citep[e.g.][]{mccaughrean96a}, many disks around low-mass stars have been observed to-date, and a host of statistical studies of the disk populations in various nearby regions have now been published \citep[e.g.][]{ansdell16a,ansdell18a,tripathi17a,andrews18a}. The disks in these surveys, which are typically around unembedded Class II sources, have masses in the range $\sim$0.0001-0.1\,M$_{\odot}$  and radii from 10s to 100s of au. 

In comparison, the existence of disks around the young but much more massive B-type stars (equivalent to 8-16\,M$_{\odot}$ zero-age main-sequence stars) was not confirmed until much later \citep[e.g.][]{cesaroni05b,patel05a}. This was caused by observational difficulties due to their relative rarity and thus larger distances, and the fact that massive forming stars are often observed in highly clustered environments, which, combined with a lack of sufficient angular resolution, made the study of these objects challenging. This was followed recently by the first discoveries of disks around O-type stars \citep[][]{kraus10a,jimenez-serra12a,wang12, johnston15a,ilee16a,cesaroni17a, zapata19a, sanna19a}. As disks provide a conduit for accretion that has a small solid angle as viewed from the star, they are a particularly essential ingredient in the theory of massive star formation, as they are needed to overcome the large amount of radiation pressure that, in the case of spherical accretion, would impede the accretion of material onto massive stars, and thus halt their formation \citep[e.g.][]{larson71,kahn74, nakano89a}. Therefore, the recent detection of disks around embedded massive stars confirmed observationally that this was a viable solution to this problem.

Given that there seem to be no unembedded disks around high-mass stars \citep[][]{williams11a}, low-mass embedded Class 0 and I sources are the more likely analogs of observed forming high-mass stars, and are therefore the objects to which we should compare high-mass disk properties. In the case of low-mass Class 0 and I sources, the number of studied objects is also comparatively small, with many studies only observing one object \citep[e.g.][]{jorgensen09, jorgensen16a, enoch11a, tobin12b, murillo13b, murillo18a, aso15a, tobin15a, yen15a, segura-cox16a, lee17a, tobin19a}. These disks are found to have a large range of masses \citep[$\sim$0.001-1\,M$_{\odot}$, e.g.][]{looney03, enoch11a, harsono14a, segura-cox16a} and disk radii \citep[$<$10\,au to 100s of au, e.g.][]{choi10a, murillo13b, aso15a, yen15a, lee16a}. Recent observations of disks around forming high-mass stars have also uncovered a large range of disk properties, including large \citep[$\sim$1000\,au,][]{johnston15a,ilee16a}, small \citep[][]{ csengeri18a, ginsburg18a, girart18a, maud19a} and non-existent disks \citep{ginsburg17a, silva17a, beuther19a}.

One explanation for the large range of observed disk properties may lie in the highly clustered environments in which young stars form \citep{lada03}. Interactions between protostars, sometimes in binary or multiple systems, should truncate and shape disks \citep[e.g.][]{bonnell03b,bate18a}. These effects will be more prominent for high-mass stars, which are more likely to form in binary systems \citep[e.g.][]{sana14a, kraus17a, koumpia19a, zhang19a}.

Although the study of chemistry in disks around low-mass stars has begun in earnest \citep[e.g.][]{henning13a, dutrey14a}, current knowledge about chemistry in disks around massive stars is far more limited, as little has been done on the theory and modeling of their chemistry and only a handful of forming massive stars have resolved observations of several molecules within their disks \citep[e.g.,][and see Section~\ref{chemHMdisk} and references within]{jimenez-serra12a,ilee16a}. However, the high densities and temperatures reached in the circumstellar environments of embedded massive stars have led to the detection of many complex molecules toward these objects \citep[i.e. hot core chemistry,][]{kurtz00a,van-der-tak00b}. It could be that each of the large numbers of complex ($>$6 atoms) but easily detected molecules observed in these sources traces a different part of the circumstellar environment and/or disk, which would prove an extremely useful toolkit in understanding the processes of disk accretion around massive stars. However, we have only begun to scratch the surface in our understanding of this field.

In this work, we present observations of AFGL\,4176 (also known as G308.918+0.123 or IRAS\,13395-6153), a massive young stellar object (MYSO) with a high luminosity \citep[$\sim10^{5}$\,L$_{\odot}$,][]{boley12}, situated at a distance of 4.2$\pm$0.9\,kpc \citep{green11a}. Although \citet{bailer-jones18a} have recently determined a distance of 3.7$^{+2.6}_{-1.6}$\,kpc for a star in close projection to AFGL\,4176, we note that the previous determination has a smaller error and is consistent within errors with the \citet{bailer-jones18a} measurement. Therefore, to be consistent with our previous work, we assume a distance of 4.2\,kpc.

The infrared source associated with AFGL\,4176 lies at the northern edge of a compact H\,\textsc{ii} region \citep[][]{caswell92,ellingsen05,shabala06} and is associated with a group of 6.7\,GHz Class II methanol masers \citep{phillips98}. It has been previously observed at 1.2\,mm with the Swedish-ESO Submillimetre Telescope (SEST) by \citet{beltran06}, who found a dense core of 0.8\,pc and 890\,M$_{\odot}$ at a distance of 4.2\,kpc. In addition, we presented initial results from ATCA NH$_3$ observations that indicated AFGL\,4176 may be embedded in a large-scale rotating toroid with a radius of $\sim$0.7\,pc \citep{johnston14a}. \citet{de-buizer09} detected shocked H$_2$ emission surrounding AFGL\,4176, possibly from knots in an outflow, and \citet{ilee13a} observed 2.3$\mu$m CO bandhead emission toward AFGL\,4176 that could be tracing the inner $\sim$10\,au of a Keplerian disk. Modeling of the spectral energy distribution (SED) and mid-IR interferometric observations was carried out by \citet{boley12,boley13a}, who found that a combination of a disk-like structure with a radius of 660\,au (inclination of 60$^{\circ}$, PA=112$^{\circ}$), as well as a spherically symmetric Gaussian halo with an FWHM of $\sim$600\,au, was able to explain the observations.

In a previous study \citep[][hereafter J15]{johnston15a}, we presented initial results and modeling of 1.2\,mm continuum and line ALMA observations of AFGL\,4176. Although there were many lines detected at 1.2\,mm, J15 presented only the results for the CH$_3$CN J=13-12 K transitions, uncovering a near-Keplerian disk with a mass of 12\,M$_{\odot}$ and a radius of 2000\,au around an O7 star. In the same paper, we also presented $^{12}$CO observations that showed the presence of a large-scale outflow perpendicular to the AFGL\,4176 mm1 disk. Using the same ALMA data, \citet{bogelund19a} published an analysis of the chemistry in AFGL 4176, with a particular focus on measuring the temperatures, column densities, and abundances of various species, deriving excitation temperatures between 120 - 320\,K for the molecules they detected. They found that the column density with respect to methanol is three times higher for O-bearing species than N-bearing species, indicating that AFGL\,4176 is likely a young source, with little evidence for significant processing by the central star. They also compared the chemistry of AFGL\,4176 mm1 to several other well-studied sources such as Orion KL and Sgr\,B2, concluding that it was most similar to the low-mass protostar IRAS\,16293-2422B, and thus there has not been significant processing of the circumstellar gas by the radiation of the central star.

In this paper, we present a detailed analysis of the circumstellar environment and disk of AFGL 4176, combining continuum and line data from ALMA with complementary ATCA and APEX observations at 1.2-1.5\,cm and 1.2\,mm, respectively. We have made these data available online\footnote{http://doi.org/10.5281/zenodo.3369188}. In Section~\ref{sec:observations}, we present the details pertaining to the observations. In Section~\ref{sec:ALMAresults}, we present the results from the ALMA data, including details of the millimeter continuum source properties and the characterization and analysis of all of the spectral lines detected in the ALMA data. In Section~\ref{sec:ATCAresults}, we present the results from the ATCA observations, including 1.2\,cm continuum and transitions of ammonia. In Section~\ref{sec:CombinedResults} we present results that utilize observations from more than one telescope: APEX and ALMA in the case of C$^{34}$S J=5-4, and ATCA and ALMA in the case of the hydrogen recombination lines H29$\alpha$, H64$\alpha$, H65$\alpha$ H67$\alpha$ and H68$\alpha$. In Section~\ref{sec:discussion} we discuss our results, including a comparison to other detected disks around O-type young stars, and a short review of how our findings fit within the current knowledge of chemistry in disks around MYSOs. We present our conclusions in Section~\ref{sec:conclusions}.

\section{Observations} \label{sec:observations}

\subsection{ALMA Observations} \label{sec:ALMAobs}

The ALMA observations of AFGL\,4176 were taken in August 2014 as a Cycle 1 Transfer project during ALMA Cycle 2. The observations are described in J15. Here we mention several further properties of the observations. The cycle time of the phase calibrator was approximately 10\,minutes, and the total time on-source across the observations was 1.37\,hr. We number each of the four observed spectral windows (spw) by increasing frequency: spw0 to spw3. Table~\ref{ALMAspws} provides further information about each of the observed ALMA spectral windows, including the frequency range covered, the imaged channel width (which is same as the observational spectral resolution), the imaged synthesized beam, and the image rms. The angular resolutions for each spw range from 0.26 to 0.30$\arcsec$, corresponding to 1100 to 1300\,au. Each of the four observed spectral windows were imaged using Briggs weighting and a robust parameter of 0.5, and the continuum was subtracted using \texttt{imcontsub} in CASA to avoid over-subtraction. The rms noise in each spectral window (given in Table \ref{ALMAspws}) was measured after removal of the line emission, which was performed by $\sigma$-clipping each cube. Except for Fig.\ref{contfig}, all figures in this paper show primary-beam-corrected images.

When imaging H29$\alpha$ and C$^{34}$S, we instead used natural weighting to recover extended emission, producing beam parameters of 0.33$\times$0.32$''$ PA=69.0$^{\circ}$  and 0.35$\times$0.30$''$ PA=-31.1$^{\circ}$, respectively. The H29$\alpha$ image had an rms noise of 1.3\,mJy\,beam$^{-1}$ in a channel width of 2.0\,km\,s$^{-1}$ (larger than the native spectral resolution of 0.33\,km\,s$^{-1}$), and the C$^{34}$S had an rms noise of 4\,mJy\,beam$^{-1}$ in a channel width of 1.4\,km\,s$^{-1}$. The ALMA and APEX C$^{34}$S images were combined using the CASA task \texttt{feather}.

\begin{table*}
\center
\small
\begin{tabular}{lllll}
\hline \hline
Spw Name & Frequency Range & Channel Width & Synthesized Beam & RMS Noise\\
& (GHz) & (km\,s$^{-1}$)& & (mJy\,beam$^{-1}$) \\  
\hline
 spw0 & 238.8376 -- 239.3064 & 0.354 & 0.30 $\times$ 0.28$''$ PA: $38.0^{\circ}$ & 3.8 \\
 spw1 & 239.6035 -- 241.4785 & 1.407 & 0.29 $\times$ 0.25$''$ PA: $-30.8^{\circ}$ & 2.0 \\
 spw2 & 253.1055 -- 254.9805 & 1.332 & 0.28 $\times$ 0.23$''$ PA: $-33.7^{\circ}$ & 2.1 \\
 spw3 & 256.1146 -- 256.5834 & 0.330 & 0.26 $\times$ 0.26$''$ PA: $-171^{\circ}$ & 3.4 \\
\hline
\end{tabular}
\caption{Summary of observed ALMA spectral windows \label{ALMAspws} }
\end{table*}

\subsection{ATCA Observations} \label{sec:ATCAobs}

We observed AFGL\,4176 with the Australia Telescope Compact Array (ATCA) under project C2646 in the 15-mm band during 2012 on April 17, September 3, and December 21. The observations were designed to target transitions of NH$_3$ and hydrogen recombination lines. Table~\ref{obs} presents a summary of the observations, including the start time, length of observation, range of baseline lengths, rms path length measured by the seeing monitor, largest angular scale (LAS), and primary beam for each of the three observing sessions. The pointing center of the observations was 13$^h$43$^m$01$^s.$9 $-62^{\circ}$08$'$55.5$''$ (J2000). Properties of the observed bands and lines on the three observing dates are given in Table~\ref{ATCAspws}. On the first two observing dates, only two continuum bands, with bandwidths of 2.112\,GHz and comprising of 33 $\times$ 64\,MHz channels, and two zoom bands, with bandwidths of 63.929\,MHz and comprising of 2049 $\times$ 31.2\,kHz channels, were available. This was expanded to two continuum bands and eight zoom bands on 2012 December 21; all of these zoom bands had a bandwidth of 63.929\,GHz and 2049 channels, apart from the 23.709 and 24.509\,GHz bands, which had bandwidths of 95.878\,MHz and 3073 $\times$ 31.2\,kHz channels.

The setup calibrator, which was used to calibrate the delays, pointing, and an initial amplitude and phase calibration, was 0537-441 in all observations, and the primary flux calibrator was 1934-638. The bandpass calibrator was 1253-055, and the phase calibrator was 1352-63, which was observed in a cycle of 1 to 3\,minutes with the target source. Calibration and imaging was carried out in the 20120419 ATNF version of MIRIAD. After flagging, the remaining continuum bandwidth was 1.8\,GHz for each band in the April and September 2012 datasets, and 1.6 and 1.5\,GHz for the 20.905 and 24.205\,GHz bands observed on December 2012. Due to poor weather on the final date of observation, we carried out self-calibration on the continuum data and applied these solutions to the line data, which improved the final images. The data were all imaged using a Briggs robust weighting parameter of 0.5. Unfortunately, none of the NH$_3$ lines observed in December 2012 were detected and therefore will not be discussed further.

\begin{table*}
\center
\begin{tabular}{ccccccc}
\hline \hline
Start of & Time & Configuration & Baseline & Seeing Monitor & LAS & Primary  \\
Observation & Observed &  & Lengths\tablenotemark{a} & Path Length rms & & Beam\tablenotemark{b} \\
 (UTC) & (hr) & & (m) & ($\mu$m) & (arcsec) & (arcmin)\\  
\hline
2012 Apr 17 08:36:14.9 & 8.79 & 1.5B & 30.6 - 4301.0 & 20 - 270 & 28.1 & 2.4 \\
2012 Sep 02 22:39:34.9 & 8.71 & 6A & 627.6 - 5938.8 & 50 - 250 & 16.6 & 2.3 \\
2012 Dec 21 14:09:34.9 & 11.38 & 1.5D & 107.1 - 4438.8 & 100 - 950 & 23.7 & 2.3 \\
\hline
\end{tabular}
\tablenotetext{a}{Baseline lengths stated are physical lengths before observation.}
\tablenotetext{b}{Calculated from the highest frequencies and thus provides the smallest primary beam for each observation.}
\caption{Summary of ATCA observations. \label{obs} }
\end{table*}

\begin{table*}
\center
\small
\begin{tabular}{llllllll}
\hline \hline
Observation & Central & Continuum & Line Rest & Imaged & Restoring & rms & Detected?\\
Date &  Frequency & or Line(s) & Frequency & Chan. Width & Beam & Noise & \\
 (UTC) &  (GHz) & & (GHz) & (km\,s$^{-1}$)& & (mJy\,bm$^{-1}$) & \\  
\hline
2012 Apr 17  & 20.434 & Continuum & \nodata & \nodata & 1.30 $\times$ 1.03$''$ PA: 8.9$^{\circ}$ & 0.74 & Y (Fig.~\ref{ATCA_contin_Apr}) \\ 
                     & 23.712 & Continuum & \nodata & \nodata & 1.20 $\times$ 0.90$''$ PA: 7.1$^{\circ}$ & 0.59 & Y (Fig.~\ref{ATCA_contin_Apr}) \\ 
                     & 20.466 & H68$\alpha$ & 20.46177 & 5 & 3.63 $\times$ 2.58$''$ PA: 12.2$^{\circ}$ \tablenotemark{a} & 1.85 & Y (Figs.~\ref{ATCA_NH3_11}, \ref{ATCA_temp}, \ref{ATCA_H68alpha}) \\ 
                     & 23.712 & NH$_3$(1,1) & 23.69450 & 0.4 & 12.64 $\times$ 11.26$''$ PA: 15.4$^{\circ}$ \tablenotemark{b} & 6.51 & Y (Fig.~\ref{ATCA_NH3_11}) \\ 
                     & \nodata & NH$_3$(2,2) & 23.72263 & 0.4 & 12.67 $\times$ 11.19$''$ PA: 16.8$^{\circ}$ \tablenotemark{b} & 6.34 & Y \\ 
2012 Sep 2  & 24.139 & Continuum & \nodata & \nodata & 0.51 $\times$ 0.36$''$ PA: 14.4$^{\circ}$ & 0.19 & Y (Fig.~\ref{ATCA_contin}) \\ 
                    & 24.533 & Continuum & \nodata & \nodata & 0.51 $\times$ 0.35$''$ PA: 14.5$^{\circ}$ & 0.19 & Y (Fig.~\ref{ATCA_contin}) \\ 
                    & 24.139 & NH$_3$(4,4) & 24.13942 & 0.8 & 0.54 $\times$ 0.38$''$ PA: 14.8$^{\circ}$ & 1.95 & Y (Fig.~\ref{ATCA_NH3_moms}) \\ 
                    & 24.533 & NH$_3$(5,5) & 24.53299 & 0.4 & 0.53 $\times$ 0.37$''$ PA: 14.6$^{\circ}$ & 2.38 & Y (Fig.~\ref{ATCA_NH3_moms}) \\ 
2012 Dec 21 & 20.905 & Continuum & \nodata & \nodata & 0.96 $\times$ 0.83$''$ PA: -88.4$^{\circ}$ & 0.71 & Y \\ 
                     & 24.205 & Continuum & \nodata & \nodata & 0.87 $\times$ 0.77$''$ PA: -88.7$^{\circ}$ & 0.59 & Y \\ 
                     & 20.457 & H68$\alpha$ & 20.46177 & 5 & 2.88 $\times$ 2.40$''$ PA: -88.6$^{\circ}$ \tablenotemark{a} & 2.74 & Y \\ 
                     & 21.385 & H67$\alpha$ & 21.38478 & 5 & 2.83 $\times$ 2.37$''$ PA: -89.5$^{\circ}$ \tablenotemark{a} & 4.42 & Y \\ 
                     & 23.405 & H65$\alpha$ & 23.40428 & 5 & 2.84 $\times$ 2.33$''$ PA: -87.9$^{\circ}$ \tablenotemark{a} & 3.93 & Y\\ 
                     & 23.709 & NH$_3$(1,1) & 23.69450 & 0.8 & 7.30 $\times$ 6.99$''$ PA: 75.4$^{\circ}$ \tablenotemark{c} & 14.25 & N \\ 
                     & \nodata & NH$_3$(2,2) & 23.72263 & 0.8 & 7.25 $\times$ 6.99$''$ PA: 74.7$^{\circ}$ \tablenotemark{c} & 14.17 & N \\ 
                     & 23.885 & NH$_3$(3,3) & 23.87013 & 0.8 & 7.20 $\times$ 7.09$''$ PA: 32.7$^{\circ}$ \tablenotemark{c} & 14.55 & N \\ 
                     & 24.141 & NH$_3$(4,4) & 24.13942 & 0.8 & 0.91 $\times$ 0.84$''$ PA: 86.2$^{\circ}$ & 7.08 & N \\ 
                     & 24.509 & NH$_3$(5,5) & 24.53299 & 0.8 & 0.90 $\times$ 0.83$''$ PA: 87.2$^{\circ}$ & 6.48 & N \\ 
                     & \nodata & H64$\alpha$  & 24.50990 & 5 & 2.86 $\times$ 2.32$''$ PA: -87.4$^{\circ}$ \tablenotemark{a} & 3.18 & Y\\ 
                     & 25.069 & NH$_3$(6,6) & 25.05603 & 0.8 & 0.88 $\times$ 0.81$''$ PA: 85.4$^{\circ}$ & 6.30 & N \\ 
\hline
\end{tabular}
\tablecomments{Notes in the final column indicate the figures in which images of the observed band are shown.}
\tablenotetext{a}{A gaussian taper of 2$''$ was applied to the data during imaging}
\tablenotetext{b}{A gaussian taper of 10$''$ was applied to the data during imaging}
\tablenotetext{c}{A gaussian taper of 5$''$ was applied to the data during imaging}
\caption{Summary of observed ATCA Bands \label{ATCAspws} }
\end{table*}

\vspace{2cm}
\subsection{APEX Observations} \label{sec:ALMAobs}

We observed AFGL\,4176 with the Atacama Pathfinder Experiment (APEX)\footnote{APEX is a collaboration between the Max-Planck-Institut f\"ur Radioastronomie, the European Southern Observatory, and the Onsala Space Observatory.} 12\,m antenna under program M0007\_97 on 2016 April 23 and 26. The precipitable water vapor on these dates was 2.0-2.4\,mm and 2.8-3.8\,mm, respectively. The pointing center of the observations was 13$^h$43$^m$01$^s.$7 $-$62$^{\circ}$08$'$51.3$''$ (J2000). We used the Swedish Heterodyne Facility Instrument (SHeFI) APEX-1 receiver to observe C$^{34}$S(J=5-4) at 241.016\,GHz in the lower sideband, which covered frequencies between 240 and 244\,GHz with a channel width of 76.3\,kHz. The corresponding beam size at this frequency is 25.9$''$. On-the-fly maps of 2$' \times$2$'$ extent were taken in two perpendicular scan directions to reduce scanning artifacts. Pointing calibration was performed every 1\,hr, the total observing time was 8.8\,hr and the total time on-source was 5.3\,hr. Data reduction was performed within GILDAS/CLASS\footnote{http://www.iram.fr/IRAMFR/GILDAS}. The achieved rms in the final C$^{34}$S(J=5-4) cube in the central part of the map is 66\,mK in a channel width of 0.28\,km\,s$^{-1}$ (three times the native resolution).

\section{ALMA Results} \label{sec:ALMAresults}

\subsection{ALMA Millimeter Continuum} \label{sec:ALMAcont}

\subsubsection{Millimeter Continuum Source Properties} \label{sec:cont_src_prop}
\begin{figure}
\epsscale{1.2}
\plotone{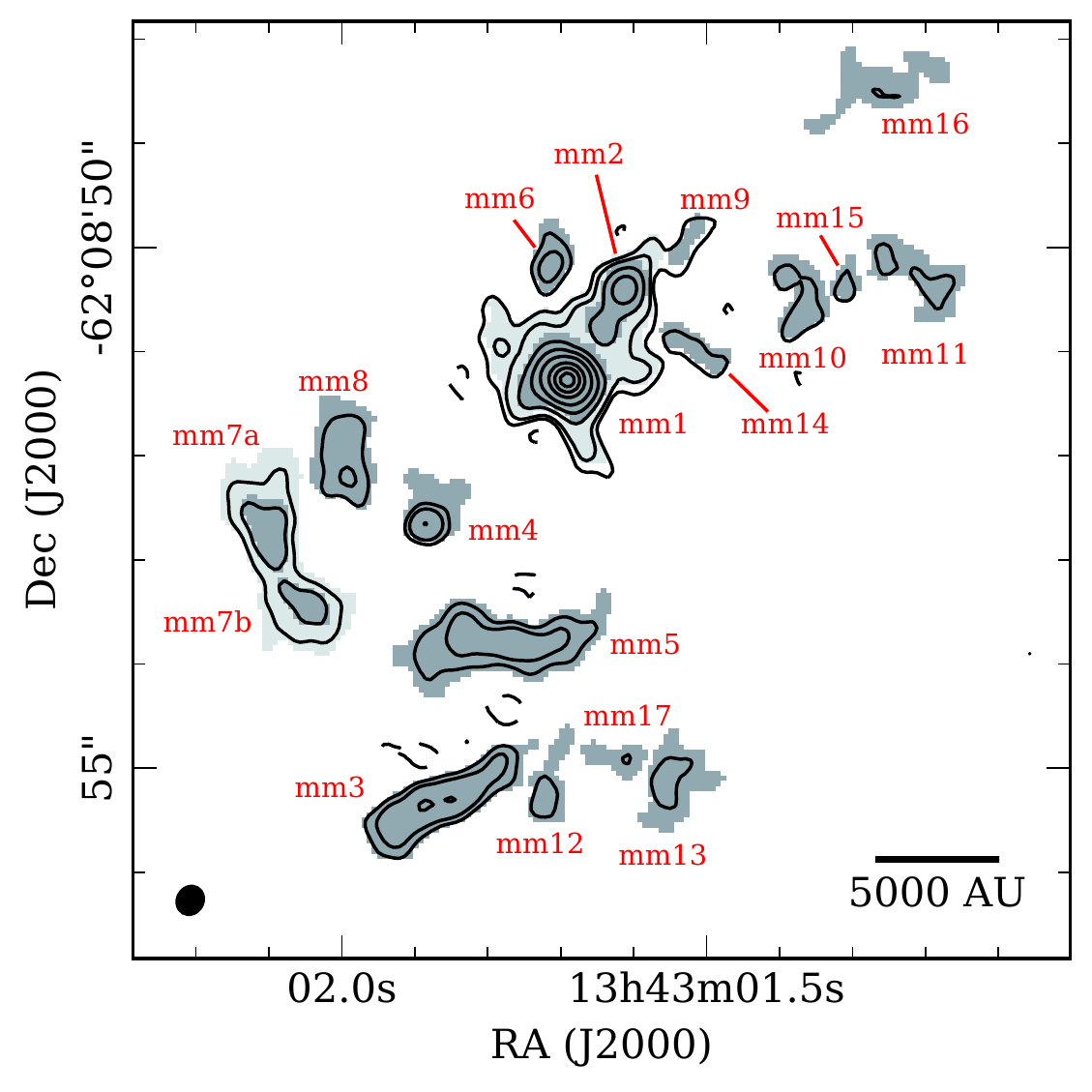}
\caption{Continuum emission toward AFGL\,4176 at 1.21\,mm observed with ALMA shown in contours (non-primary-beam corrected, $\sigma$ = 78$\mu$Jy\,beam$^{-1} \times$ -5, 5, 10, 25, 50, 100, 200, 300, 400). The area of each detected ALMA 1.21\,mm source is shown in dark gray, as well as combined sources mm1 branch (containing mm1 and mm2) and mm7 in light gray (see Section~\ref{sec:cont_src_prop} for details). The 1.21\,mm sources, which are listed in Table~\ref{dendro_contin_table},  are labeled in red. The synthesized beam, with size 0.28$\times$0.24$''$, PA = -30.2$^{\circ}$, is shown in the bottom left corner. \label{contfig}}
\end{figure}

In this section, we determine the observed properties of the 1.2\,mm continuum sources detected with ALMA, as initially presented in Figure 1 of J15. In J15, only the observed properties of the brightest source mm1 were given; here we present the properties of all of the continuum sources within 5$''$ of mm1. We measured the positions and flux densities of these sources using the \texttt{astrodendro} software\footnote{http://www.dendrograms.org}. In this process, we used the non-primary-beam-corrected image and set the peak flux required for a detection to be 5$\sigma$, as the noise across this image is constant. The measured properties for these identified sources were then extracted from the primary-beam-corrected image. The flux limit above which flux densities are measured in the primary-beam-corrected image is 2$\sigma$ in the non-primary-beam-corrected image, and the required flux difference between embedded structures to be assigned as a separate source is 1$\sigma$. Table~\ref{dendro_contin_table} provides the properties of these sources, which include source name, peak position, peak flux $S_{\rm peak}$, integrated flux $S_{\rm int}$, mass $M$, peak column density $N_{H_2}$, and the geometric mean diameter $D$, calculated from the area covered by each dendrogram structure. Figure~\ref{contfig} presents the non-primary-beam-corrected ALMA 1.21\,mm continuum emission from AFGL\,4176 as contours; the sources are labeled with the names given in Table~\ref{dendro_contin_table}, and the area of each source shown in grayscale. We detect 17 sources in the field. The peak and integrated flux densities measured for mm1 are slightly larger than that quoted in J15, which was determined using a Gaussian fit. This is mainly because the Gaussian source lies on a background of more diffuse emission, which was not included in the Gaussian fit, but is included when the source fluxes are determined using \texttt{astrodendro}, which assigns groups of pixels to a source. The uncertainties on the flux densities given in Table~\ref{dendro_contin_table} only give the random uncertainties, whereas an absolute flux calibration error of 10-20\% is also expected.

The mass $M$ and peak column density $N_{H_2}$ for each source were determined using,

\begin{equation} M = \frac{g S_{\rm int} d^2}{B(\nu,T) \kappa_{\nu}} \end{equation}

and 

\begin{equation} N_{H_2} = \frac{g S_{\rm peak}}{2.8 m_H  \Omega B(\nu,T) \kappa_{\nu}}, \end{equation}

\noindent where $g$ is the gas-to-dust ratio, $d$ is the distance, $B(\nu,T)$ is the blackbody function (a function of frequency $\nu$ and temperature $T$), $\kappa_{\nu}$ is the frequency-dependent opacity, 2.8 is the mean molecular weight, $m_H$ is the mass of a hydrogen atom, and $\Omega$ is the ALMA continuum beam area. As in J15, we assume a distance of 4.2\,kpc. We assume a gas-to-dust ratio measured for the solar neighborhood of 154 \citep{draine11} and a temperature of 100\,K, except for mm1, where we use a temperature of 190\,K, which was the average temperature of the disk derived from modeling CH$_3$CN line emission using CASSIS\footnote{CASSIS is developed by IRAP-UPS/CNRS (http://cassis.irap.omp.eu)} (J15).  To be consistent with J15, the dust opacity at 1.21\,mm was assumed to be 0.24\,cm$^2$\,g$^{-1}$, which was determined from the \citet[][]{draine03a,draine03b} Milky Way dust model with $R_V=5.5$. This is a model that fits the dust opacities for the ISM in star forming regions well, and in the rest of this paper, we use this model when referring to ISM dust. In comparison, the \citeauthor{ossenkopf94} (\citeyear{ossenkopf94}, hereafter OH94) dust opacity at 1.21\,mm for dust with thin ice mantles, subject to grain growth and with densities of $n$=10$^{6}$\,cm$^{-3}$ is 1\,cm$^2$\,g$^{-1}$, which if used instead would reduce the masses and column densities by a factor of $\sim$4. We could not use this opacity for the radiative-transfer dust-continuum modeling presented in J15, because the OH94 dust models do not have the associated scattering properties required as input to the radiative-transfer dust code \textsc{Hyperion} \citep{robitaille11}. If dust grain growth has occurred in AFGL\,4176, then we would expect higher opacities than typical ISM values. The higher densities associated with massive star formation would lead to fast grain growth, but massive stars have short formation timescales \citep[the lifetime of the MYSO phase is $\sim1-3\times10^5$yr,][]{kuiper10a, kuiper15a, kuiper16a, davies11a, kuiper18a}. \citet{boley13a} found evidence for an enhanced population of 1.5\,$\mu$m grains in AFGL\,4176, suggesting grain growth has started, but this would not be enough to change the millimeter opacities significantly.

\begin{table*}
\center
\small
\begin{tabular}{lllllll}
\hline \hline \\
Source & Peak Position & S$_{\rm peak}$ & S$_{\rm int}$ & Mass\tablenotemark{a} & Column Density\tablenotemark{a} & Diameter \\
name & (J2000) & (mJy\,beam$^{-1}$) & (mJy) & (M${_\odot}$) & (cm$^{-2}$) & (au) \\
\hline
mm1 & 13:43:01.693 -62:08:51.25 &     38.49 $\pm$  0.08 &     61.22 $\pm$  0.04 &      9.4 \tablenotemark{b} & 8.3$\times 10^{24}$ \tablenotemark{b} & 3500  \\  
mm1 branch\tablenotemark{c}& 13:43:01.693 -62:08:51.25 &     38.49 $\pm$  0.08  &     86.21 $\pm$  0.07 &     13.3 \tablenotemark{b} & 8.3$\times 10^{24}$ \tablenotemark{b} & 6500  \\ 
mm2 & 13:43:01.614 -62:08:50.40 &      6.79 $\pm$  0.08  &     11.87 $\pm$  0.03 &      3.6 & 2.9$\times 10^{24}$ & 2600  \\ 
mm3\tablenotemark{d} & 13:43:01.886 -62:08:55.35 &      2.07 $\pm$  0.08  &      9.41 $\pm$  0.05 &      2.8 & 8.7$\times 10^{23}$ & 4200  \\  
mm4 & 13:43:01.886 -62:08:52.65 &      2.05 $\pm$  0.08  &      2.38$\pm$  0.03 &      0.7 & 8.7$\times 10^{23}$ & 2700  \\ 
mm5 & 13:43:01.764 -62:08:53.80 &      1.54 $\pm$  0.08  &      9.03 $\pm$  0.05 &      2.7 & 6.5$\times 10^{23}$ & 5100  \\ 
mm6 & 13:43:01.714 -62:08:50.20 &      1.30 $\pm$  0.08 &      1.80 $\pm$  0.02 &      0.5 & 5.5$\times 10^{23}$ & 2100 \\ 
mm7 & 13:43:02.100 -62:08:52.65 &      1.21 $\pm$  0.08 &      9.68 $\pm$  0.06 &      2.9 & 5.1$\times 10^{23}$ & 5400  \\ 
mm7a & 13:43:02.100 -62:08:52.65 &      1.21 $\pm$  0.08 &      2.76 $\pm$  0.02 &      0.8 & 5.1$\times 10^{23}$ & 2200  \\ 
mm7b & 13:43:02.042 -62:08:53.40 &      1.06 $\pm$  0.08 &      1.61 $\pm$  0.02 &      0.5 & 4.5$\times 10^{23}$ & 1800  \\
mm8 & 13:43:01.993 -62:08:52.20 &      0.94 $\pm$  0.08 &      3.36 $\pm$  0.04 &      1.0 & 4.0$\times 10^{23}$ & 3500  \\ 
mm9 & 13:43:01.536 -62:08:50.10 &      0.71 $\pm$  0.08 &      0.71 $\pm$  0.08 &      0.2 & 3.0$\times 10^{23}$ &1300 \\ 
mm10 & 13:43:01.365 -62:08:50.65 &      0.66 $\pm$  0.08 &      1.88 $\pm$  0.03 &      0.6 & 2.8$\times 10^{23}$ & 2600  \\ 
mm11 & 13:43:01.179 -62:08:50.40 &      0.65 $\pm$  0.08 &      2.18 $\pm$  0.03 &      0.7 & 2.7$\times 10^{23}$ & 2900  \\ 
mm12\tablenotemark{d} & 13:43:01.721 -62:08:55.30 &      0.62 $\pm$  0.08  &      1.13 $\pm$  0.02 &      0.3 & 2.6$\times 10^{23}$ & 2300  \\ 
mm13\tablenotemark{d} & 13:43:01.557 -62:08:55.15 &      0.60 $\pm$  0.08  &      1.84 $\pm$  0.03 &      0.6 & 2.5$\times 10^{23}$ & 3100  \\ 
mm14 & 13:43:01.514 -62:08:51.00 &      0.60 $\pm$  0.08 &      1.02 $\pm$  0.02 &      0.3 & 2.5$\times 10^{23}$ & 1900  \\ 
mm15 & 13:43:01.308 -62:08:50.40 &      0.57 $\pm$  0.08  &      0.57 $\pm$  0.02 &      0.2 & 2.4$\times 10^{23}$ &1400 \\ 
mm16 & 13:43:01.265 -62:08:48.50 &      0.53 $\pm$  0.08  &      1.79 $\pm$  0.03 &      0.5 & 2.2$\times 10^{23}$ & 3100  \\ 
mm17\tablenotemark{d} & 13:43:01.607 -62:08:54.90 &      0.41 $\pm$  0.08  &      0.57 $\pm$  0.02 &      0.2 & 1.7$\times 10^{23}$ & 1800  \\ 

\hline
\end{tabular}
\tablenotetext{a}{The masses and column densities are derived using opacities from \citet{draine03a,draine03b}. To obtain masses and column densities using opacities from OH94 for dust with thin ice mantles at densities of $n$=10$^{6}$\,cm$^{-3}$, the values should be divided by $\sim4$.}
\tablenotetext{b}{A temperature of 190\,K was assumed, as determined in J15.}
\tablenotetext{c}{mm1 branch includes sources mm1 and mm2.}
\tablenotetext{d}{As discussed in Section~\ref{sec:cont_src_prop}, these sources are likely to be dominated by free-free emission from ionized gas.}
\caption{Measured and calculated properties of the ALMA 1.2\,mm continuum sources. \label{dendro_contin_table}}
\end{table*}

The sources mm3, mm12, mm13, and mm17 are likely dominated by free-free emission (see Section~\ref{sec:cmcont}); thus, their calculated masses and column densities should be viewed with caution.

\begin{figure}
\epsscale{1.2}
\plotone{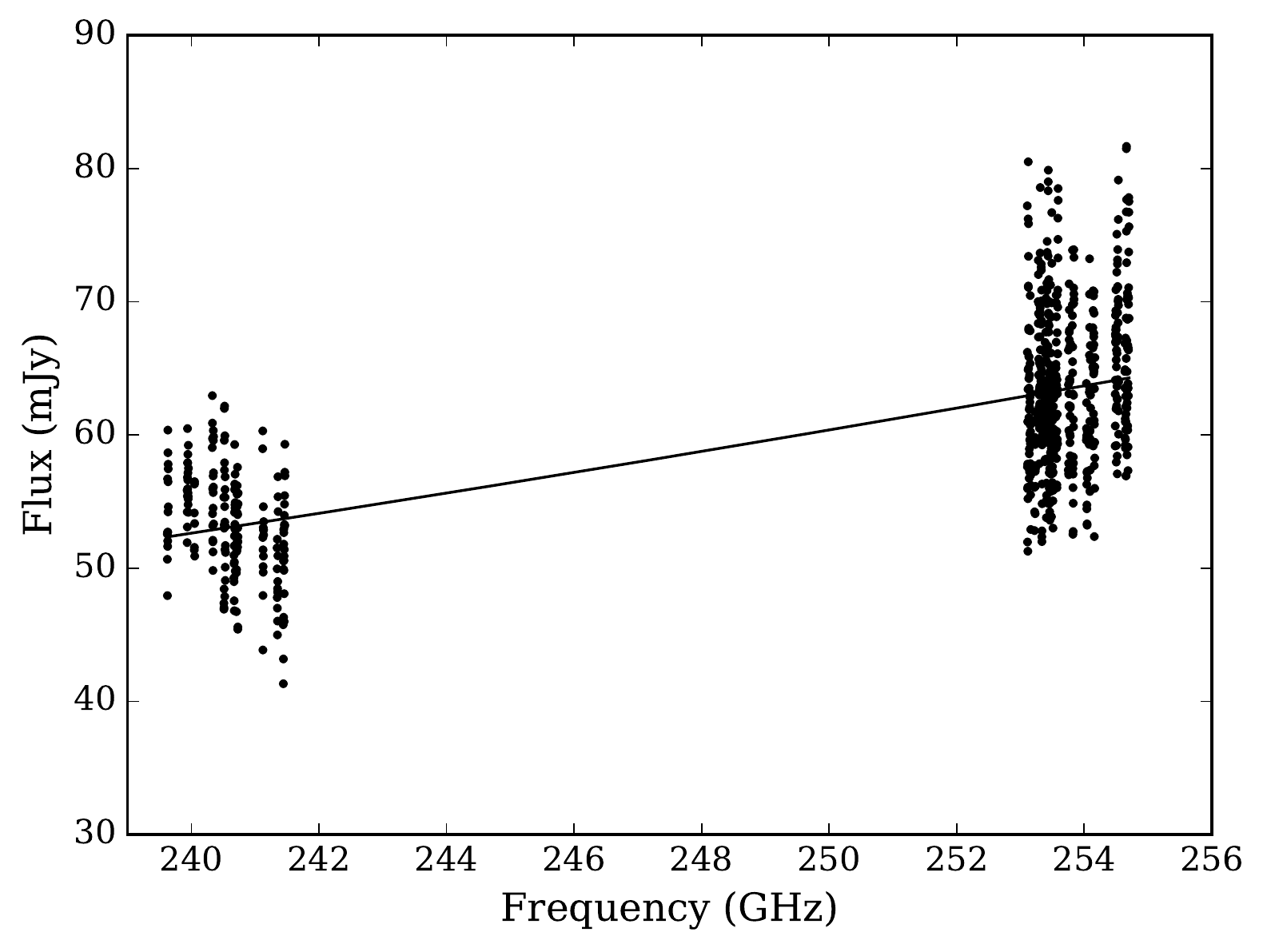}
\caption{Spectrum of mm1 constructed from the continuum-only channels in the ALMA wide spws (spw1 and spw2). Each channel is shown as a black point. The black solid line shows the best-fitting spectral index of 3.36$\pm$0.14. \label{ALMAspecindex}}
\end{figure}

We determined the spectral index of mm1 between spw1 and spw2 (between 240.541 and 254.043\,GHz) by fitting the continuum data in these spectral windows. We applied the phase self-calibration (i.e. no amplitude self-calibration) derived from the continuum data to the wide spectral windows spw1 and spw2 and selected the channels without line emission to estimate the continuum. Spectral windows spw0 and spw3 were not used, as they contained only a small number of line-free channels. We then found the integrated flux in each line-free channel within a 1\arcsec\, diameter aperture around the mm1 peak, and subsequently fitted these continuum channel fluxes by a power law, shown in Fig.~\ref{ALMAspecindex}. The resulting spectral index is 3.36$\pm$0.14, where the uncertainty is determined from the scatter in the data. To estimate other sources of uncertainty in our determination of the spectral index, we compared the spectral index derived during our calibration for our phase calibrator J1308-6707 (found to be $-0.60$ between spw1 and spw2) to that derived from the fluxes in the ALMA Calibrator Source Catalog\footnote{https://almascience.eso.org/sc/} ($-0.48$ between 91.5 and 343.5\,GHz with both frequencies observed on the same date), which provided an estimate of the uncertainty: $-0.12$. Thus, including the contribution from the scatter in the fit, the combined error in the spectral index is $\sim$0.2. To confirm our result, we also imaged the continuum emission in the same wide spectral windows using \texttt{clean} and \texttt{nterms=2} and obtained a similar average spectral index of 3.35 over pixels within 0.15$''$ of the continuum peak (where the estimated error in the spectral index in these pixels was $<$2).

The measured spectral index of 3.4$\pm$0.2 can be explained by dust emission, with a contribution of $\nu^2$ from the Rayleigh-Jeans tail of the Planck function (expected at 1.21\,mm for dust hotter than 30\,K) and the remainder from the frequency dependence of the opacity $\nu^{\beta}$. In this case, we would infer $\beta=1.4\pm0.2$. ISM dust has values of $\beta\sim1.6$ \citep{draine03a,draine03b}, which is within 1$\sigma$ of our measured spectral index. In the following section we assess the contribution to mm1 from free-free emission at 1.21\,mm. 

\subsubsection{Contribution of Free-free Emission to mm1} \label{sec:freefreecontrib}
In Section~\ref{sec:cmcont}, we find centimeter continuum emission associated with mm1, which cannot be explained by dust emission alone. Therefore, the millimeter continuum emission from mm1 may also contain free-free emission. The spectral index estimated between 1.21\,mm and 1.23\,cm is 1.56$\pm$0.15 (Section~\ref{sec:cmcont}), which might be explained by almost-optically thick free-free emission with the same spectral index. Assuming a hypercompact HII region with a size of 2000\,au is producing the centimeter continuum emission (Section~\ref{sec:cmcont}), and the temperature of the HII region is 10$^4$\,K, the electron density would have to be $\gtrsim~5\times10^6$\,cm$^{-3}$ to ensure that the turnover frequency was above 24.328\,GHz \citep[corresponding to 1.23\,cm;][]{mezger67}. Although high, this would in theory be possible, as the hydrogen molecular number density derived from molecular line fitting is $>8\times10^7$\,cm$^{-3}$ (J15), which is larger than this value. However, the spectral index at 1.21\,mm is 3.4$\pm$0.2, indicating dust emission is present, and the morphology and extent of the emission at the two wavelengths is dissimilar, implying that the emission at each wavelength is not tracing the same material. Therefore, it is likely that both ionized gas and dust emission contribute to the continuum emission at both wavelengths, but that dust emission dominates at 1.21\,mm and ionized gas at 1.23\,cm.

We now derive the fraction of ALMA continuum emission due to dust. When combining the contribution from dust and ionized gas, the total flux density $F_{\nu}$ at a given frequency $\nu$ within the ALMA spws is given by
\begin{equation}
F_{\nu} = F_{\nu_0} \left[ f \left( \frac{\nu}{\nu_0}\right)^{\alpha_{d}} + (1-f)\left(\frac{\nu}{\nu_0}\right)^{\alpha_{\rm ff}} \right],
\label{eqnFnu}
\end{equation}

\noindent where $F_{\nu_0}$ is the total flux density at ${\nu_0}$, $f$ is the fraction of ALMA continuum emission due to dust at $\nu_0$, and $\alpha_{d}$ and $\alpha_{\rm ff}$ are the spectral indices of the dust and ionized gas emission, respectively.

Starting from Equation~(\ref{eqnFnu}), the spectral index can be expressed as

\begin{eqnarray}
\alpha_{\rm obs} &&= \frac{d \log_{10}{F_{\nu}}}{d \log_{10}{\nu}} \\
&&= \frac{F_{\nu_0}}{F_{\nu}}  \left[ f \alpha_{d} {\left(\frac{\nu}{\nu_0}\right)}^{\alpha_d} + (1-f)\alpha_{\rm ff} {\left(\frac{\nu}{\nu_0}\right)}^{\alpha_{\rm ff}} \right] 
\end{eqnarray}

As the ALMA spws cover a small range of frequencies, we assume that the observed spectral index $\alpha_{\rm obs}$ across these spws is constant. Therefore, by evaluating $\alpha_{\rm obs}$ at $\nu_0$, we find
\begin{equation}
\alpha_{\rm obs}  = f \alpha_{d} + (1-f) \alpha_{\rm ff}.
\end{equation}
and thus,
\begin{equation}
f = \frac{\alpha_{\rm obs} - \alpha_{\rm ff}}{\alpha_{d} - \alpha_{\rm ff}}.
\end{equation}

Assuming an upper limit for the spectral index of dust emission of 3.6 (derived from the dust model used in J15), an observed (i.e. combined) spectral index $\alpha_{\rm obs}$ of 3.4$\pm$0.2, and an upper limit to the ionized gas spectral index at 1.21\,mm of 2, we find a lower limit to the dust emission fraction $f$ of 87$\pm$13\%. Thus, dust emission dominates at 1.21\,mm, even in the case of the largest possible contribution from free-free emission.

As we do not know the exact contribution to the emission from ionized gas at 1.21\,mm, and because the relative astrometry is not sufficiently accurate and the lower resolution of the ATCA data precludes our knowledge of the morphology of the 1.23\,cm emission, we do not attempt to remove the free-free contribution from the ALMA continuum data. However, we have determined it only contributes $<$13\% and thus the majority of the emission arises from dust. 

\subsubsection{Comparison to Maser Positions} \label{sec:cont_masers}
\begin{figure}
\epsscale{1.2}
\plotone{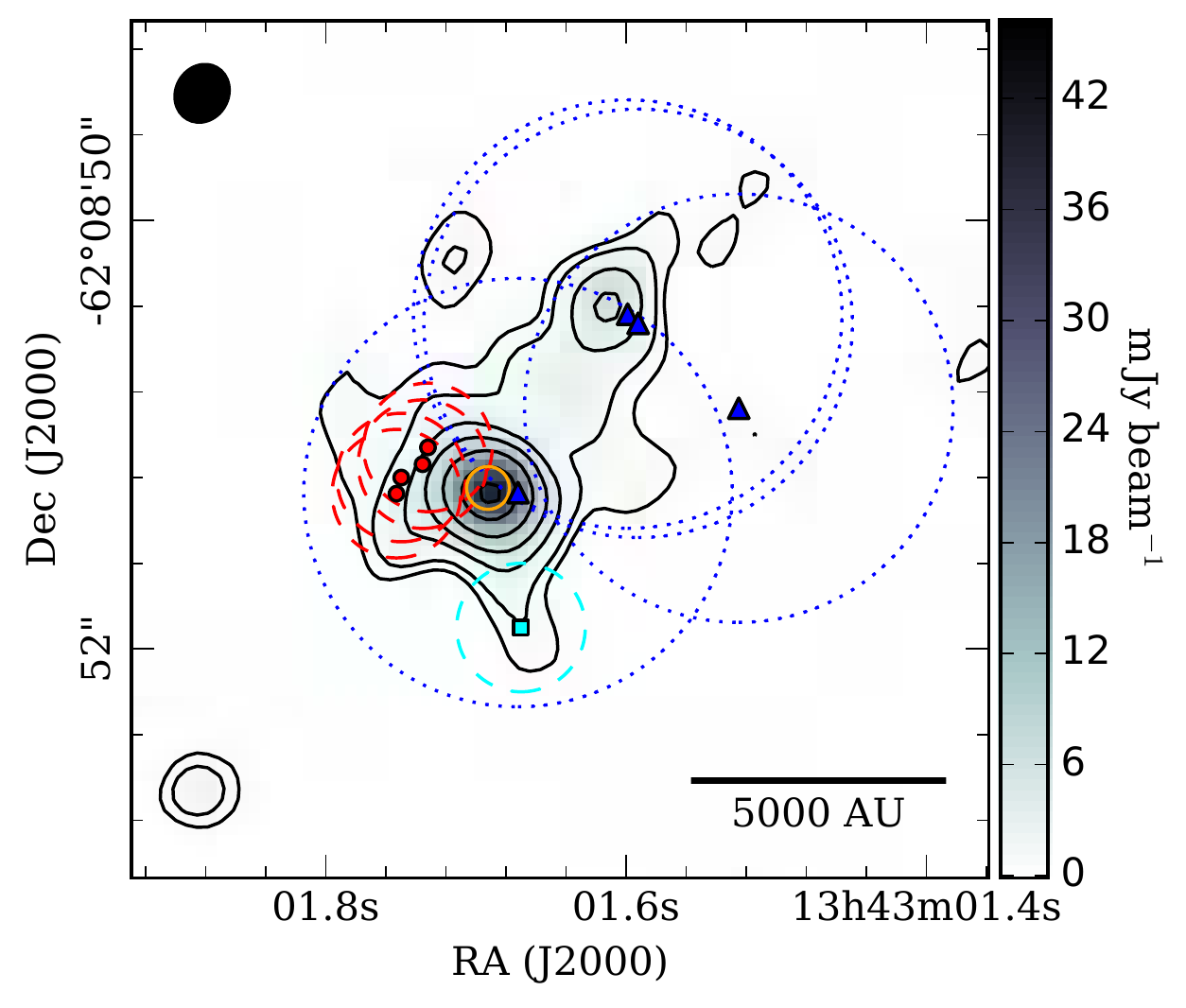}
\caption{Continuum emission toward AFGL\,4176 at 1.21\,mm observed with ALMA shown in grayscale and contours (local $\sigma$ = 0.12\,mJy\,beam$^{-1} \times$ -3, 3, 5, 10, 25, 50, 100, 200, 300, 400) with maser positions and their error circles overplotted. The red circles show Class II methanol masers from \citet{phillips98}, the cyan square shows the position of the main line ground-state OH masers detected by \citet{caswell98a}, and the blue triangles show the positions of water masers detected by \citet{walsh14b}. The colored dashed or dotted circles represent the absolute positional uncertainties on the maser positions. The ALMA beam is shown in the top left corner. The positional uncertainty of the ALMA data is shown as an orange circle at the continuum peak position. \label{masers}}
\end{figure}

Figure~\ref{masers} shows the positions of several maser species detected toward AFGL\,4176, including four Class II methanol masers from \citet{phillips98}, the position of ground-state OH masers at 1665 and 1667\,MHz detected by \citet{caswell98a}, and four water masers detected by \citet{walsh14b}. Although the spatially linear methanol maser group (red circles) appears to not be coincident with the continuum peak of mm1, the absolute positional uncertainty of the ALMA data (orange circle, $\sim$0.1$''$ or $\sim$1/3 $\times$ the beamwidth) would nearly allow the methanol maser group to be coincident with the peak of mm1. However, in Section~\ref{sec:bluedominant} we show that the thermal CH$_3$OH emission is also offset to the east of the millimeter continuum peak of mm1, so that the maser emission is probably occurring in the areas of brightest thermal CH$_3$OH emission. The OH maser (cyan square) is coincident with an extension of 5$\sigma$ emission to the southwest of mm1. There are similar but less-prominent extensions to the northwest and southeast. Given the orientation of the disk as determined by J15, OH masers could potentially be tracing the outflow cavity walls, which is similar to that seen in G35.20-0.74 \citep{de-buizer06a}. Finally, although the positions of the water masers (blue triangles) are stated to be less accurate than the other masers shown, there is one maser slightly offset to the west of mm1, and another two masers that are similarly offset from mm2. This near-coincidence and similar offset from the two brightest millimeter continuum sources in the field may indicate that the water masers are in fact coincident with these two millimeter continuum sources.

\subsubsection{Comparison to Mid-IR observations} \label{sec:MIR}
It is also illuminating to compare previous estimates of the source orientation via mid-IR observations. Using spectropolarimetry of the silicate absorption feature in AFGL\,4176, \citet{smith00a} found a polarization position angle (PA) of 55$\pm$1$^{\circ}$. Interestingly, this compares well to the disk PA determined from our 1.2\,mm ALMA continuum observations (59$\pm$17$^{\circ}$, J15). In comparison, \citet{boley12,boley13a} determined the PA of the mid-IR emission using MIDI observations at 10.6\,$\mu$m, finding a value of 112$^{\circ}$. However, as discussed in J15, this difference may be because the MIDI observations trace the heated dust in the outflow cavity instead of the disk emission itself \citep[e.g., for W33A,][]{de-wit10a}.

\vspace{2cm}
\subsection{ALMA Line Emission} \label{sec:ALMA_line}

In this Section, we present the spectra and detected lines toward AFGL\,4176, and we investigate the various categories of spectral and spatial morphology seen across the detected species.

\subsubsection{Spectra and Detected Lines} \label{sec:spectra}

\begin{table*}
\center
\begin{tabular}{cc}
\hline \hline
Group Designation & Molecules \\
\hline
O-bearing & HCOOH, CH$_2$CO, CH$_3$OH, CH$_3$CHO, CH$_3$OCHO, HCOCH$_2$OH, C$_2$H$_5$OH \\
& CH$_3$OCH$_3$, CH$_3$COCH$_3$, aGg'-(CH$_2$OH)$_2$ \tablenotemark{a} \\
N-bearing & HC$_3$N, H$_2$CCN, NH$_2$CN, CH$_3$CN, C$_2$H$_3$CN, C$_2$H$_5$CN, NS, HNCO, HC(O)NH$_2$ \\
S-bearing & C$^{34}$S, H$_2$CS, SO, SO$_2$, OCS\\
Other & CH$_3$CCH \\
\hline
\end{tabular}
\tablenotetext{a}{ aGg' conformer of ethylene glycol}
\caption{List of Molecules Detected Toward AFGL\,4176 mm1 with ALMA \label{moldetect} }
\end{table*}

\begin{figure*}
\begin{center}
\includegraphics[width=22.4cm,angle=270]{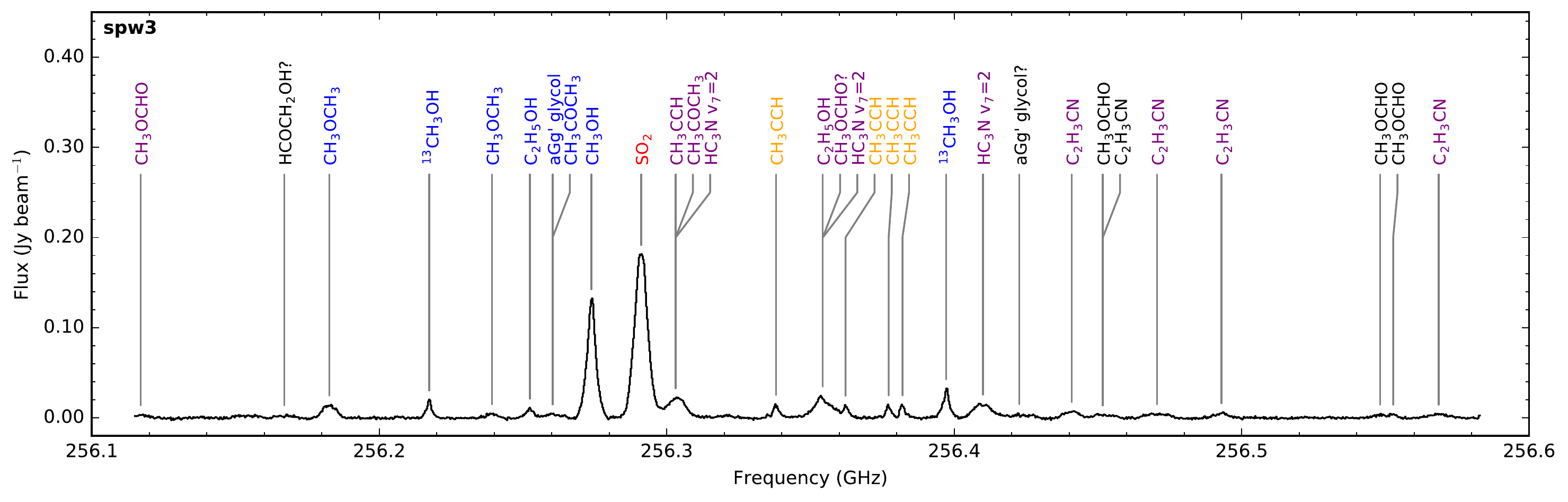}
\includegraphics[width=22cm,angle=270]{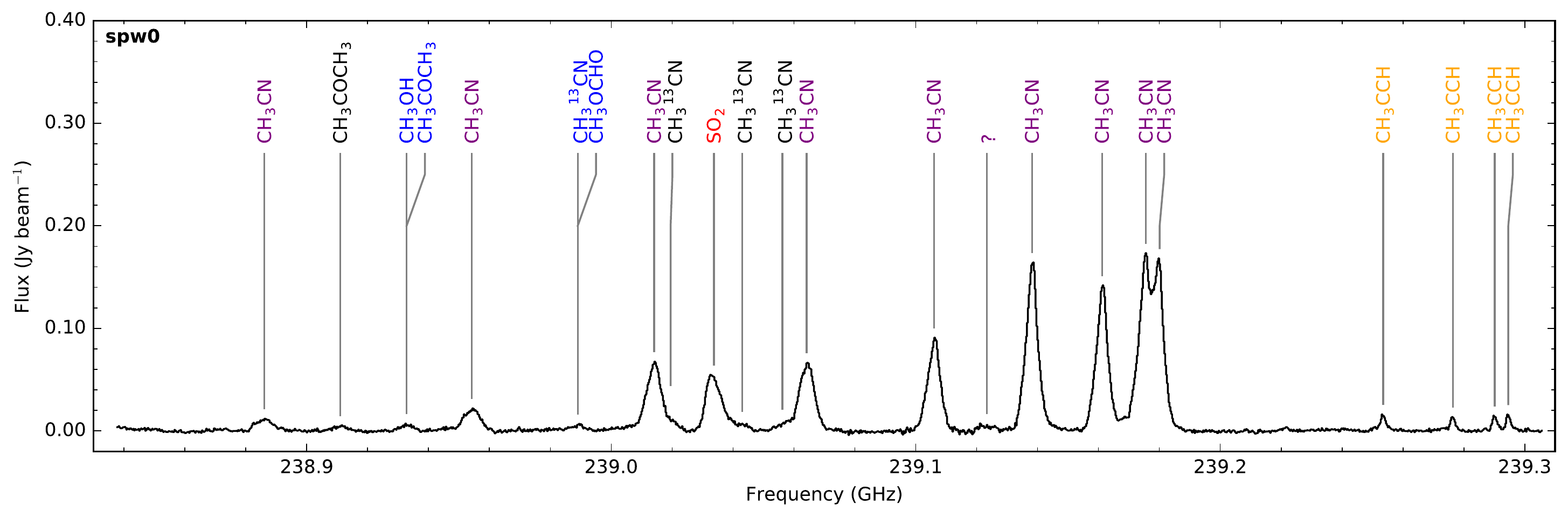}
\caption{Spectrum of the two ALMA narrow spectral windows covering 238.8376 -- 239.3064 and 256.1146 -- 256.5834\,GHz. The different line label colors denote different types of line morphology: purple is disk-tracing, blue is blue-dominant, red is red-dominant, and orange is outflow-tracing. Line labels are black when no classification was possible.\label{figspectrumnarrow}}
\end{center}
\end{figure*}

\begin{figure*}
\begin{center}
\includegraphics[width=22.0cm, angle=270]{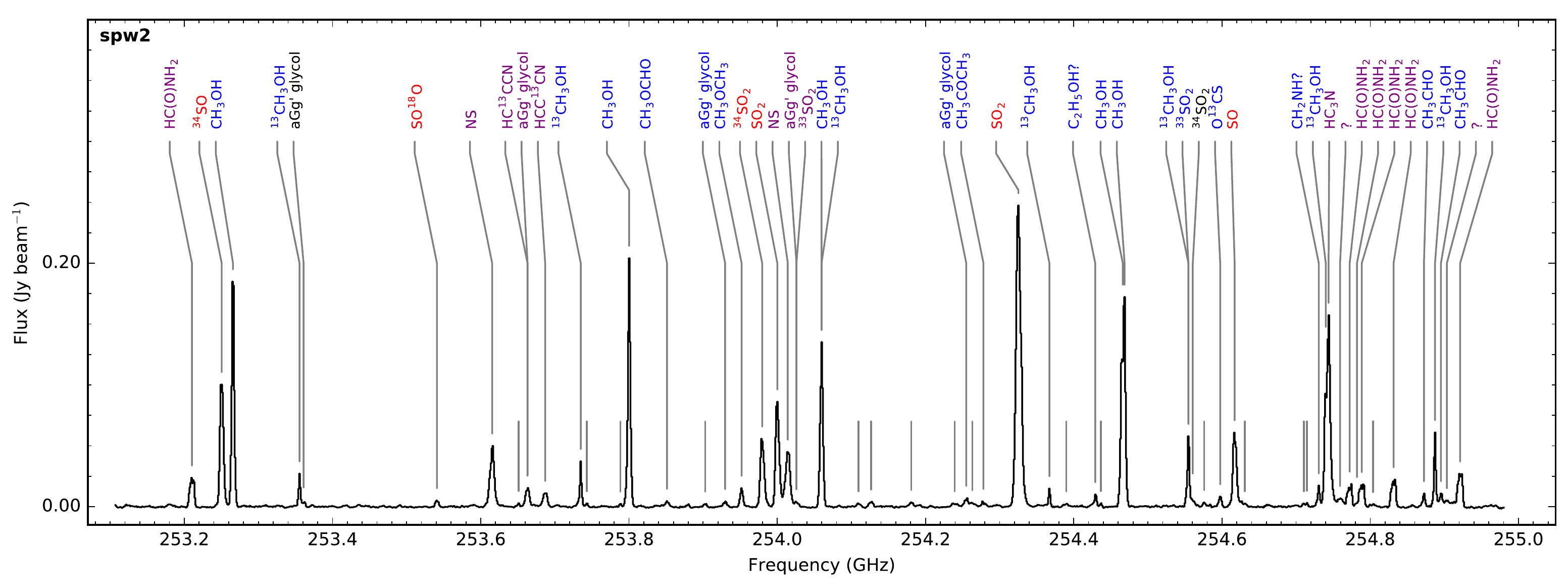}
\includegraphics[width=22.0cm, angle=270]{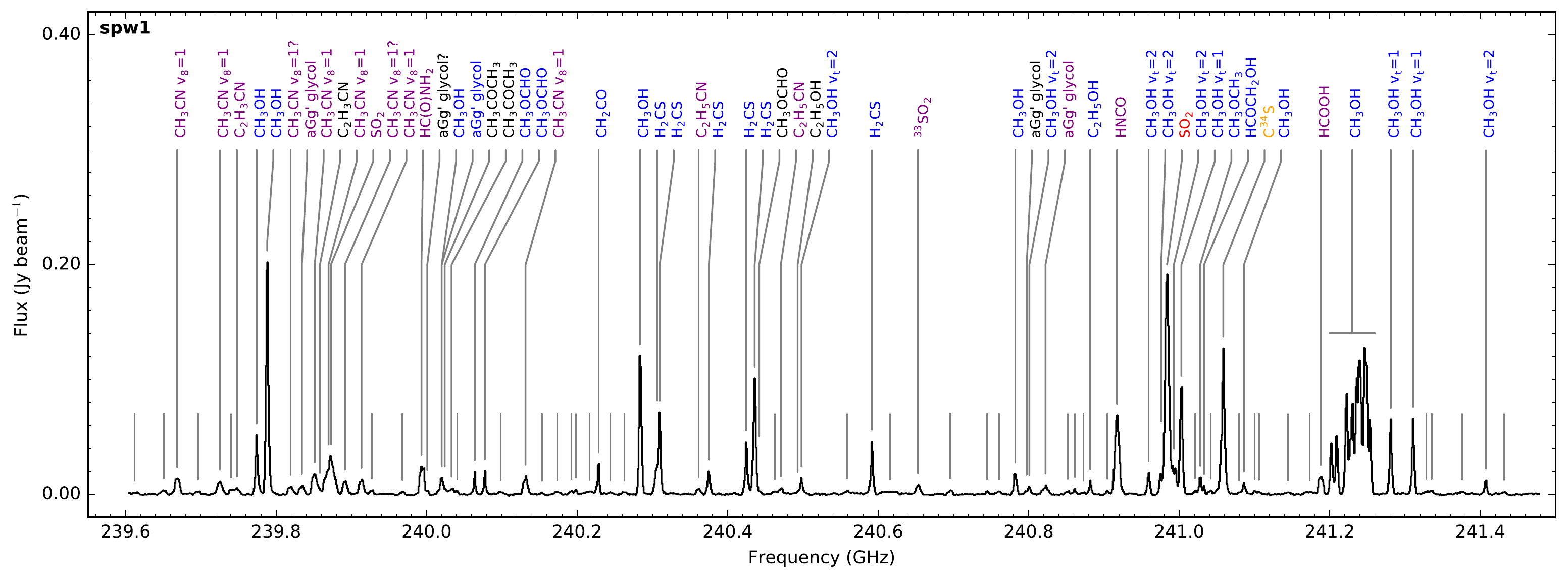}
\end{center}
\caption{Spectrum of the two ALMA wide spectral windows covering 239.6035 -- 241.4785 and 253.1055 -- 254.9805\,GHz. The different line label colors denote different types of line morphology: purple is disk-tracing, blue is blue-dominant, red is red-dominant, and orange is outflow-tracing. Line labels are black when no classification was possible. \label{figspectrumwide}}
\end{figure*}

In this Section, our aim is to provide a complete census of the lines present in the spectra from AFGL\,4176 mm1. We do not aim to determine the physical properties associated with each species from fitting all lines with a global physical and chemical model, but instead concentrate on using this as a method to identify the lines detected in the spectra. As our aim is to obtain a census of the lines present across the entire area in which AFGL\,4176 mm1 is emitting, and not only at one point in the maps, we produced mean spectra averaged over a circular aperture of 0.5$''$ in radius centered at the peak position of the continuum emission: 13$^h$43$^m$01$^s.$693 $-$62$^{\circ}$08$'$51.25$''$ (FK5 J2000). Figures~\ref{figspectrumnarrow} and \ref{figspectrumwide} display the spectra for the narrow and wide spectral windows, respectively. The 0.5$''$ aperture also allows us to determine the bulk properties of the gas associated with AFGL\,4176 mm1 averaged over all of its emission, which we will analyze in Section~\ref{sec:morphologies}.

The line identification was carried out as follows. The noise was measured in an empty section of each spectrum and was found to be 0.968, 0.307, 0.410, and 0.703\,mJy\,beam$^{-1}$, respectively, for spw0-3. All lines that have a peak flux larger than 5$\sigma$ were identified in the spectra. 

Using the spectral analysis package CASSIS,\footnote{CASSIS is developed by IRAP-UPS/CNRS (http://cassis.irap.omp.eu)} we first fit an LTE model to the CH$_3$OH and $^{13}$CH$_3$OH lines in spw2 to determine an approximate set of physical parameters that we could expect for each detected line. To do this, we started from the parameters at the position of the peak continuum emission found from the CH$_3$CN CASSIS line fitting in J15 (N$_{H_2}$ = 2.8$\times$10$^{24}$\,cm$^{-2}$, v$_{_{\rm LSR}}$=-53\,km\,s$^{-1}$, $\Delta v _{\rm_{FWHM}}$=8.4\,km\,s$^{-1}$, T=290\,K, and a size of 0.5$''$), and interactively adjusted the parameters to obtain a good fit by eye. CH$_3$OH was used because it had a large number of lines that were clearly detected in the spectra; other lines and molecules were not usable until they were correctly identified. We note that the fitted parameter values were not required to be exact for fitting other lines, as they were only used as initial estimates that were then varied for each molecule.

The values that provided a reasonable fit to the data were N$_{H_2}$ = 4$\times$10$^{25}$\,cm$^{-2}$, v$_{_{\rm LSR}}$=-52\,km\,s$^{-1}$, $\Delta v _{\rm_{FWHM}}$=5\,km\,s$^{-1}$, T=160\,K, and a size of 0.5$''$. We assumed an abundance relative to hydrogen of 10$^{-8}$ for CH$_3$OH \citep[found for Hot Molecular Cores,][]{gerner14a} and find that an isotopic ratio of $\sim$20 is required to simultaneously fit the $^{13}$CH$_3$OH lines. We note that the column density of N$_{H_2}$ = 4$\times$10$^{25}$\,cm$^{-2}$ is slightly higher than that found for mm1 in Table~\ref{dendro_contin_table} (8.3$\times10^{24}$\,cm$^{-2}$), but given the uncertainty in abundance, this value is in reasonable agreement. In addition, we find later in the paper (Section~\ref{sec:opticaldepth}) that several of the CH$_3$OH lines are optically thick, which would indicate that the isotopic ratio is higher and therefore the overall density of CH$_3$OH is also higher. 

For each detected line, we used the Cologne Database for Molecular Spectroscopy \citep[CDMS,\footnote{https://cdms.astro.uni-koeln.de/classic/entries/]}][]{muller05a} and Jet Propulsion Laboratory \citep[JPL,\footnote{https://spec.jpl.nasa.gov/ftp/pub/catalog/catdir.html}][]{pickett98a} line databases to determine a list of transitions that could explain the line. We then manually interactively adjusted the abundance (keeping the column density of molecular hydrogen fixed), excitation temperature, velocity, and linewidth for each possible species, using the fitted parameters for CH$_3$OH as initial estimates, to gauge whether a given species was responsible for emitting a specific spectral line. If the spectrum predicted by CASSIS produced bright lines that were not in the spectrum, we discarded this identification. If the species in question produced other lines present in the spectrum with the correct relative brightnesses, this increased the likelihood that this was the correct identification. CASSIS produced a spectrum that included the effects of blending; thus, as line identifications were added, the resultant combined spectrum could be checked for agreement with the observed one.

Once the line identification was complete, all line properties were determined via a simultaneous spectral fit with Gaussians to all of the detected lines using \texttt{astropy.modeling} with the Sequential Least Squares Programming (SLSQP) fitter, leaving flux density, velocity, and linewidth as free parameters. As input to this algorithm, an estimate for the central frequency of each line was first determined from a separate fit by hand to each line in CASSIS  (i.e., specifying the frequency range in which to conduct an automated Gaussian fit for the line), which was fed to the SLSQL fitter, along with the measured flux at this frequency as an estimate for the line amplitude, and twice the channel width as an estimate for the linewidth. The peak flux was allowed to vary within 0.7-1.5 times the estimated peak flux; the central frequency was allowed to range between $\pm 4$ and $\pm 1$ times the channel width for the narrow and wide spws, respectively; and the linewidth was allowed to vary within $3-10$ and $1-4$ times the channel width for the narrow and wide spws, respectively. In busy portions of the wide spectral windows (spws 1 and 2), we fit the lines by hand with a set of Gaussians in CASSIS and included their fixed properties in the global fit. This was necessary because the automated fitting did not work well in these portions of the spectra. As we conducted a simultaneous fit of the observed lines, the properties of any blended lines were appropriately determined. The combined fit of all the best-fitting Gaussians for each line is shown overplotted on the observed spectra in Figs~\ref{figspectrumnarrow_wfit} and \ref{figspectrumwide_wfit}. A list of identified lines and their measured properties, including their fitted fluxes, velocities, and linewidths, is given in the Appendix in Tables \ref{spw37}-\ref{spw15}. The line databases used for each species are given in Table~\ref{linedats}. The molecular tags given in Table~\ref{linedats} provide, for each species, the unique identifier used to identify it within each database.

Not including isotopologues, we detect a total of 25 different molecules, which are listed in Table \ref{moldetect}, from a total of 203 lines, summarized in Table~\ref{linedats}. The line identifications are generally in good agreement with \citet{bogelund19a}, who find 23 molecules in their spectra from a total of 354 lines. In addition to the 23 molecules detected by \citet{bogelund19a}, we also find lines of H$_2$CCN and NH$_2$CN, bringing our total to 25, which is likely due to the larger aperture used to obtain our spectra. We did not find sufficient evidence to confirm the detection of gGg' ethylene glycol. We detect seven lines that we cannot identify (marked with a ? in Tables \ref{spw37}-\ref{spw15}), which are also not identified in \citet{bogelund19a}. We suspect that the fact \citet{bogelund19a} detect 354 lines in their spectra whereas we only detect 203 is mainly due to their lower detection threshold (3$\sigma$). For instance, 228 lines are labeled in their Figs. 2, C.1, C.2, and C.3 (likely to be the brightest unblended ones in their Figures), similar to the number we detect. There are two other potential explanations for the discrepancy in the number of detected lines. The first is that we define a line as a detected peak in the observed spectra, and thus in the case of blending associate several possible identifications with one detected line. However, in \citet{bogelund19a}, each contribution to the blended observed line constitutes one line, thus resulting in more lines detected. Second, there may be some lines that have morphologically compact emission. In this case, obtaining an average spectrum within a 0.5$''$ radius aperture would wash these out, thus removing them from our measured average spectra in comparison to the \citet{bogelund19a} spectra, which were measured at the continuum peak position.

\subsubsection{Line morphological categories \label{sec:morphologies}}
By inspecting the channel maps for each line, we found that the lines could be grouped into several morphological types, which will be reviewed in detail in the following sections. These types are as follows: disk-tracing, blue-dominant, red-dominant, and outflow-tracing lines. There is also a fifth category labeled as ``unknown'', which covers lines that did not clearly fit into one of the above categories, often because the emission from these lines did not have a high signal-to-noise. Tables \ref{spw37}-\ref{spw15} list the morphological group for each detected line; Table~\ref{linedats} summarizes the number of identified lines for each morphological type for each species.

\begin{figure*}
\begin{center}
\includegraphics[width=18cm,angle=0]{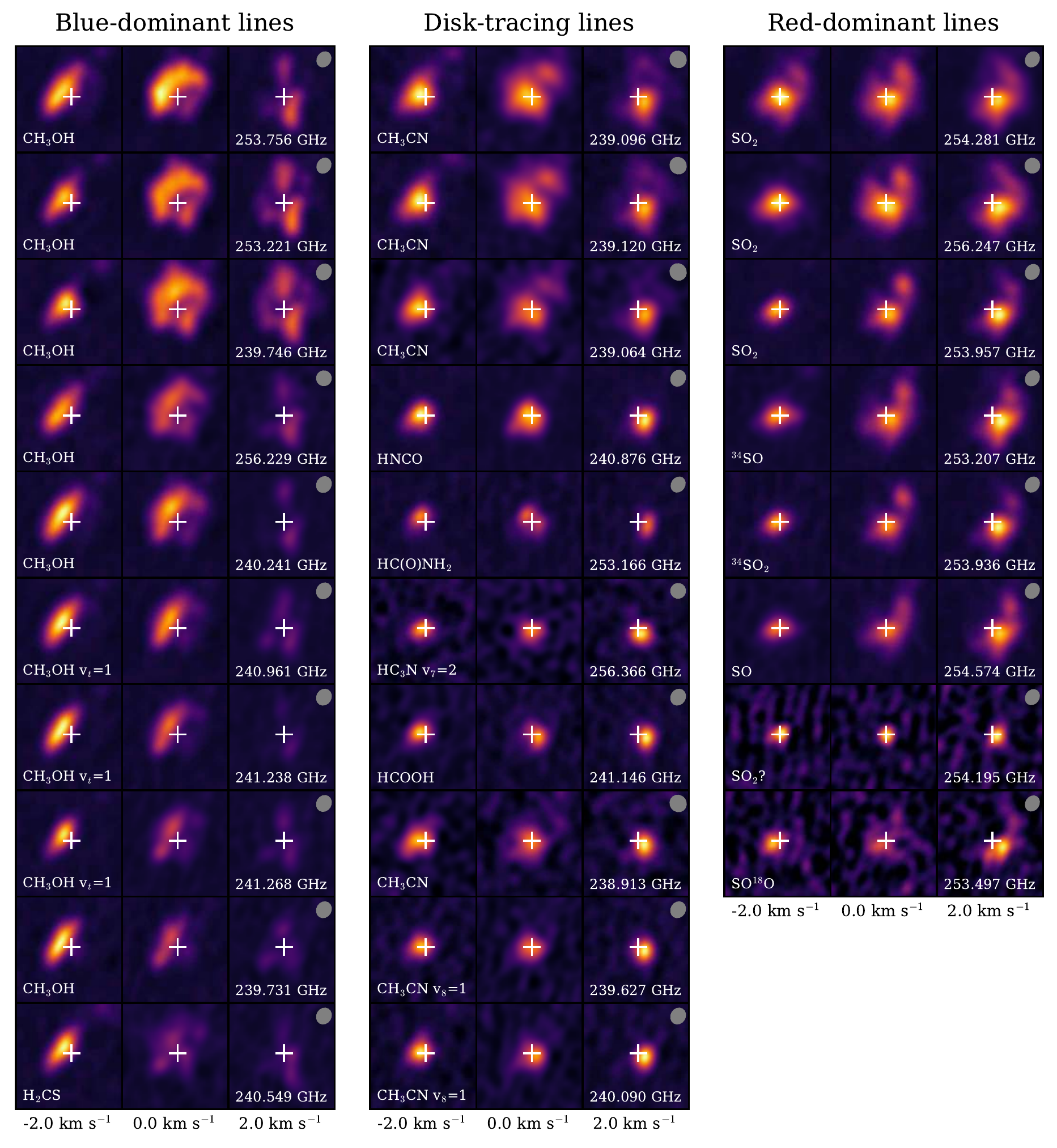}
\caption{Comparison of the channel maps at -2, 0, and 2\,km\,s$^{-1}$ of the line center, v$_{LSR}$ = -52\,km\,s$^{-1}$, for the 10 brightest blue-dominant, disk-tracing, and red-dominant lines (eight for red-dominant). Each panel is a 2$''$ (8400\,au at 4.2\,kpc) square centered on 13$^h$43$^m$01$^s.$71 $-$62$^{\circ}$08$'$51.35$''$ (J2000). The peak position of
the 1.2 mm continuum (mm1) is shown as a white plus sign in each panel. The molecule and rest frequency for each line are shown in the bottom corners of the left and right panels, respectively. The beam is shown in the top right of the right panel for each line. The emission for each line has been normalized to the line map peak flux density. \label{compare_morphologies}}
\end{center}
\end{figure*}

In Figure~\ref{compare_morphologies} we compare the emission from the 10 brightest lines (eight for the red-dominant lines) that had a unique identification and had no other detected line within 10\,MHz ($\sim$12\,km\,s$^{-1}$) for the three morphologies with compact emission: blue-dominant, disk-tracing, and red-dominant. Figure~\ref{compare_morphologies} visually shows the difference between these morphology groupings. The blue-dominant lines show a prominent bar of emission to the east of the peak of mm1 in the blueshifted and sometimes central channels. They also have fainter emission in the redshifted channel. The disk-tracing lines are more kinematically symmetric, showing a velocity gradient progressing from east to west. We will show in Section~\ref{sec:disk-tracing} that these are morphologically and kinematically similar to the upper K-ladder transitions of CH$_3$CN and therefore trace the disk. The red-dominant lines are more similar to the disk-tracing lines, but are generally more extended in the redshifted channel, with most possessing an arc of emission reaching up to the northwest.

To quantitatively determine the similarity and differences within and between the compact morphology groups, we calculated average reduced $\chi^2$ values between the different lines. For each pair of lines, we calculated a $\chi^2$ value using three channels at $-2$, $0$, and $+2$\,km\,s$^{-1}$ from the v$_{\rm LSR}$ ($-52$\,km\,s$^{-1}$), considering pixels in a $2\times2\arcsec$ field of view centered on the peak continuum position of mm1, as shown in Fig.~\ref{compare_morphologies}. To make the comparison uniform, we added noise to each of the three channels, so that all channels had a peak signal-to-noise of 15. We also only considered lines with a peak signal-to-noise that was above 15 in all three channels before adding noise. The reduced $\chi^2$ was calculated by comparing these three channel maps separately to the channel maps of another line at the same velocities, and an average $\chi^2$ for a pair of lines was then determined by combining the $\chi^2$ over these three velocity channels.

With a reduced $\chi^2$ for each pair of lines in hand, we computed an overall average reduced $\chi^2$ (with associated error in the mean) between morphological types, including comparing each line morphological type with itself. The results presented in Table~\ref{chi2_morphologies} show that the reduced $\chi^2$ values when comparing the line morphologies with themselves are close to one, indicating similarity, as expected (if the images were identical, modulo noise, this would produce a reduced $\chi^2$ of $\sim$1), but that the average reduced $\chi^2$ is statistically significantly higher when the $\chi^2$ is calculated from comparison between line types, thus showing that the morphology groups are quantitatively different from one another. 

\begin{table}
\center
\begin{tabular}{cccc}
\hline \hline
 & B & D & R \\
\hline
B & 1.78$\pm$0.05 & 3.33$\pm$0.08 & 3.55$\pm$0.07 \\
D & 3.33$\pm$0.08 & 1.63$\pm$0.05 & 2.14$\pm$0.05 \\
R & 3.55$\pm$0.07 & 2.14$\pm$0.05 & 1.35$\pm$0.05 \\
\hline
\end{tabular}
\caption{Average reduced $\chi^2$ values for comparison between the three compact morphological types. B stands for blue-dominant, D for disk-tracing, and R for red-dominant. \label{chi2_morphologies}}
\end{table}

\begin{figure}
\begin{center}
\includegraphics[width=7.8cm,angle=0]{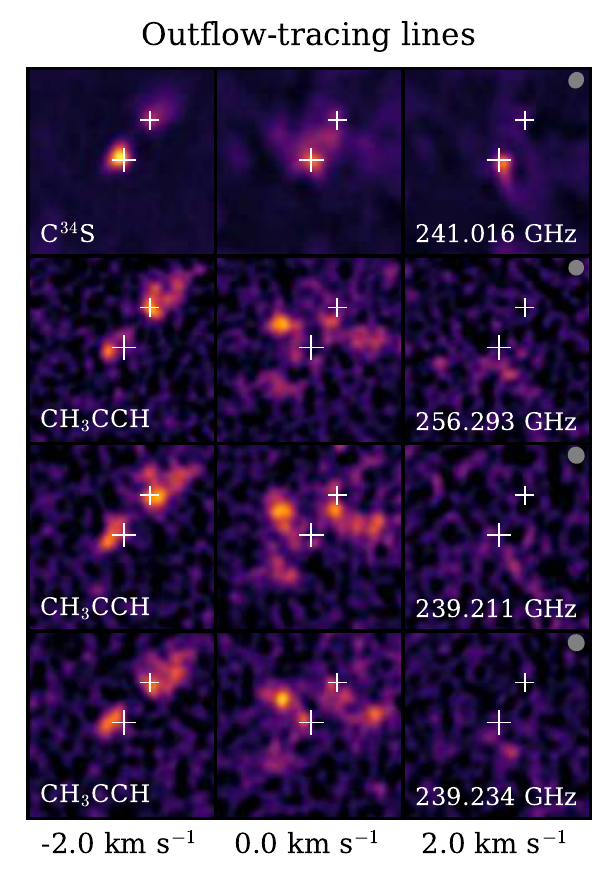}
\caption{Comparison of the channel maps at $-2$, $0$, and $2$\,km\,s$^{-1}$ of the line center, v$_{LSR}$ = -52\,km\,s$^{-1}$, for the brightest outflow-tracing lines with unambiguous identifications and without any lines detected within 10\,MHz ($\sim$12\,km\,s$^{-1}$). Each panel is 4$''$ (16800\,au at 4.2\,kpc) square and centered on 13$^h$43$^m$01$^s.$71 $-$62$^{\circ}$08$'$51.35$''$ (J2000). The peak positions of the 1.2\,mm continuum for mm1 and mm2 are shown as white plus signs. The molecule and rest frequency for each line is shown in the bottom corner of the left and right panels, respectively. The beam is shown in the top right of the right panel for each line. The emission for each line has been normalized to the line map peak flux density. \label{outflow_channel_maps}}
\end{center}
\end{figure}

As the emission from the outflow-tracing lines is more extended and clearly different to the first three morphological types, we show these in Figure~\ref{outflow_channel_maps}, which has a larger field of view of 4$\arcsec$. The outflow-tracing lines are similar in that they all have secondary extended emission near the position of mm2, not seen in the other lines, which is most prominent in the $-2$\,km\,s$^{-1}$ channel. This emission extends away from mm1 in a direction perpendicular to the disk midplane in mm1.

In terms of chemistry, the 10 O-bearing molecules we detect have a mixture of morphologies, but for the most part have predominantly blue-dominant morphologies. In addition, we detect nine N-bearing molecules, all of which morphologically trace the disk (i.e. are disk-tracing lines), except H$_2$CCN, whose lines are too faint to determine a morphology. Lastly, there are five S-bearing molecules and one further molecule (CH$_3$CCH) detected, which have a mixture of morphologies. 

\begin{figure*}
\begin{center}
\includegraphics[height=5.5cm]{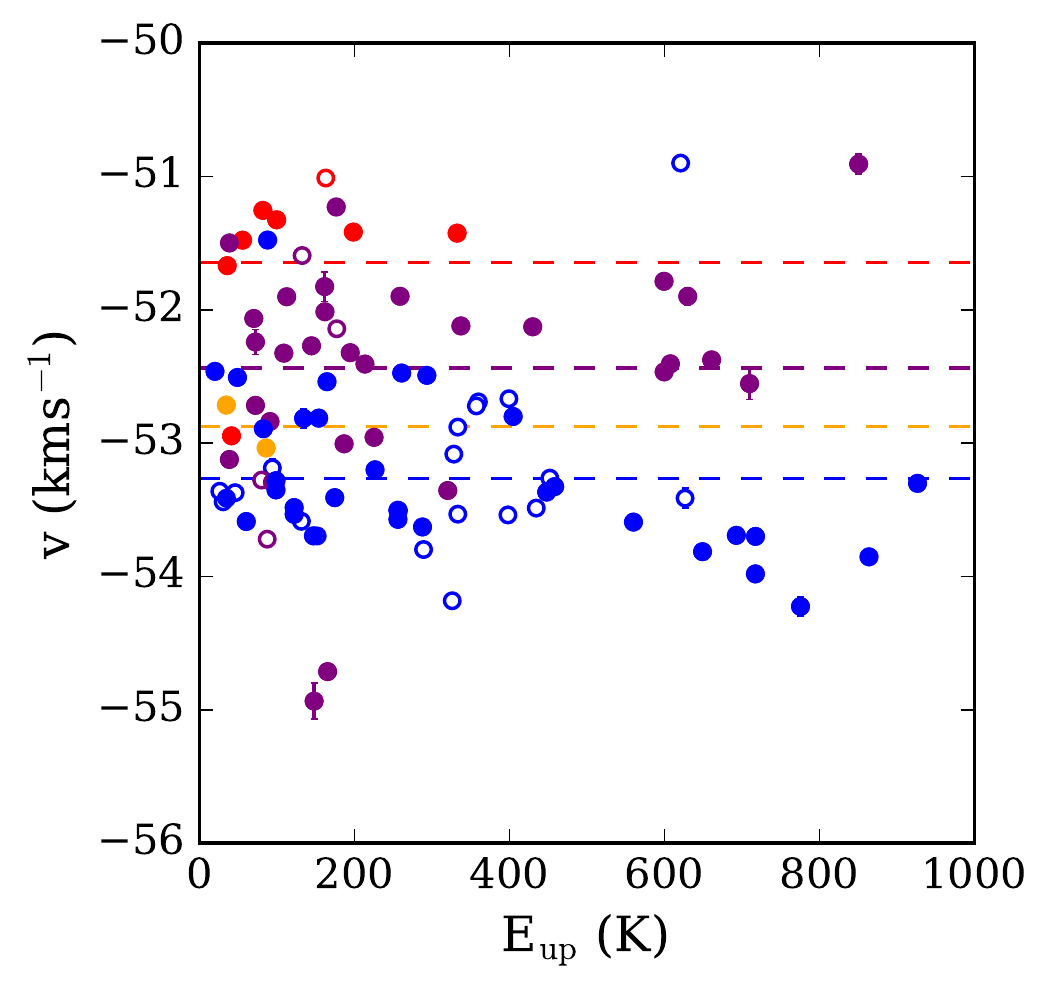}
\includegraphics[height=5.5cm]{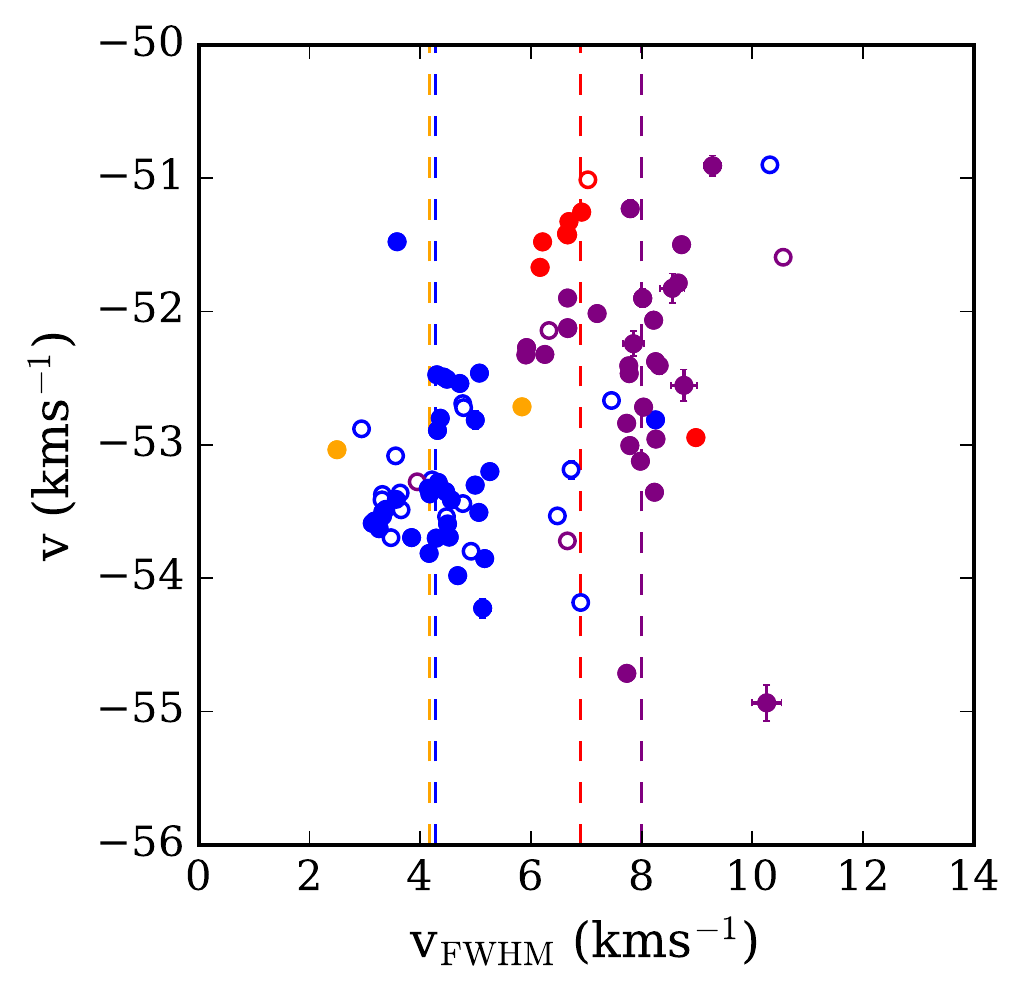}
\includegraphics[height=5.5cm]{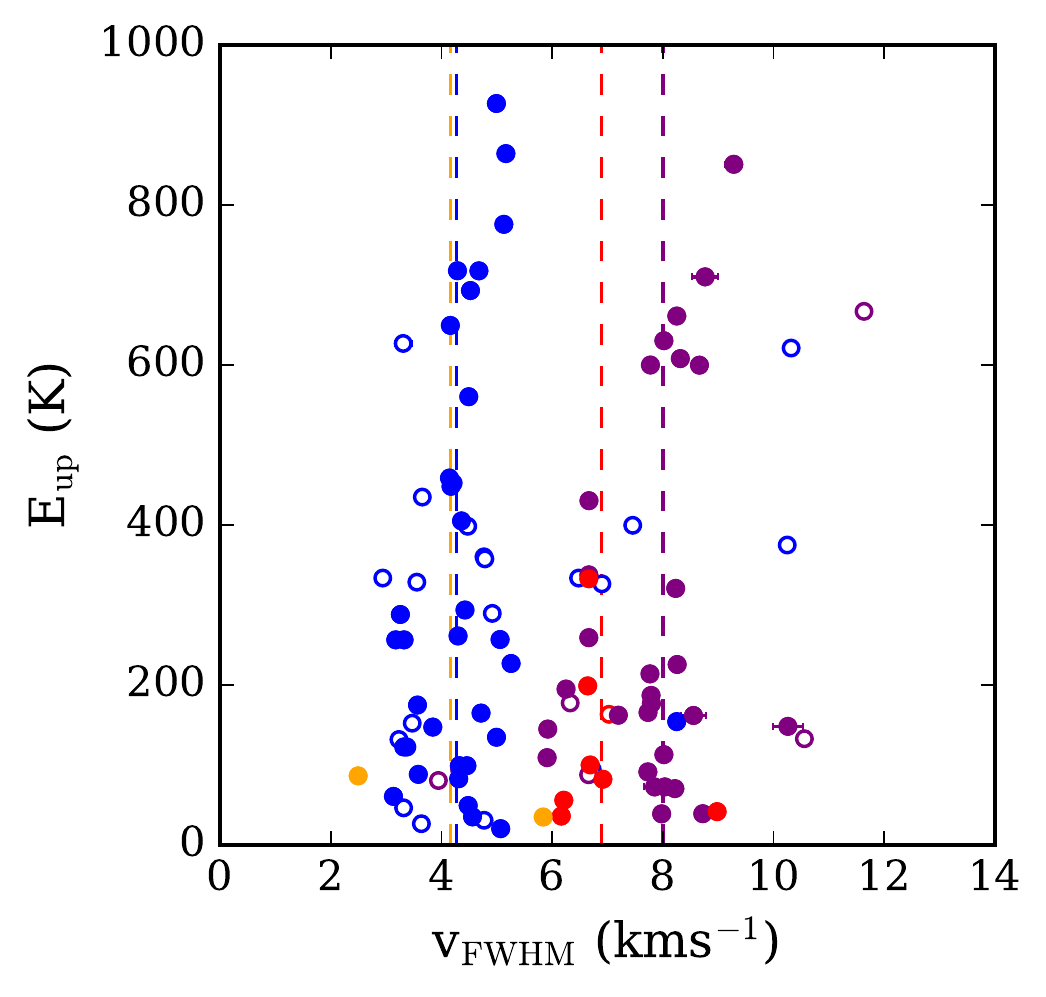}
\end{center}
\caption{Fitted peak velocity of lines with detections $>$20\,$\sigma$ plotted against the energy of the upper level (left panel) and linewidth (middle panel). The right panel shows the energy of the upper level against the linewidth. The red, purple, blue, and orange dots represent red-dominant, disk-tracing, blue-dominant, and outflow-tracing transitions, respectively. The empty points show lines that have ambiguous identifications or have another line within 10\,MHz ($\sim$12\,km\,s$^{-1}$) and therefore may be blended. Dashed horizontal lines show the mean peak velocity. The dashed vertical lines show the mean linewidth, for each line morphology type. \label{vel_vfwhm_trends}}
\end{figure*}

Figure \ref{vel_vfwhm_trends} displays the variation of the fitted peak velocity of the detected molecules as a function of the upper transitional level energy (left panel) and linewidth (middle panel), as well as their fitted upper transitional level energy as a function of linewidth (right panel). All molecules with a detection $>$20\,$\sigma$ are shown. The different line morphological types (disk-tracing, blue-dominant, red-dominant, and outflow-tracing) are shown as points of different colors with error bars. Since the fitting algorithm used for the simultaneous fit with hundreds of free parameters did not provide errors for the fitted parameters, the errors were instead estimated by testing the fitting algorithm used with Monte Carlo simulations of synthetic data to determine the relationship between the error in the velocity and linewidth of each line to the signal-to-noise and linewidth of that line. The error bars are for the most part smaller than the data points.

Each line type displays velocities or linewidths scattered around different values. A dashed line showing the mean velocity and linewidth for each line type is shown in each panel (ambiguously identified or possibly blended lines are shown as empty points). The mean velocities for the solid points for each line type are -51.64$\pm$0.21, -52.44$\pm$0.15, -52.88$\pm$0.11, and -53.27$\pm$0.10\,km\,s$^{-1}$ for red-dominant, disk-tracing, outflow-tracing, and blue-dominant lines, respectively. The left and right panels of Fig.~\ref{vel_vfwhm_trends} show no trend in the line velocities or linewidths with $E_{up}$, indicating that the velocities and linewidths are dominated by kinematics. The middle panel of Fig.~\ref{vel_vfwhm_trends} shows that the line morphologies also cluster by linewidth, with the blue-dominant lines clustering around a mean linewidth of 4.27$\pm$0.11\,km\,s$^{-1}$, whereas the red-dominant lines cluster around a mean of 6.90$\pm$0.34\,km\,s$^{-1}$. The disk-tracing lines have the largest linewidths, with values clustering around a mean of 8.00$\pm$0.22\,km\,s$^{-1}$, which is not unexpected if they trace fast-rotating gas close to the (proto)star. The fact that the linewidths of the blue- and red-dominant lines are less than those of the disk-tracing lines could be explained if these lines are predominantly tracing the emission on the blue- and redshifted side of the circumstellar structure, respectively.

\begin{figure*}
\begin{center}
\includegraphics[width=10.cm]{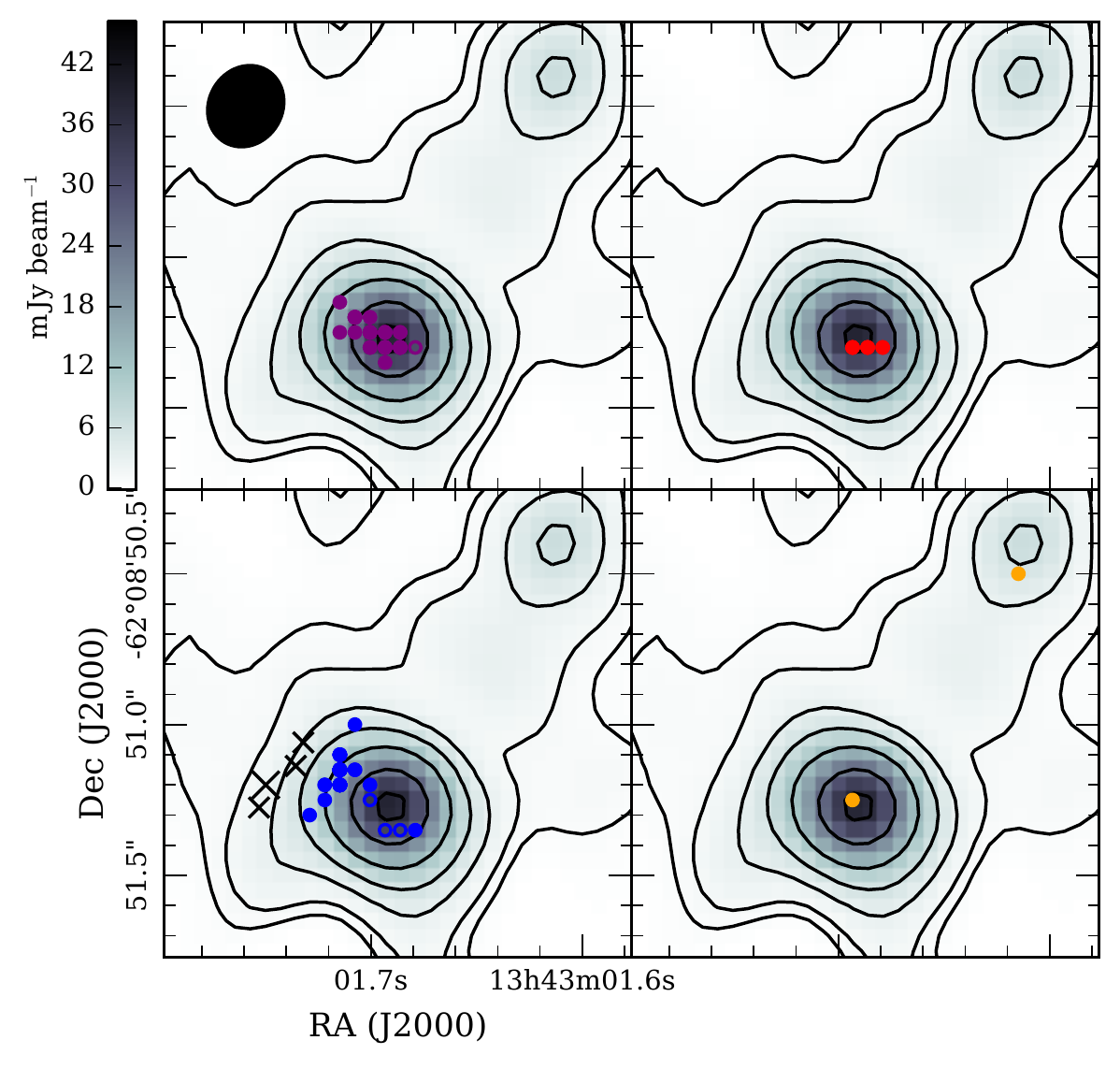}
\end{center}
\caption{Peak position in integrated zeroth-moment maps for lines with detections $>$20\,$\sigma$. Each sub-panel shows the positions for a given line morphological type: Purple, red, blue, and orange dots represent the peak positions of the disk-tracing, red-dominant, blue-dominant, and outflow-tracing transitions, respectively. The empty points show lines that have ambiguous identifications or have another line within 10\,MHz ($\sim$12\,km\,s$^{-1}$) and therefore may be blended. The error in these positions is on the order of the pixel size. The beam is shown in the top left panel. Grayscale and contours show the 1.2\,mm continuum emission, with contours at 5, 10, 25, 50, 100, 200, 300, 400 $\times$ 0.12\,mJy\,beam$^{-1}$. Black crosses in the bottom left panel show Class II methanol masers from \citet{phillips98}. \label{peak_mom0_fig}}
\end{figure*}

Figure \ref{peak_mom0_fig} shows the peak positions within integrated zeroth-order moment maps that were made for all lines with detections $>$20\,$\sigma$. The zeroth-moment maps were made by integrating a spectral slab centered on $-52$\,km\,s$^{-1}$ with a width of 20\,km\,s$^{-1}$. The peaks of the zeroth-moment maps are distributed differently for each line type. In the case of the disk-tracing lines, the peaks line up close to the disk midplane, slightly shifted to the blueshifted, northeast side of the disk. Most of the blue-dominant lines peak far into the blueshifted side of the disk, $\sim$0.22$''$ or $\sim$920\,au from the continuum peak to the northeast along the disk midplane at PA$\sim$60$^{\circ}$ (J15). At this angular separation from the continuum peak along the disk midplane, these points also appear to show a bar-like distribution that has a PA perpendicular to that of the disk midplane. The peaks of the unblended and unambiguous lines in the zero moment maps for the red-dominant lines (Fig. \ref{peak_mom0_fig}) lie in a tight cluster slightly southwest of the continuum peak (on the redshifted side of the disk).

\subsubsection{Disk-tracing Lines} \label{sec:disk-tracing}

In J15, we presented the observed CH$_3$CN J=13-12 K ladder emission from AFGL\,4176, which was well-modeled by a disk in Keplerian rotation. 
However, including CH$_3$CN, we find a total of 55 lines detected within the same observation that have a similar symmetric velocity gradient and therefore are also likely to be tracing the disk. These molecules include eight of the nine (excluding H$_2$CCN, which is too faint) N-bearing molecules listed in Table \ref{moldetect}, and HCOOH (formic acid). Therefore, given that all of the molecules that include nitrogen that are bright enough to determine their morphology are disk-tracing, the presence of nitrogen in the molecules of these observations appears to indicate that it is a good disk tracer. 

To quantify the similarity between the CH$_3$CN J=13-12 K ladder lines, which we previously determined traced the disk via radiative-transfer modeling in J15, and the remaining lines that we refer to as disk-tracing, we calculated a similar combined $\chi^2$ to that produced in Section~\ref{sec:morphologies}. In this case, we compared the CH$_3$CN J=13-12 K=7 line that traces the inner disk to the non-CH$_3$CN lines (which included the vibrationally excited CH$_3$CN lines not analyzed in J15). The resulting average reduced $\chi^2$ value is 1.31$\pm$0.06, demonstrating the similarity of the non-CH$_3$CN lines (as well as vibrationally excited CH$_3$CN) to the CH$_3$CN lines known to trace the disk. If the CH$_3$CN K=7 line is compared to all of the non-CH$_3$CN lines (excluding the vibrationally excited CH$_3$CN lines), the average reduced $\chi^2$ value is 1.32$\pm$0.10, confirming that excluding the vibrationally excited CH$_3$CN lines does not significantly change the result.

Comparing with the results of \citet{bogelund19a}, we note that they find the velocity gradients seen in four of our eight disk-tracing lines: NH$_2$CHO or HC(O)NH$_2$, CH$_3$CN, C$_2$H$_3$CN, and C$_2$H$_5$CN, as well as in CH$_3$OCHO. We categorize several CH$_3$OCHO lines as disk-tracing, and several others as blue-dominant; therefore, we find that some of these lines trace the disk. 

\begin{figure*}
\begin{center}
\includegraphics[width=18cm]{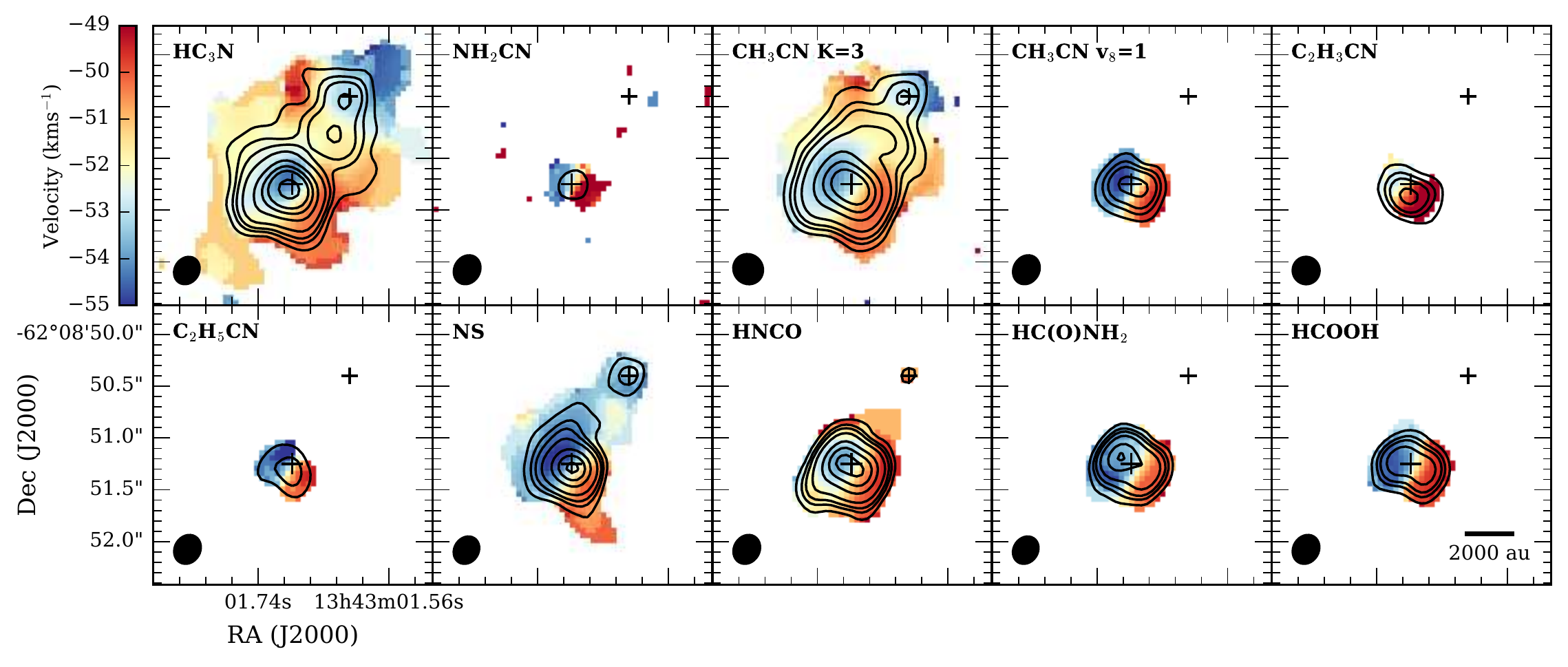}
\end{center}
\caption{Zeroth- (contours) and first-order (colorscale) moment maps for the nine disk-tracing molecules observed in the ALMA data, which are shown in bold in Tables \ref{spw37}-\ref{spw15}. Except for CH$_3$CN K=3, the brightest, unblended line was chosen if possible. The velocity ranges were chosen to cover the full range of the line emission for each line. The contours are shown at 5, 10, 15, 25, 50, 75, 100, and 150 $\times$ the local rms noise, which is 0.021, 0.016, 0.017, 0.016, 0.008, 0.012, 0.019, 0.018, 0.014, and 0.013\,Jy\,beam$^{-1}$ km\,s$^{-1}$ for each panel, respectively, read from left to right, then top to bottom. The only available HC$_3$N line is blended with a CH$_3$OH line and was therefore only integrated between -60 and -49\,km\,s$^{-1}$. The positions of mm1 and mm2 are shown as large and small plus signs, respectively.\label{momentsfig}}
\end{figure*}

Figure \ref{momentsfig} presents zeroth- and first-moment maps of representative lines for each of the nine disk-tracing molecules listed above, as well as the brightest unblended line for CH$_3$CN v$_8$=1. The first-moment maps were created using a 5$\sigma$ cut, except for the faint NH$_2$CN line, for which we used 3$\sigma$. The example lines shown in Fig. \ref{momentsfig} are shown in bold in Tables \ref{spw37}-\ref{spw15}. 

\begin{figure*}
\begin{center}
\includegraphics[width=18.cm,angle=0]{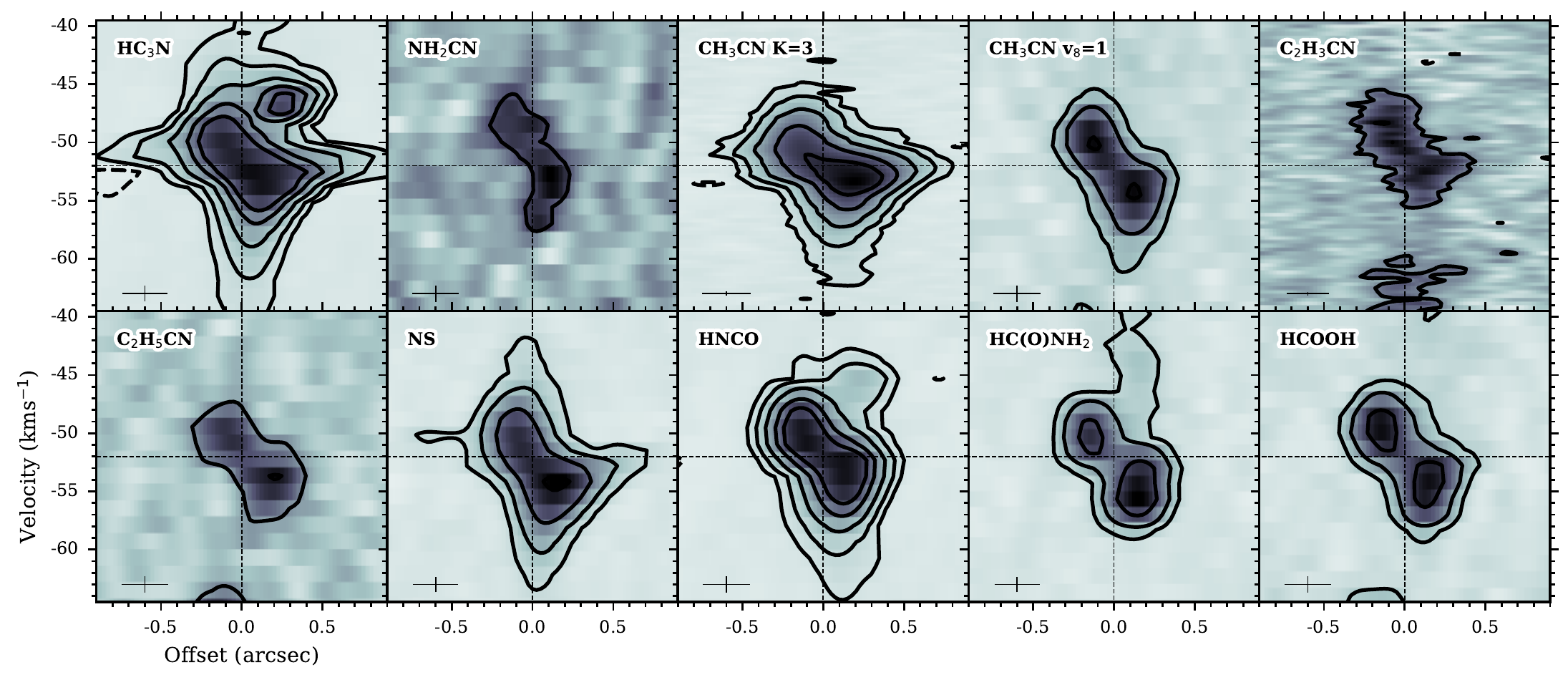}
\caption{Position-velocity diagrams of the disk-tracing lines shown in Fig.~\ref{momentsfig}, averaged along a cut centered on the mm1 continuum peak position, with PA = 61.5$^{\circ}$ and width = 1$\arcsec$. The horizontal and vertical dashed lines mark the position of the continuum peak and a velocity of $-$52\,km\,s$^{-1}$. The crosses in the bottom left of each panel show the observational spatial and spectral resolution. Contour levels are -3, 3, 10, 25, 50, and 75 $\times$ $\sigma$, which is 2.3, 1.0, 1.7, and 1.4\,mJy\,beam$^{-1}$ for spws 0-3, respectively. \label{pv_disktracing}}
\end{center}
\end{figure*}

Similar to that found for CH$_3$CN, all the lines shown in Fig. \ref{momentsfig} display a velocity gradient across the source with blueshifted emission in the northeast and redshifted emission in the southwest. In Figure~\ref{pv_disktracing}, we present the PV diagrams of the lines shown in Fig. \ref{momentsfig}, which also shows their similarity to the CH$_3$CN line, providing further evidence that this group of lines is indeed tracing the disk. For instance, each PV diagram in Fig.~\ref{pv_disktracing} shows a velocity gradient from blue- to red-shifted from east to west, and resembles one of the simulated model PV diagrams for a Keplerian disk shown in J15. A few lines, such as HC$_3$N and NS, also clearly show a curve tending to higher velocities at smaller radii in the lowest contours in the top left and bottom right quadrants of the PV diagram, as expected in the case of Keplerian rotation.

\begin{figure}
\begin{center}
\includegraphics[width=8.5cm]{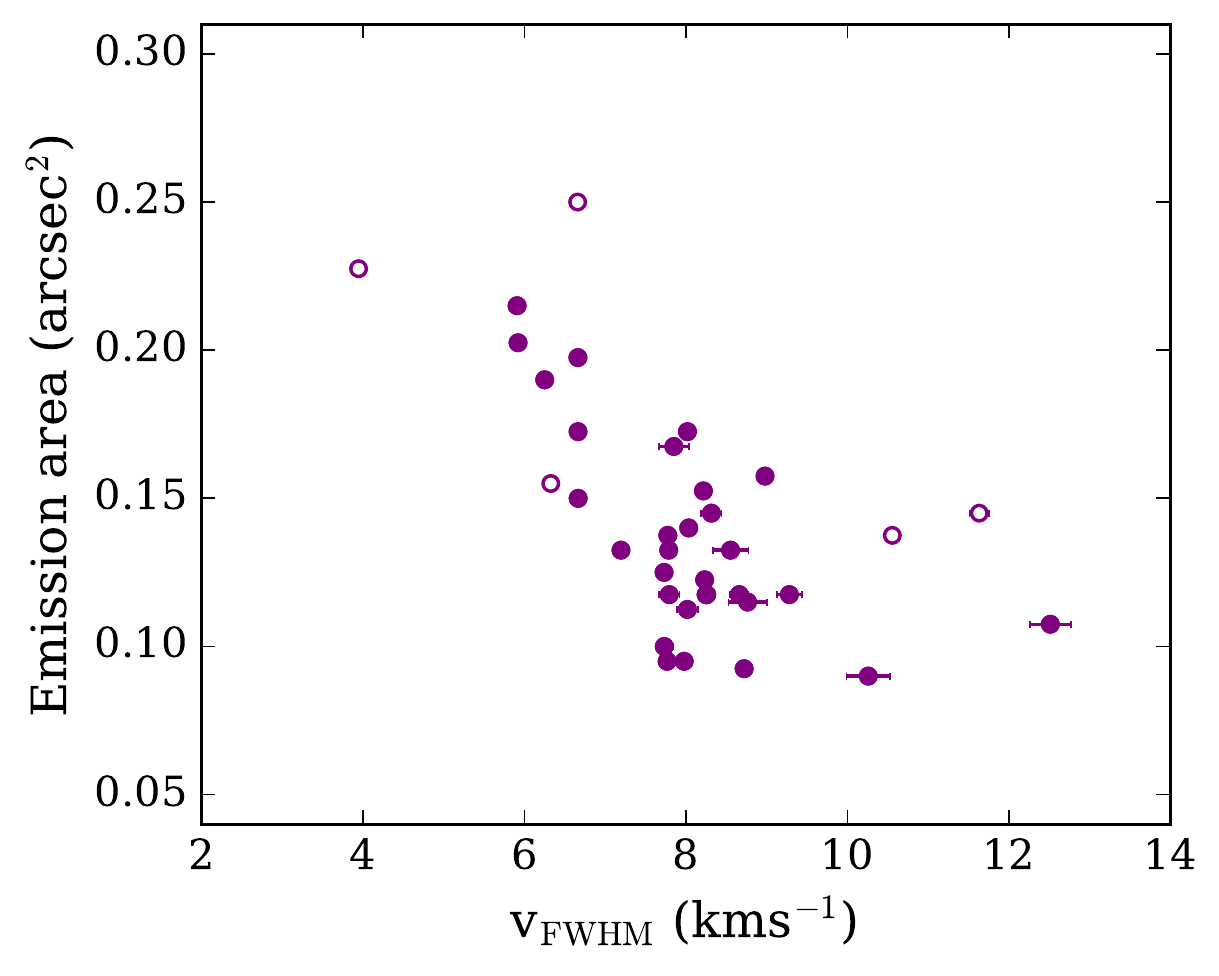}
\end{center}
\caption{The area of all pixels above half the peak flux density in their zeroth-moment map against linewidth for all disk-tracing lines. The empty points show lines that have ambiguous identifications or have another line within 10\,MHz ($\sim$12\,km\,s$^{-1}$) and therefore may be blended. \label{area_linewidth}}
\end{figure}

Both J15 and \citet{bogelund19a} found that more excited lines, which trace hotter gas close to the central source, have smaller extents and steeper velocity gradients. Therefore, it is likely that the spatially compact lines trace the disk while the extended lines (e.g. low K transitions of CH$_3$CN and HC$_3$N) may also trace part of the envelope. This is supported by Figure~\ref{area_linewidth}, where we plot the area of all pixels brighter than half the peak flux density in the zeroth-moment map for the disk-tracing lines against their linewidth, showing a trend of more compact emission for lines with larger linewidths.

Due to the fact that it traces extended emission and is blended with a CH$_3$OH line on the redshifted side, HC$_3$N is the least clear-cut of the disk tracers. However, HC$_3$N evaporates off dust grains at a similar temperature to CH$_3$CN \citep{collings04a, jaber-al-edhari17a}, the HC$_3$N J=28 -- 27 transition we detect has a similar critical density to that of CH$_3$CN J =13-12 K=3 ($>$1$\times$10$^6$ and 1.5$\times$10$^6$\,cm$^{-3}$ at 100\,K, respectively), and the energies of the upper levels of the transition are similar (177.26 and 144.63\,K). Therefore, assuming similar abundances, it is not surprising that the extent of the HC$_3$N emission shown in Fig. \ref{momentsfig} is similar to that of CH$_3$CN. 

Several of the disk-tracing lines trace only the inner several hundred to thousand astronomical units of the source, namely NH$_2$CN, CH$_3$CN v$_8$=1, C$_2$H$_3$CN and C$_2$H$_5$CN. These lines have reasonably high excitation temperatures (205.25, 607.71, 278.02, and 169.27\,K), although lower abundances of these molecules also probably play a role in their smaller extent.

\begin{figure}
\begin{center}
\includegraphics[width=9.0cm,angle=0]{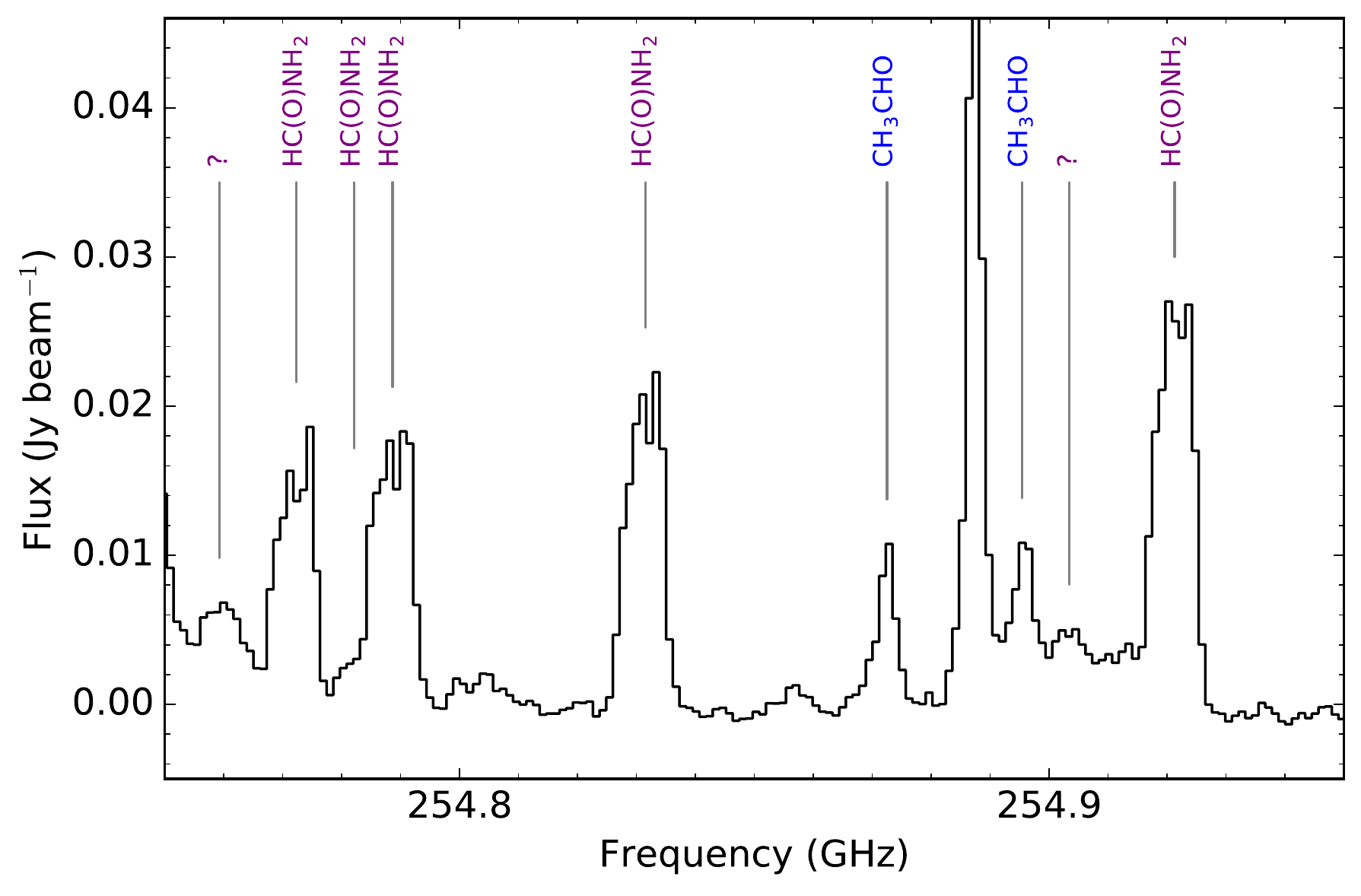}
\caption{Spectrum showing several example formamide lines in the upper portion of spw2. The different line label colors denote different types of line morphology: purple is disk-tracing and blue is blue-dominant. \label{formamide}}
\end{center}
\end{figure}

Formamide or HC(O)NH$_2$ is particularly interesting, in that the lines show double-peaked line structures (see spw2 in Fig. \ref{figspectrumwide} and Fig.~\ref{formamide}), which may indicate a lack of lower-velocity envelope emission from these lines, which would ``fill in'' the line profile close to the line center. In Section~\ref{sec:opticaldepth} we discuss how the majority of lines (including those for HC(O)NH$_2$) are likely to be optically thin, so optical-depth effects will not be important. However, missing flux issues due to interferometric filtering of emission at larger scales may complicate this interpretation. Nevertheless, the moment maps shown in Fig.~\ref{momentsfig} show that, along with vibrationally exited CH$_3$CN and HCOOH, the HC(O)NH$_2$ emission displays a clear velocity gradient, is confined to the inner region of the disk, and is thus a good disk tracer.

\subsubsection{Blue-dominant Lines} \label{sec:bluedominant}

We find a total of 85 lines that are categorized as blue-dominant, and that molecules categorized as blue-dominant are predominantly oxygen-bearing. There are five molecules that only have lines with a blue-dominant morphology in Tables~\ref{spw37}-\ref{spw15}, which are CH$_2$CO, CH$_3$OH, HCOCH$_2$OH, CH$_3$OCH$_3$, and H$_2$CS. There are four molecules that have lines that are designated mostly blue-dominant but with a few categorized as disk-tracing: CH$_3$OCHO, C$_2$H$_5$OH, CH$_3$COCH$_3$, and aGg'-(CH$_2$OH)$_2$. The only oxygen-bearing molecule that does not have lines that are categorized as blue-dominant is HCOOH. Conversely, H$_2$CS is the only line that is not oxygen-bearing that is categorized as blue-dominant.

\begin{figure*}
\begin{center}
\includegraphics[width=18.cm]{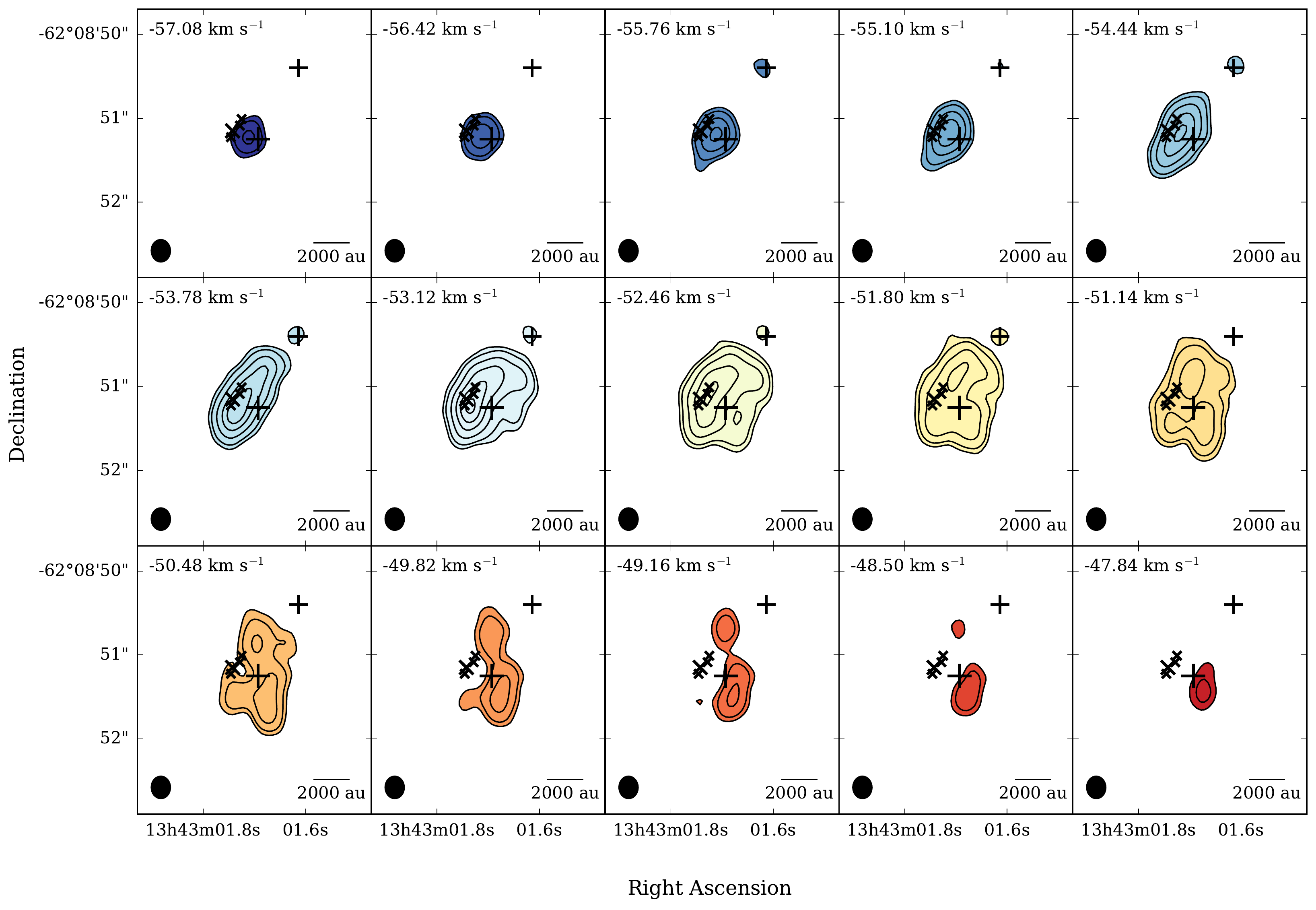}
\end{center}
\caption{Channel map of the CH$_3$OH 17(3,15) $\rightarrow$ 17(2,16) A$^{+-}$ line imaged at 0.66\,km\,s$^{-1}$ spectral resolution between -57.08 and -47.84\,km\,s$^{-1}$. The contours show 5, 10, 25, 50, 75, and 100 $\times$ 3.4\,mJy\,beam$^{-1}$ (the local rms noise), and the contours are filled with a colorscale to represent the channel velocity. The positions of mm1 and mm2 are shown as large and small plus signs, respectively, and the crosses represent the Class II methanol maser components found by \citet{phillips98}. \label{CH3OH_chanmap}}
\end{figure*}

Figure \ref{CH3OH_chanmap} provides an example of the morphology of a blue-dominant line, showing the CH$_3$OH 17(3,15) $\rightarrow$ 17(2,16) A$^{+-}$ line with a rest frequency of 256.228714\,GHz. As introduced in Section~\ref{sec:morphologies}, this example methanol line, as well as the other blue-dominant lines, shows a blue-shifted bar-like morphology that is perpendicular to the disk plane. In this example, this is especially noticeable in the -53.78\,km\,s$^{-1}$ channel. At higher, redshifted velocities, the emission reaches around the peak continuum position toward the southwest, suggesting a possible ring or shell-type structure.

Notably, the distribution of the brightest emission bears resemblance to the distribution of the four known Class II methanol masers in the region \citep{phillips98}, which are shown as crosses in Fig.~\ref{CH3OH_chanmap} and in the bottom left panel of Fig.~\ref{peak_mom0_fig}. The brightest maser component, component C in \citet{phillips98}, shown as a slightly larger cross in Fig.~\ref{CH3OH_chanmap}, has a velocity of $-54.5$\,km\,s$^{-1}$, which is blueshifted compared to the systemic velocity ($\sim$-52\,km\,s$^{-1}$), and is slightly blueshifted compared to the mean velocity of the blue-dominant lines (-53.3\,km\,s$^{-1}$). As seen in figure 4(a) of \citet{phillips98}, the four maser points are all blueshifted, with velocities between -56 and -52\,km\,s$^{-1}$. Given the spatial and velocity coincidence of the masers and thermal emission, it is likely that these are tracing the same physical structure.

Due to the low inclination of the system ($i\sim30^{\circ}$ between the rotation axis of the disk and the line of sight, J15), it is unlikely that inclination and thus optical-depth effects would be causing the asymmetry of emission (see also Section~\ref{sec:opticaldepth}). Instead, there may be something in the disk or circumstellar environment that is desorbing the oxygen-rich blue-dominant molecules from the dust grains at this position to increase their local abundance, such as a shock.

Previously, \citet{bogelund19a} found that some of the O-bearing species they detect toward AFGL\,4176 peaked 0.2$''$ offset from the continuum peak; however, they also state there are no large differences in the spatial distribution of N- and O-bearing species. In contrast, as discussed above, we find that the blue-dominant lines, which show an asymmetric morphology often peaking in the blue-shifted side of the disk and are mostly O-bearing, have a very different morphology to that of the disk-tracing or N-bearing species. Further, the association of the blue-dominant line emission with the methanol masers also indicates that different physical processes are involved with the production of the emission from disk-tracing and blue-dominant lines, and thus N- and O-bearing lines. The difference in our findings could be attributed to the fact that \citet{bogelund19a} fit Gaussians to the zeroth-moment maps of each line and determine the peak position from these fits, whereas we inspected the channel maps for each line (e.g., Fig.~\ref{CH3OH_chanmap}) to determine its morphological type. We also note that \citet{bogelund19a} find a different velocity gradient in the first-moment maps of CH$_3$OH, C$_2$H$_5$OH, CH$_3$OCH$_3$, and H$_2$CS compared to the molecules they found trace the disk rotation. However, the channel maps of the blue-dominant lines (e.g., Fig.~\ref{CH3OH_chanmap}) show that these molecules do show a similar velocity gradient to that of the disk-tracing lines, but the emission is often brighter and more extended on the blue-shifted side of the disk.

\subsubsection{Red-dominant Lines} \label{sec:reddominant}

\begin{figure*}
\begin{center}
\includegraphics[width=18.cm]{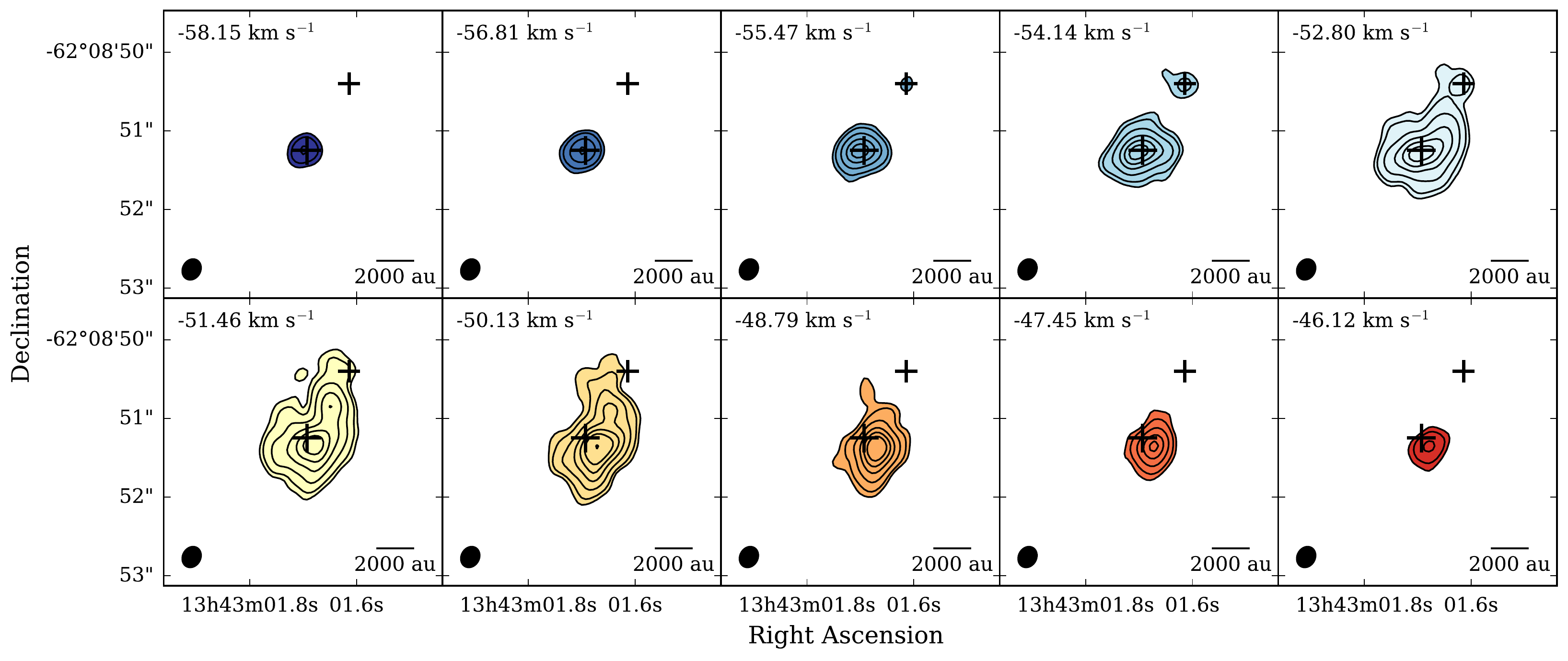}
\end{center}
\caption{Channel map of the  $^{34}$SO $^{3}\Sigma$ N,J = 6,6 $\rightarrow$ 5,5 line imaged at 1.33\,km\,s$^{-1}$ spectral resolution between -58.15 and -46.12\,km\,s$^{-1}$. Contours show 5, 10, 25, 50, 75, 100, 150 $\times$ 2.1\,mJy\,beam$^{-1}$, and the contours are filled with a colorscale to represent the channel velocity. The positions of mm1 and mm2 are shown as large and small plus signs, respectively. \label{SO_chanmap}}
\end{figure*}

A subset of the detected lines exhibits a red-dominant morphology; these are the lines SO and SO$_2$ and their isotopologues. A channel map of an example line, $^{34}$SO $^{3}\Sigma$ N,J = 6,6 $\rightarrow$ 5,5 is shown in Fig.~\ref{SO_chanmap}. This figure shows that $^{34}$SO, like other lines of this type, is extended in the redshifted channels and has bright redshifted emission on the southwest side of the disk. Further, both mm1 and mm2 are traced by the $^{34}$SO emission.

Similarly to the blueshifted lines, it is difficult to explain the asymmetry of the emission in these lines by optical-depth effects, and thus there may be something present or occurring in the disk or surroundings that has liberated SO and SO$_2$ from the dust grains. One explanation would be a shock, similar to that seen in SO at the centrifugal barrier toward low-mass protostars \citep[e.g.][]{sakai17a}; however, it is not clear what would differentiate it from the shocks that may be producing the emission on the blueshifted side of the disk in the blue-dominant lines.

\subsubsection{Outflow-tracing Lines} \label{sec:outflowtracing}

The molecules whose lines trace extended emission on scales of $\sim$2-10$''$ (e.g., Fig.~\ref{outflow_channel_maps}), which we have interpreted as emission from the outflow or outflow cavity walls (see Section~\ref{sec:c34s}), are the sulfur-bearing molecules C$^{34}$S, H$_2$CS, as well as CH$_3$CCH. We note that while H$_2$CS exhibits blue-dominant morphology on the scale of the disk ($\sim$1$''$), there is also larger-scale H$_2$CS emission, which we interpret as from the outflow. As C$^{34}$S is the best example of this morphology type, but we have complementary APEX data, we postpone a discussion of the combined ALMA+APEX C$^{34}$S emission to Section~\ref{sec:CombinedResults}. We note that \citet{bogelund19a} also find that the emission from CH$_3$CCH is morphologically double-peaked.

\subsubsection{Effect of optical depth on line morphologies} \label{sec:opticaldepth}
To estimate whether some of the line emission was optically thick, we took the spectral cubes in brightness temperature for each spectral window and compared it to the map of excitation temperature $T_{\rm ex}$ derived from CH$_3$CN by J15. We found that the maximum optical depths across the image cube for each line were $<$1 for all but the brightest lines: the brightest line of methanol in spw1, and the four brightest methanol and the brightest SO$_2$ lines in spw2. We note that the lines shown in Figs.~\ref{CH3OH_chanmap} and \ref{SO_chanmap} are both optically thin. 

Although some of the brightest lines may be optically thick, the same asymmetry and blue/red-shifted morphology is seen in the line emission in their corresponding optically thin isotopologue lines (e.g., $^{13}$CH$_3$OH and $^{34}$SO$_2$, the latter of which is shown in Fig.~\ref{compare_morphologies}), indicating that the observed morphologies are not due to optical-depth effects. This is further seen in Fig.~\ref{vel_vfwhm_trends}, where there is no trend in velocity seen with E$_{\rm up}$. For any pair of lines of the same species with different E$_{\rm up}$, the higher E$_{\rm up}$ lines trace hotter and denser gas closer to the source compared to the lower E$_{\rm up}$ lines. In the case where both lines are optically thin, they will trace all of the material/molecule along the line of slight that is hot enough to produce emission in that transition. Thus, in the optically thin case the columns are symmetrical around the source along the line of sight, and therefore the two lines will be centered at the same velocity. In the case where the column is high enough to make the lines optically thick, they will only trace material up to an optical depth of $\sim$1, thus missing different amounts of material from the back side of the circumstellar structure. Therefore, given we expect that the envelope is infalling, this difference in the material traced by both lines will lead to a difference in the average velocity of the two lines. As this velocity difference is not seen between different E$_{\rm up}$ lines in Fig.~\ref{vel_vfwhm_trends} we can deduce that optical-depth effects are not important.

\newpage
\subsubsection{The Nature of mm2} \label{sec:methanol}
With a total mass inferred from its dust emission of 3.6\,M$_{\odot}$, mm2 is the second most massive core in the field. The column density of mm2 derived from the dust emission in Section~\ref{sec:ALMAcont} is 2.9$\times 10^{24}$\,cm$^{-2}$. In comparison, the column density of CH$_3$CN molecules derived in J15 also peaks toward mm2 but is 1.6$\times10^{15}$cm$^{-2}$, which assuming a CH$_3$CN abundance of $10^{-8}$ corresponds to an H$_2$ column density of 1.6$\times10^{23}$cm$^{-2}$. Given the discrepancy of an order of magnitude, this probably indicates that the CH$_3$CN abundance in mm2 is much lower than $10^{-8}$.

Figure \ref{figmm2} provides average spectra in the four spectral windows centered on the position of mm2 in a circular aperture of 0.5$''$ in radius. The most obvious difference between the spectra of mm2 and mm1 is that lines of CH$_3$CCH and C$^{34}$S are comparatively brighter in the mm2 spectra. There is also a lack of emission from more complex molecules. Methanol and H$_2$CS are present, as well as many sulfur-bearing lines such as SO, SO$_2$, and O$^{13}$CS. Several lines categorized as disk-tracing lines, such as NS, HC$_3$N, HNCO, and CH$_3$CN, are also present, although they do not appear to trace a disk at the position of mm2 (see Fig.~\ref{momentsfig}). This is likely because these lines, in addition to tracing the disk of mm1, also trace part of the circumstellar envelope or outflow around mm1.

The presence of bright CH$_3$CCH and C$^{34}$S emission at the position of mm2, which appears to form part of the emission from the outflow or outflow cavity walls, suggests that mm2 is possibly a knot in the blueshifted side of the jet or outflow from mm1. This is supported by the fact mm2 is blueshifted in comparison to mm1 (figure 3 in J15) and by the location of mm2, which lies along the axis of the mm1 disk. Further, the presence of jet shocks would increase the density, temperature, and linewidth at the position of the shock, supported by the elevated temperature and linewidth found at the position of mm2, with values of $\sim$-53\,km\,s$^{-1}$, $\sim$200\,K and $\sim$5\,km\,s$^{-1}$ respectively (J15). Thus, although we cannot rule out that mm2 is another protostar, which would make this system a binary with a separation of 1$''$ or 4200\,au, there is reasonable evidence that mm2 is associated with shocks in the outflow or jet from mm1.

\section{ATCA Results} \label{sec:ATCAresults}
The section below details the most important results derived from the ATCA observations. Given the large number of observed ATCA bands and thus resultant images (see Table~\ref{ATCAspws}), we do not present all of these here, although the images can be obtained in the accompanying online data.\footnote{http://doi.org/10.5281/zenodo.3369188} We list the ATCA bands with accompanying figures in the final column of Table~\ref{ATCAspws}.

\subsection{Centimeter Continuum} \label{sec:cmcont}

\begin{figure}
\begin{center}
\includegraphics[width=9.cm]{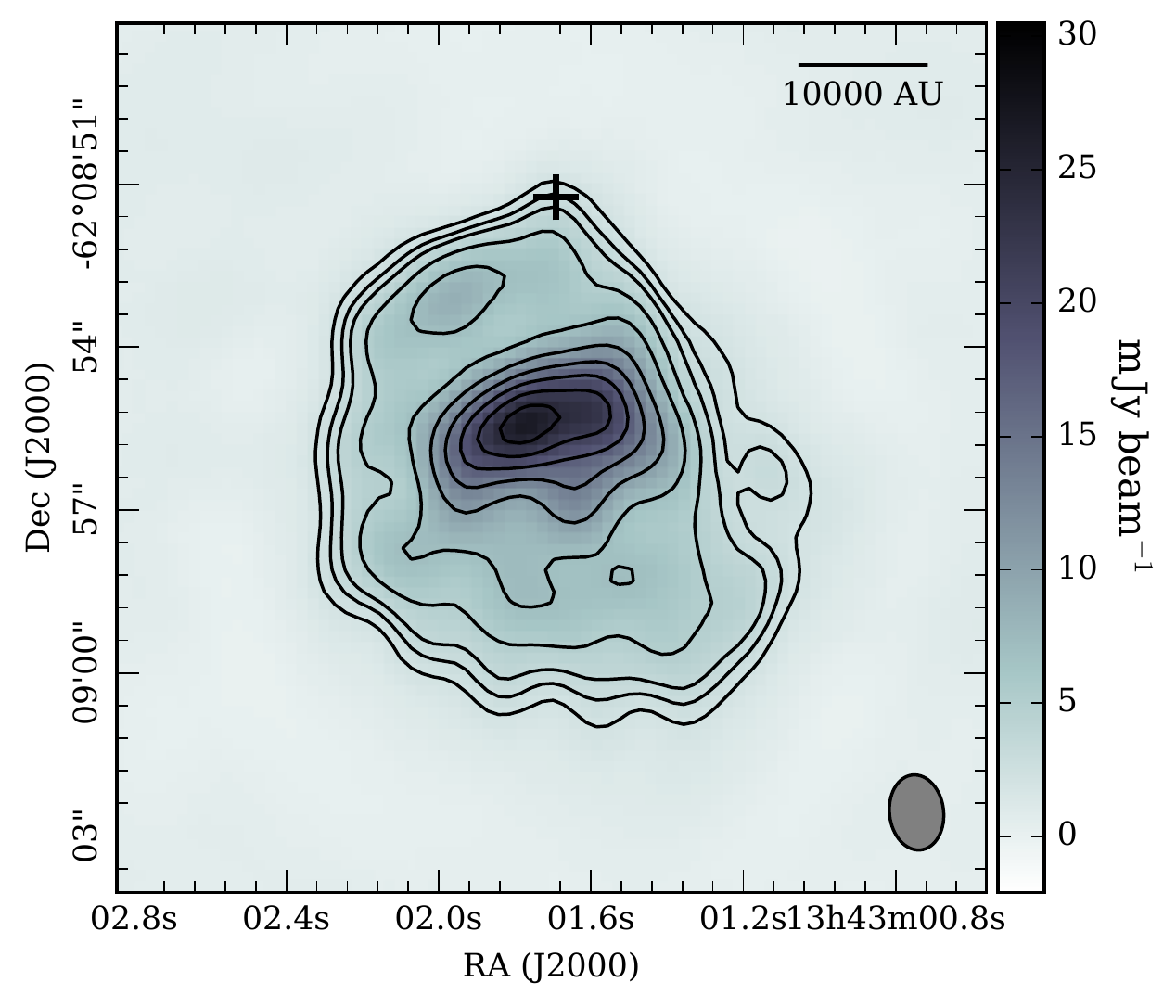}
\end{center}
\caption{ATCA continuum emission toward AFGL\,4176 at an effective frequency of 22.004\,GHz or 1.36\,cm (combined image from the two 20.434 and 23.712\,GHz bands observed in April 2012) shown in grayscale and black contours (local $\sigma$ = 0.7\,mJy\,beam$^{-1} \times$ -3, 3, 4, 5, 7, 10, 15, 20, 25, 30, 35). The position of mm1 is shown as a plus sign. The ATCA synthesized beam is 1.39$\times$1.00$''$, PA = 6.53$^{\circ}$, and is shown in the bottom right corner. \label{ATCA_contin_Apr}}
\end{figure}

\begin{figure}
\begin{center}
\includegraphics[width=9.cm]{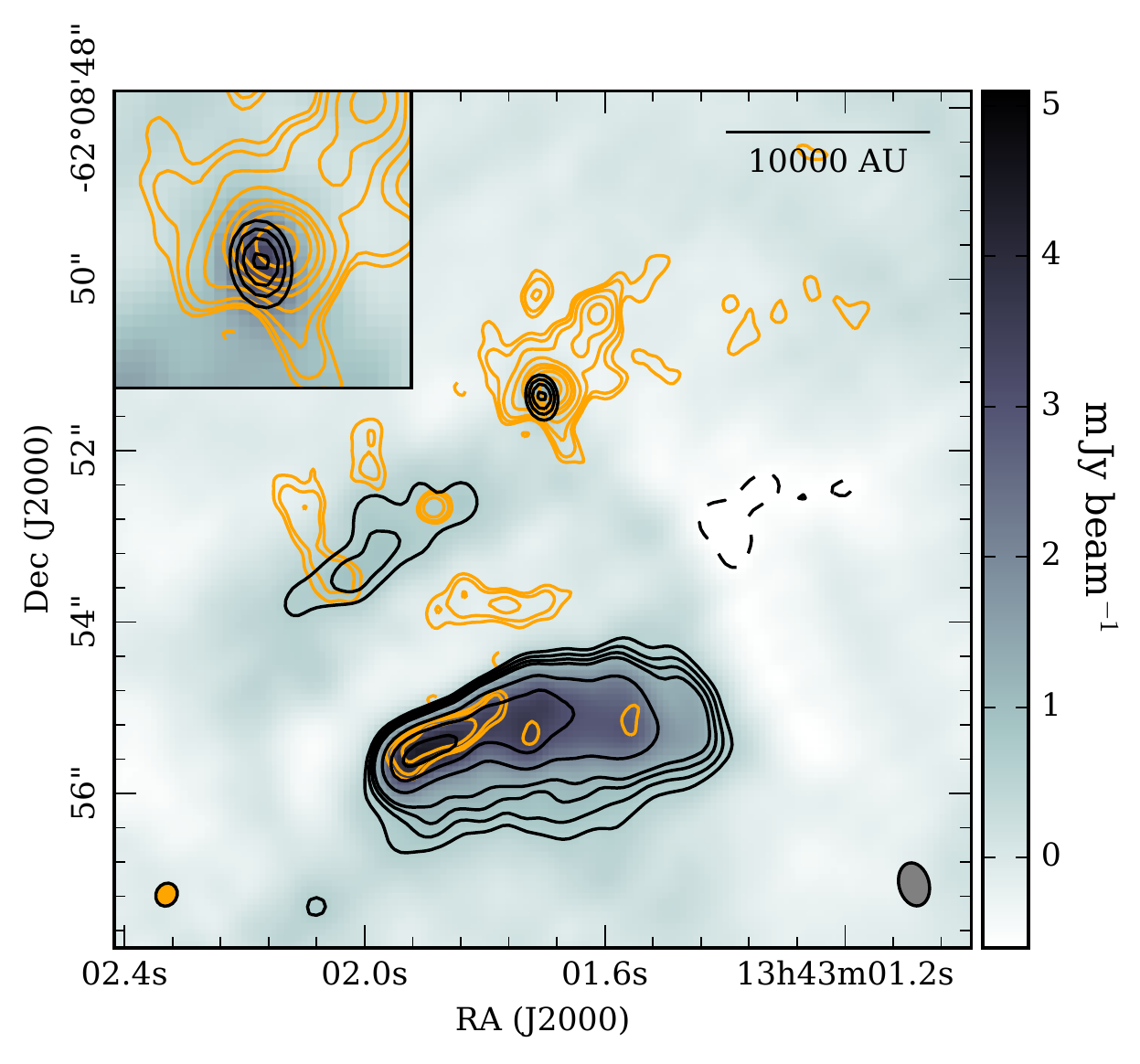}
\end{center}
\caption{ATCA continuum emission toward AFGL\,4176 at an effective frequency of 24.328\,GHz or 1.23\,cm (combined image from the two 24.139 and 24.533\,GHz bands observed in September 2012) shown in grayscale and black contours (local $\sigma$ = 0.2\,mJy\,beam$^{-1} \times$ -3, 3, 4, 5, 6, 10, 25). Orange contours show the ALMA 1.21\,mm continuum emission (local $\sigma$ = 0.12\,mJy\,beam$^{-1} \times$ -4, 4, 6, 10, 20, 40, 60, 100, 200, 400). The ATCA and ALMA beams are shown in the bottom right and left corners, respectively. The ATCA synthesized beam is 0.51$\times$0.35$''$, PA = 14.45$^{\circ}$. The inset panel shows a zoom-in of the area surrounding mm1 (stretch: -0.6 to 2\,mJy\,beam$^{-1}$). \label{ATCA_contin}}
\end{figure}

In Fig.~\ref{ATCA_contin_Apr} we present the $\sim$1$''$ resolution ATCA 22\,GHz or 1.36\,cm continuum emission observed in April 2012. The emission shows a large HII region extending $\sim$10$''$ to the south of mm1, and is very similar in morphology to that presented by \citet{ellingsen05} at 8.59\,GHz with ATCA in the 6A array. The source mm1 itself is coincident with the northwest part of this extended emission, but there is no specific peak at this position. \citet{ellingsen05} measured an integrated flux density of 363\,mJy at 8.59\,GHz for the extended HII region with the 6A array. The corresponding flux density of the emission we measure at 23.712\,GHz with the 1.5B array, the image with the lowest noise, is 300$\pm$30\,mJy (assuming a 10\% absolute flux calibration error). Given these observations have similar beam sizes, this would indicate a spectral index $\alpha$ of $-0.19\pm$0.14, which is consistent with an optically thin extended HII region. 

The higher-resolution 24.328\,GHz or 1.23\,cm continuum image taken with the 6A array in September 2012 is shown in Fig.~\ref{ATCA_contin}, along with the ALMA 1.21\,mm continuum as orange contours. There is a bright bar of 1.23\,cm emission in the south of the image, which is coincident with the millimeter sources mm3, mm12, mm13, and mm17. Therefore, these millimeter sources likely form part of the extended free-free emission from the aforementioned bar seen at longer wavelengths, and the masses and column densities given in Table~\ref{dendro_contin_table} for these sources should be viewed with caution as they may be contaminated or wholly explained by ionized gas emission.

There is a second bar of 1.23\,cm continuum emission to the north, associated with mm4 and mm7 and pointing radially away from mm1, which forms the brightest emission in a faint arc seen at $<$3$\sigma$. The source mm1 lies on the edge of this arc, suggesting the centimeter emission could be tracing the ionized edges of a cavity formed by the feedback from mm1. An alternative interpretation would be that the radially positioned bar forms part of a jet from mm1, in which mm4 is a knot.

Finally, there is an unresolved 1.23\,cm continuum source associated with mm1, which is shown in further detail in the inset panel of Fig.~\ref{ATCA_contin}. A gaussian fit to the source gives peak and integrated flux densities of 1.25$\pm$0.24\,mJy\,beam$^{-1}$ and 1.33$\pm$0.44\,mJy, respectively. Given the source is unresolved, its dimensions are $<0.48''\times0.18''$ (PA$\sim$14.45$^{\circ}$), corresponding to a physical size of $<2000\,$au$\times760\,$au. The integrated flux density from a gaussian fit to the ALMA emission at 1.21\,mm is 50 $\pm$4\,mJy (J15). The spectral index between 1.23\,cm and 1.21\,mm is therefore 1.56$\pm$0.15. As the smallest spectral index for the emission that can be produced by dust emission is $\alpha$=2 in the case of extreme grain growth (for which $\beta$ would tend to 0), this confirms that the emission seen at 1.23\,cm cannot be explained by dust emission alone and that some of the emission is due to ionized gas. 

We now determine the minimum contribution of ionized gas to the 1.23\,cm continuum, assuming that the spectral index of the dust does not change with wavelength. As found in Section~\ref{sec:freefreecontrib}, the spectral index observed at 1.21\,mm is 3.4$\pm$0.2. Since the free-free spectral index cannot be larger than 2, corresponding to optically thick emission, we know that the dust spectral index therefore has to be $>3.4\pm0.2$. Assuming a dust spectral index of $3.4\pm0.2$ as a lower limit, and that all of the 1.21\,mm emission is from dust (as an upper limit), we obtain a contribution of $<0.019^{+0.011}_{-0.007}$mJy from dust at 1.23\,cm, meaning that the dust contributes $<1.4^{+0.8}_{-0.5}$\% to the total flux at this wavelength. Therefore, the 3$\sigma$ upper limit to the contribution from dust is 3.8\%.

In Section~\ref{sec:freefreecontrib}, we determine that the ionized gas emission contributes $<$13\% of the total flux at 1.21\,mm, and above we have shown that the ionized gas contributes $>$96\% of the flux at 1.23\,cm. Thus, the resulting ionized gas spectral index between 1.23\,cm and 1.21\,mm is $<$0.7.

\subsection{Ammonia Emission} \label{sec:nh3}
\begin{figure}
\begin{center}
\includegraphics[width=9.cm]{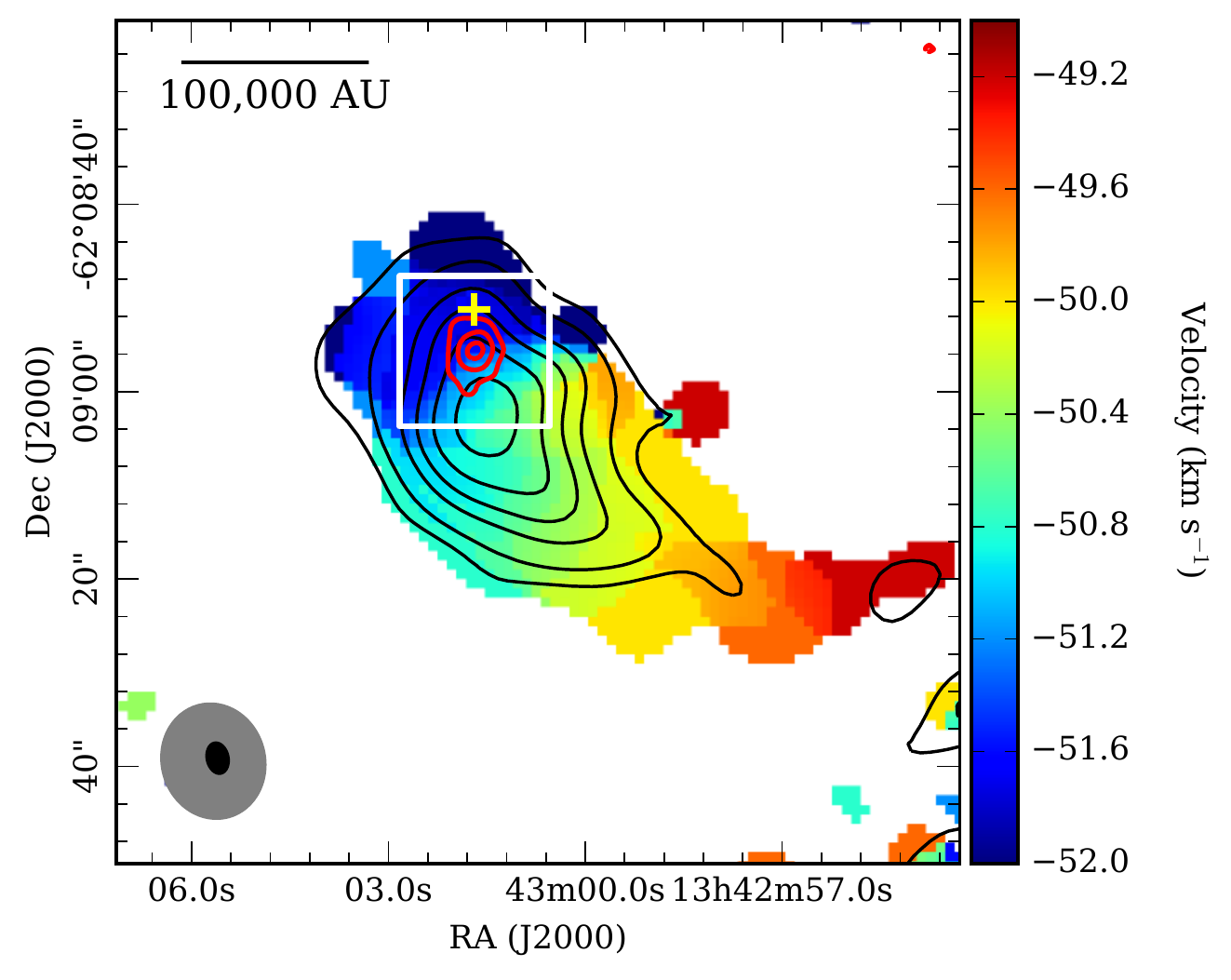}
\end{center}
\caption{First-moment map of the NH$_3$(1,1) emission observed with ATCA in colorscale. Black contours show the integrated NH$_3$(1,1) emission (local $\sigma$ = 15\,mJy\,beam$^{-1}$\,km\,s$^{-1} \times$ -3, 3, 5, 8, 10, 12, 14). Red contours show the integrated H68$\alpha$ emission (local $\sigma$ = 30\,mJy\,beam$^{-1}$\,km\,s$^{-1} \times$ -5, 5, 10, 15). The NH$_3$(1,1) and H68$\alpha$ beams are shown, respectively, in gray and black in the bottom left corner, and their properties are given in Table~\ref{ATCAspws}. The yellow plus sign marks the position of AFGL\,4176 mm1. The white box shows the area covered by Fig~\ref{ATCA_H68alpha}. \label{ATCA_NH3_11}}
\end{figure}

Figure~\ref{ATCA_NH3_11} presents the first- and zeroth-moment maps of the ATCA NH$_3$(1,1) emission from the region surrounding AFGL4176 on parsec scales. To aid the detection of the extended emission present in the data, we imaged the NH$_3$(1,1) line with a Gaussian taper of 10$''$, which was a compromise to ensure detection and reasonable resolution. As previously shown in \citet{johnston14a}, there is a clear velocity gradient across the NH$_3$(1,1) emission, oriented from northeast to southwest, which is in the same sense as the velocity gradient seen in the molecular gas detected with ALMA on smaller scales (e.g. Fig.~\ref{momentsfig}). Thus, we may be observing the large-scale rotation of the envelope or toroid in which the AFGL 4176 disk lies. Alternatively, the toroid may instead be a filament with a velocity gradient (possibly due to infall) from which the AFGL\,4176 mm1 disk has inherited its sense of rotation. However, only one low-surface brightness clump \citep[AGAL 308.944+00.121,][]{urquhart14a} lies close to the ATLASGAL source associated with AFGL\,4176 mm1 (AGAL 308.917+00.122). Therefore, the cloud is not clearly linked to another high-mass clump that would cause such a large-scale mass flow. At a flux level of 45\,mJy\,km\,s$^{-1}$, the diameter of the zeroth-moment emission spans $\sim$50$''$ or $\sim$1\,pc, giving a radius of $\sim$0.5\,pc or 100,000\,au. 

Fig.~\ref{ATCA_NH3_11} also shows the H68$\alpha$ emission in red contours (which we discuss in more detail in Section\,\ref{sec:recombline}), with a white box delineating the area covered by Figure~\ref{ATCA_H68alpha}. The peak of the H68$\alpha$ emission is roughly coincident with the center of the bar of continuum emission seen at 1.23\,cm (Fig.~\ref{ATCA_contin}). Therefore, the millimeter emission observed with ALMA, including AFGL\,4176 mm1, also lies at the far blueshifted end of the toroid, where the velocities are $>$-51\,km\,s$^{-1}$, which in fact agrees with the velocities of the molecular lines detected with ALMA (see, e.g., Fig.~\ref{vel_vfwhm_trends}). The peak position of NH$_3$(1,1) -- 13$^h$43$^m$01$^s.$47 $-$62$^{\circ}$09$'$02.5$''$ (J2000) -- is offset 11.4$''$ or 48,000\,au from the position of AFGL\,4176 mm1. This large offset could suggest that parts of the rotating envelope may be locally unstable to fragmentation, and that fragmentation is occurring on the blueshifted side of the envelope. It could also suggest that, in addition to compact HII region and AFGL\,4176 mm1, there may be other forming stars within the envelope. However, we detect no ALMA 1.21\,mm continuum emission at the position of the NH$_3$(1,1) peak (although as this lies 7.5$''$ from the pointing center, the sensitivity here is decreased by 20\%). 

We determined the temperature in the large-scale envelope by fitting the NH$_3$(1,1) and NH$_3$(2,2) emission. We fit Gaussians to the main component, and hyperfine components in the case of NH$_3$(1,1), for each pixel in the NH$_3$(1,1) and NH$_3$(2,2) cubes and determined the rotation temperature T$_{\rm rot}$ via 

\begin{equation} T_{\rm rot} = \frac{- T_{0}}{\ln\{ \frac{-0.283}{\tau} \ln{[1 - \frac{S_{\rm peak}{(2,2)}}{S_{\rm peak}{(1,1)}} (1 - e^{-\tau})]}\}} \end{equation}

\noindent \citep{ho79a}, where T$_0 \simeq$41.5\,K, $\tau$ is the optical depth of the main line, and $S_{\rm peak}{(1,1)}$ and $S_{\rm peak}{(2,2)}$ are the fitted peak fluxes associated with each line. The kinetic temperature T$_{\rm k}$ was then found by iteration using the relationship given by \citet{walmsley83a} and updated by \citet{swift05a},

\begin{equation} T_{\rm k} = T_{\rm rot} \left( 1 + \frac{T_{\rm k}}{T_0} \ln{\left[1 + 0.6 \exp{ \left( \frac{-15.7}{T_{\rm k}} \right)} \right] } \right) \label{Tk}.\end{equation}

Only pixels where the peak flux density in the main line was $>7\sigma$ in both transitions were fit. When fitting for the optical depth $\tau$, we found that the fitted values were close to zero near the center of the toroid or clump, while they were higher at the edges. Given that this effect is likely due to the lower signal-to-noise at the clump edges, we instead chose to hold $\tau$ at zero during the fitting, assuming therefore that the NH$_{3}$ clump is optically thin.

We also accounted for the effects of depopulation of the lower two energy levels at higher temperatures by applying a correction factor determined by comparing Equation \ref{Tk} with the more exact results of \citet{maret09a}. The resulting correction can be approximated by the relation (with both temperatures in kelvin):

\begin{equation} T_{\rm k,corr} = 1.09 \,T_{\rm k} - 1 \end{equation} 

\begin{figure}
\begin{center}
\includegraphics[width=8.2cm]{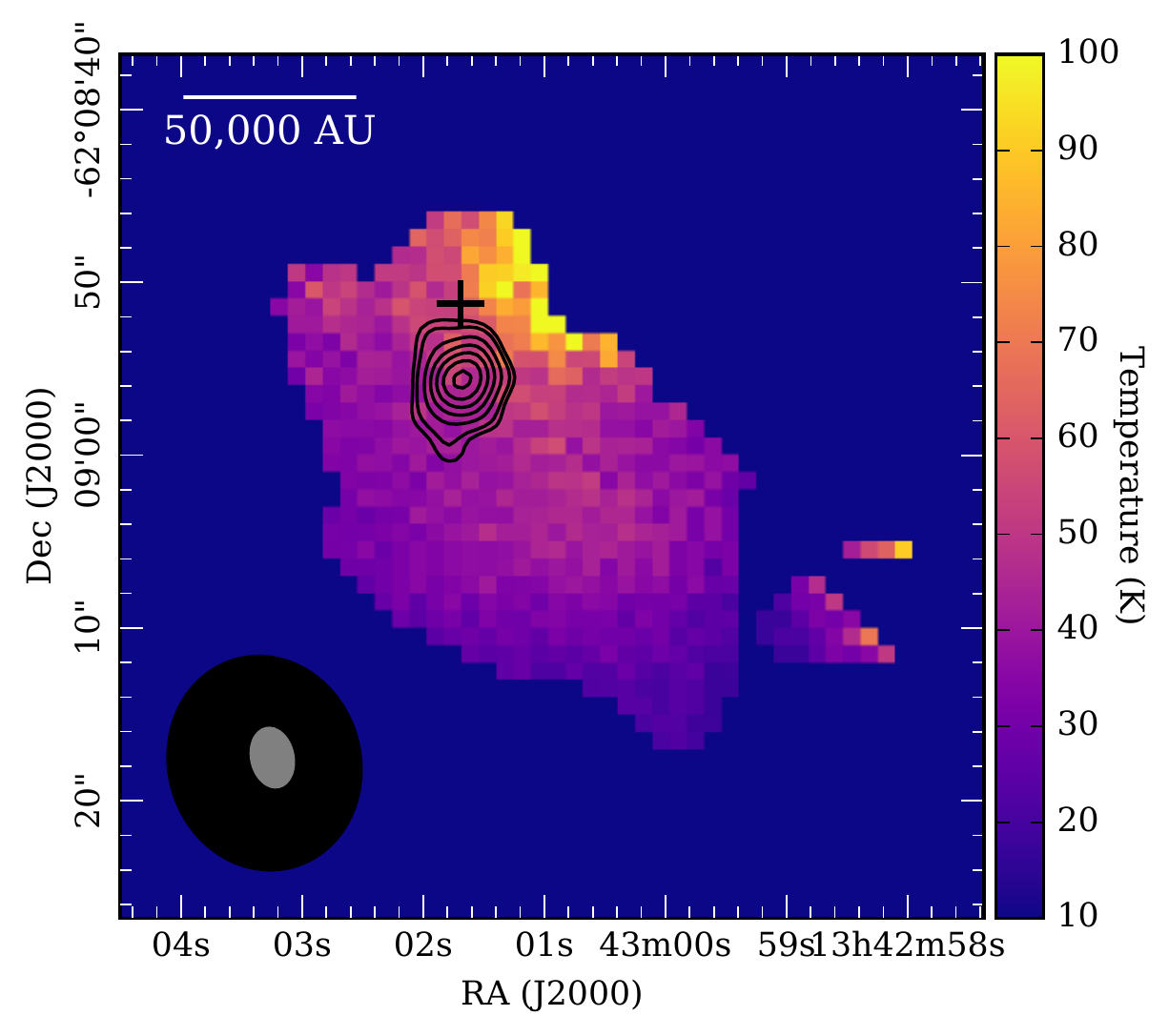}
\end{center}
\caption{Temperature map ($T_{\rm k,corr}$) derived from the observed ATCA NH$_3$(1,1) and NH$_3$(2,2) emission. Black contours show the integrated ATCA H68$\alpha$ emission (local $\sigma$ = 30\,mJy\,beam$^{-1}$\,km\,s$^{-1} \times$ -5, 5, 6, 8, 10, 12, 14, 16). The NH$_3$(1,1) and H68$\alpha$ beams are shown, respectively, in black and gray in the bottom left corner, and their properties are given in Table~\ref{ATCAspws}. The plus sign marks the position of AFGL\,4176 mm1. \label{ATCA_temp}}
\end{figure}

The resulting temperature map is shown in Fig.~\ref{ATCA_temp}, with the integrated H68$\alpha$ emission shown by black contours. The map shows a gradient in temperature, which increases from south to north, peaking at the northwestern edge of the clump. Although this temperature peak does not coincide with the position of mm1, it is interesting to note that it is reasonably nearby. Thus, this temperature gradient may be evidence of heating by mm1 and the compact HII region in the north of the clump. 

We determined the total H$_2$ mass of the toroid or clump from the NH$_3$ emission using Equations (15) and (16) of \citet{rosolowsky08a} in the optically thin limit and assuming an NH$_3$ abundance of 10$^{-7.5}$ \citep{urquhart15a}. The maps of the linewidth and the flux of the NH$_3$(1,1) derived from the NH$_3$ line fitting described above were used to determine maps of the parameters $\sigma_{\nu}$ and the brightness temperature (which can replace $T_{ex} \tau$ in equation 16 of \citet{rosolowsky08a} in the optically thin limit), respectively. Via this method, we found a map of the column density across the clump and find the total H$_2$ mass to be 9200\,M$_{\odot}$.

The integrated ATLASGAL flux density at 870\,$\mu$m is 17.32\,Jy \citep{urquhart14a}. Using Equation 1 and assuming a dust opacity at 870\,$\mu$m of 0.43\,cm$^2$\,g$^{-1}$ \citep[][with $R_V=5.5$]{draine03a,draine03b}, a temperature of 35.7\,K (the median temperature determined in the analysis above), and the remaining assumptions laid out in Section~\ref{sec:cont_src_prop}, we determine the mass of the associated dust and gas clump to be 5100\,M$_{\odot}$. If we instead assume an opacity of 1.85\,cm$^2$\,g$^{-1}$ \citep[e.g.][]{urquhart13a} derived from the opacities of the 10$^6$\,cm$^{-3}$ thin ice mantle dust model in \citet{ossenkopf94}, we obtain a dust-plus-gas mass of 1200\,M$_{\odot}$. The mass of the clump determined from the Draine opacities is within a factor of two of the mass determined from NH$_3$, confirming that there is a large reservoir of mass available for star formation around the AFGL\,4176 mm1 disk and its surrounding millimeter sources.

\begin{figure}
\begin{center}
\includegraphics[width=9.cm]{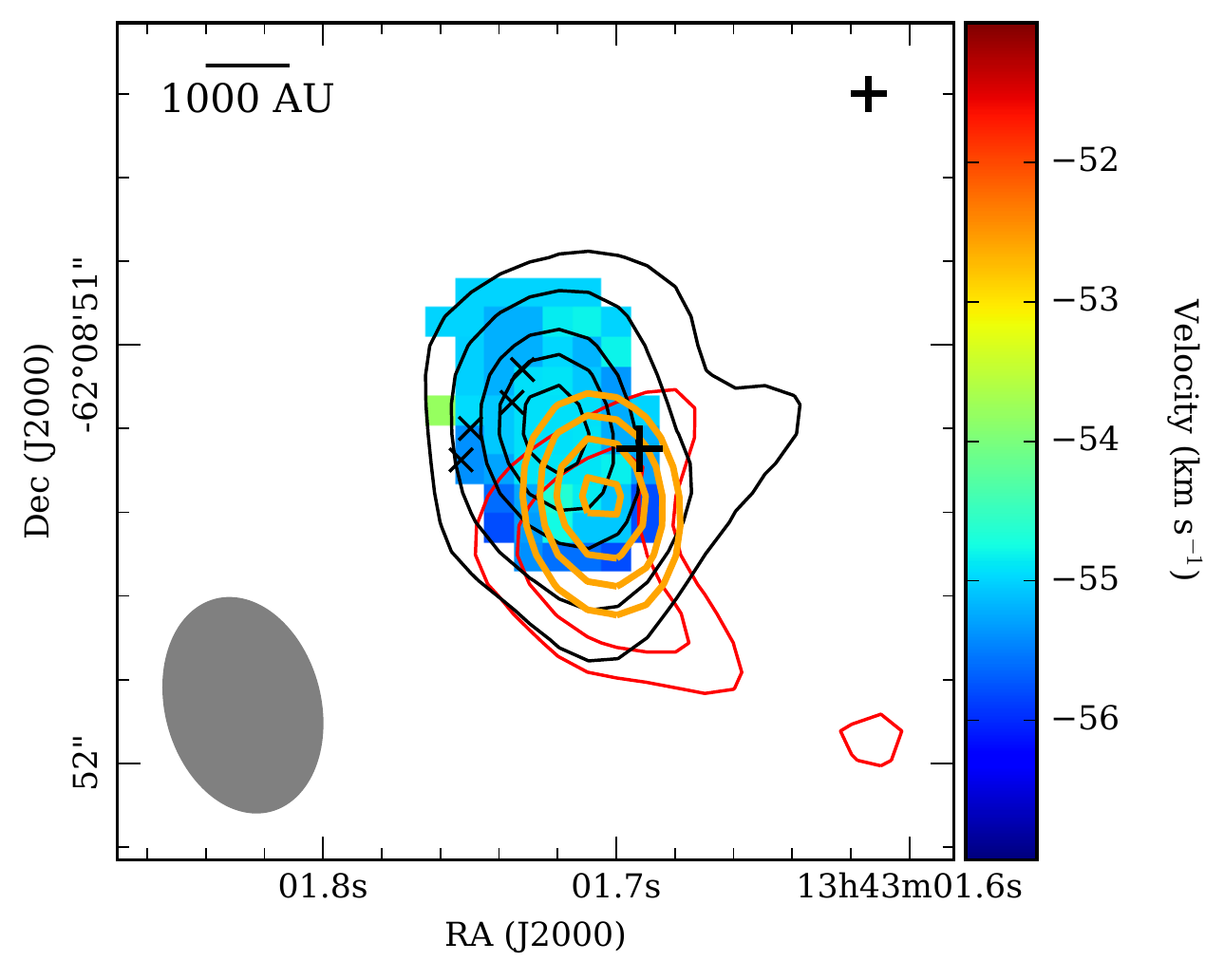}
\end{center}
\caption{First-moment map of the NH$_3$(5,5) emission observed with ATCA in colorscale. Black contours show the integrated NH$_3$(5,5) emission (local $\sigma$ = 6\,mJy\,beam$^{-1}$\,km\,s$^{-1} \times$ -3, 3, 5, 8, 10, 12, 14, 16). Red contours show the integrated NH$_3$(4,4) emission (local $\sigma$ = 6\,mJy\,beam$^{-1}$\,km\,s$^{-1} \times$ -3, 3, 4, 5), and orange contours show the 1.23\,cm continuum as in Fig.\,\ref{ATCA_contin}. The NH$_3$(5,5) beam is shown in gray in the bottom left corner, which has similar properties for the NH$_3$(4,4) and continuum images (see Table~\ref{ATCAspws}). The positions of mm1 and mm2 are shown as large and small plus signs, respectively, and the crosses mark the position of the Class II methanol masers detected by \citet{phillips98}. \label{ATCA_NH3_moms}}
\end{figure}

Figure~\ref{ATCA_NH3_moms} shows the first-moment map of the NH$_3$(5,5) line observed with ATCA, overplotted with contours of the integrated emission of both NH$_3$(4,4) and NH$_3$(5,5), as well as 1.23\,cm continuum. In contrast to NH$_3$(1,1) and NH$_3$(2,2), the emission from these more highly excited transitions lies close to mm1, marked by the large plus sign in Fig.~\ref{ATCA_NH3_moms}. The NH$_3$(5,5) emission is brighter than that of the NH$_3$(4,4) and peaks 0.2$''$ (840\,au) to the east of mm1, whereas the NH$_3$(4,4) emission peaks to the southeast. Correspondingly, the velocities shown in the NH$_3$(5,5) first-moment map are blueshifted ($\sim-$55\,km\,s$^{-1}$) compared to most of the lines observed with ALMA (e.g., Fig.\,\ref{vel_vfwhm_trends}). The upper energy levels of these two transitions lie at 201.092 and 295.942\,K for NH$_3$(4,4) and NH$_3$(5,5), respectively. Although no trend in velocity is seen with upper energy level for the lines detected with ALMA, it may be that the NH$_3$(5,5) emission is tracing a patch of gas hotter than $\sim$200\,K within the blueshifted part of the circumstellar material that is not detectable in NH$_3$(4,4). It is unexpected that the hottest gas is not found at the peak position and velocity of the continuum source mm1, where heating from the young star would excite the gas. This observation may be indicating that some process, such as shocks, could be heating the gas in the blueshifted side of the disk. Thus, it appears that similar physical processes are exciting both the NH$_3$(5,5) and the blue-dominant lines such as methanol that are presented in Section \ref{sec:bluedominant}.

\section{Combined Results} \label{sec:CombinedResults}
\subsection{Outflow-associated Emission from C$^{34}$S} \label{sec:c34s}
\begin{figure*}
\begin{center}
\includegraphics[width=18.cm]{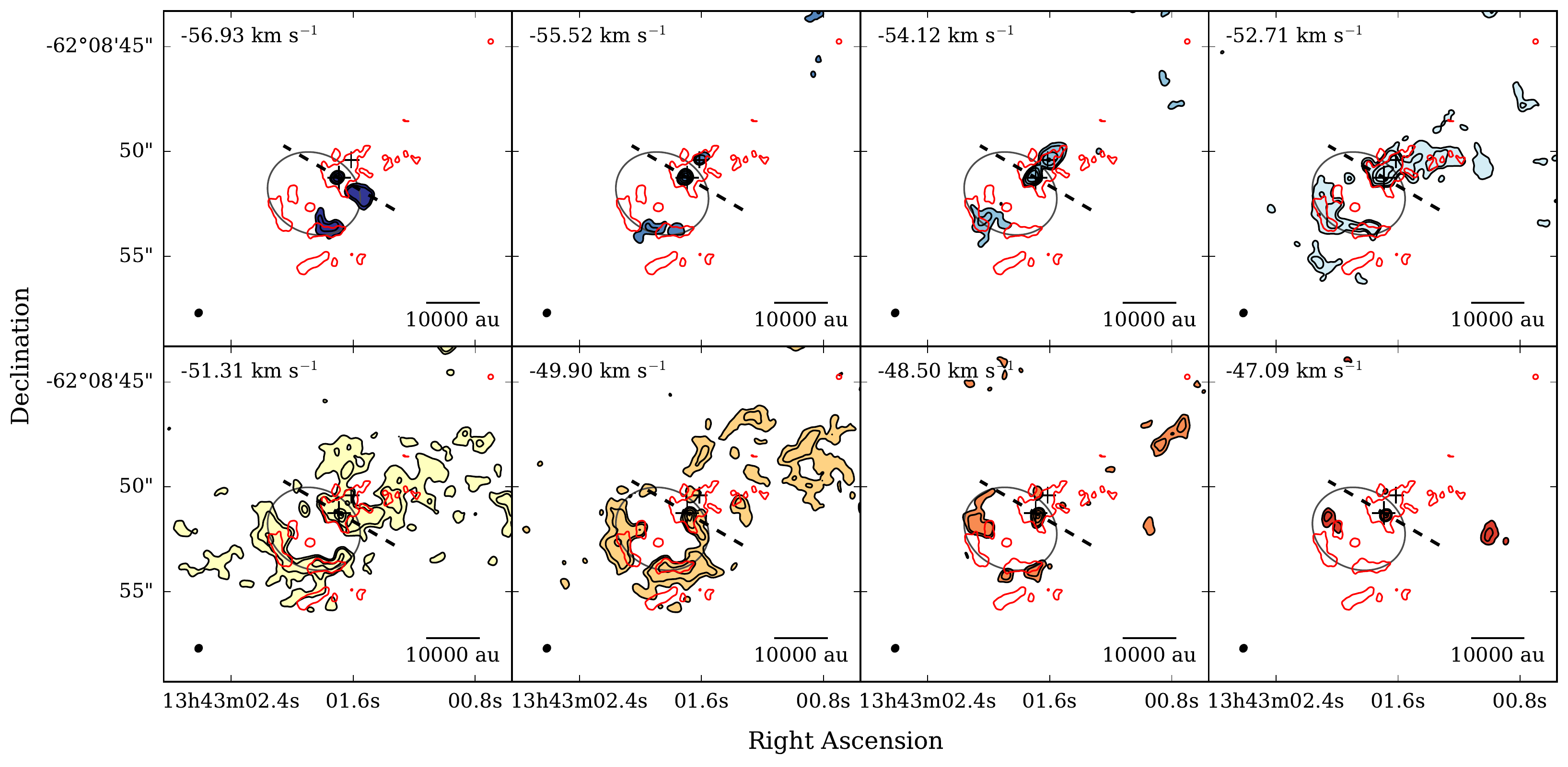}
\end{center}
\caption{Channel map of the C$^{34}$S (5-4) line, made using a combination of ALMA and APEX data, imaged at 1.41\,km\,s$^{-1}$ spectral resolution between -56.93 and -47.09\,km\,s$^{-1}$. Contours show 5, 10, 25, and 50 $\times$ 4\,mJy\,beam$^{-1}$ (the local rms noise), and the contours are filled with a colorscale to represent the channel velocity. The positions of mm1 and mm2 are shown as large and small plus signs, respectively, and the dashed line shows the position angle of the disk seen in 1.21\,mm dust continuum (59$^{\circ}$). The red contour shows the 5$\sigma$ 1.2\,mm continuum emission. The gray ellipse traces the loops seen in the SE side of the outflow (see Section~\ref{sec:c34s}). \label{C34S_chanmap}}
\end{figure*}

While other molecules have comparatively compact emission, there are three tracers detected with ALMA that show extended emission that reaches distances $>2''$ from mm1: CH$_3$CCH, H$_2$CS, and C$^{34}$S. Figure~\ref{C34S_chanmap} presents the combined ALMA + APEX C$^{34}$S (5-4) channel maps, which display emission extended perpendicular to the position angle of the major axis of the disk (shown as a dashed line). There is an arc of emission to the southeast of mm1, seen most clearly in the {-51.31\,km\,s$^{-1}$} channel, coincident with the 1.21\,mm continuum sources mm5, mm7, and mm8 (which themselves also follow the same arc, see Fig.~\ref{contfig}). In the same channel, there is also a U-shaped arc of emission pointing to the northwest of mm1, with mm1 positioned at the base of the U. Given the small linewidth of the C$^{34}$S~(5-4) emission (5.8\,km\,s$^{-1}$) compared to the velocities measured for the CO (3-2) outflow presented in J15 (13-16\,km\,s$^{-1}$ from the $v_{LSR}$), C$^{34}$S is unlikely to trace the fast part of the outflow, but instead a slower wide-angle wind, or dense material in the outflow cavity walls. Interestingly, another example of where C$^{34}$S was found to be tracing a wide-angle outflow was the Class 0 protostar HH212 \citep{codella14a}, a source that also has signatures of a Keplerian disk. \citet{codella14a} find that the C$^{34}$S emission becomes more collimated at higher velocities, which can be also be seen to the northwest of mm1 in Fig.~\ref{C34S_chanmap} at -52.71\,km\,s$^{-1}$ and again more collimated at -54.12\,km\,s$^{-1}$. They suggest this demonstrates that the wide-angle outflow of HH212 has an onion-like structure, with higher-velocity material closer to the jet axis, which may also be what we are seeing hints of here for AFGL\,4176. 

Assuming that the C$^{34}$S emission traces the walls or outer shell of a parabolic outflow described as $z = z_0 (\varpi/\varpi_0)^{1.5}$, where $z$ is the height from the x-y midplane, $\varpi$ is the cylindrical radius, and $z_0$ and $\varpi_0$ are constants, we can find the half-opening angle of the structure. Taking the inclination of 30$^{\circ}$ found from our modeling of AFGL\,4176 carried out in J15, we measured the size of the southeast outflow in the -51.31 and -49.90\,km\,s$^{-1}$ channels by approximating the emission by an ellipse with a fixed aspect ratio (shown in Fig.~\ref{C34S_chanmap}). This ellipse was measured to have semi-major and -minor axes of $\sim$2.2 $\times$ 1.9$''$ or $\sim$9200 $\times$ 8000\,au, respectively. The distance on the sky of mm1 to the center of the ellipse is $\sim$1.4$''$. Assuming an inclination of 30$^{\circ}$, the true distance between the center of the ellipse and mm1 is therefore $z_0\sim$2.8$''$ or 12000\,au at a cylindrical radius of $\varpi_0=$9200\,au. From this, we could determine that the half-opening angle of the outflow or cavity wall at 1.5$\times 10^{5}$\,au is $\sim$19$^{\circ}$. This is in reasonable agreement with the half-opening angle of 10$^{\circ}$ at 1.5$\times 10^{5}$\,au that was assumed for our previous modeling, as this provided a good fit to the SED.

\subsection{Hydrogen Recombination Line Emission} \label{sec:recombline}

\begin{figure}
\begin{center}
\includegraphics[width=9.cm]{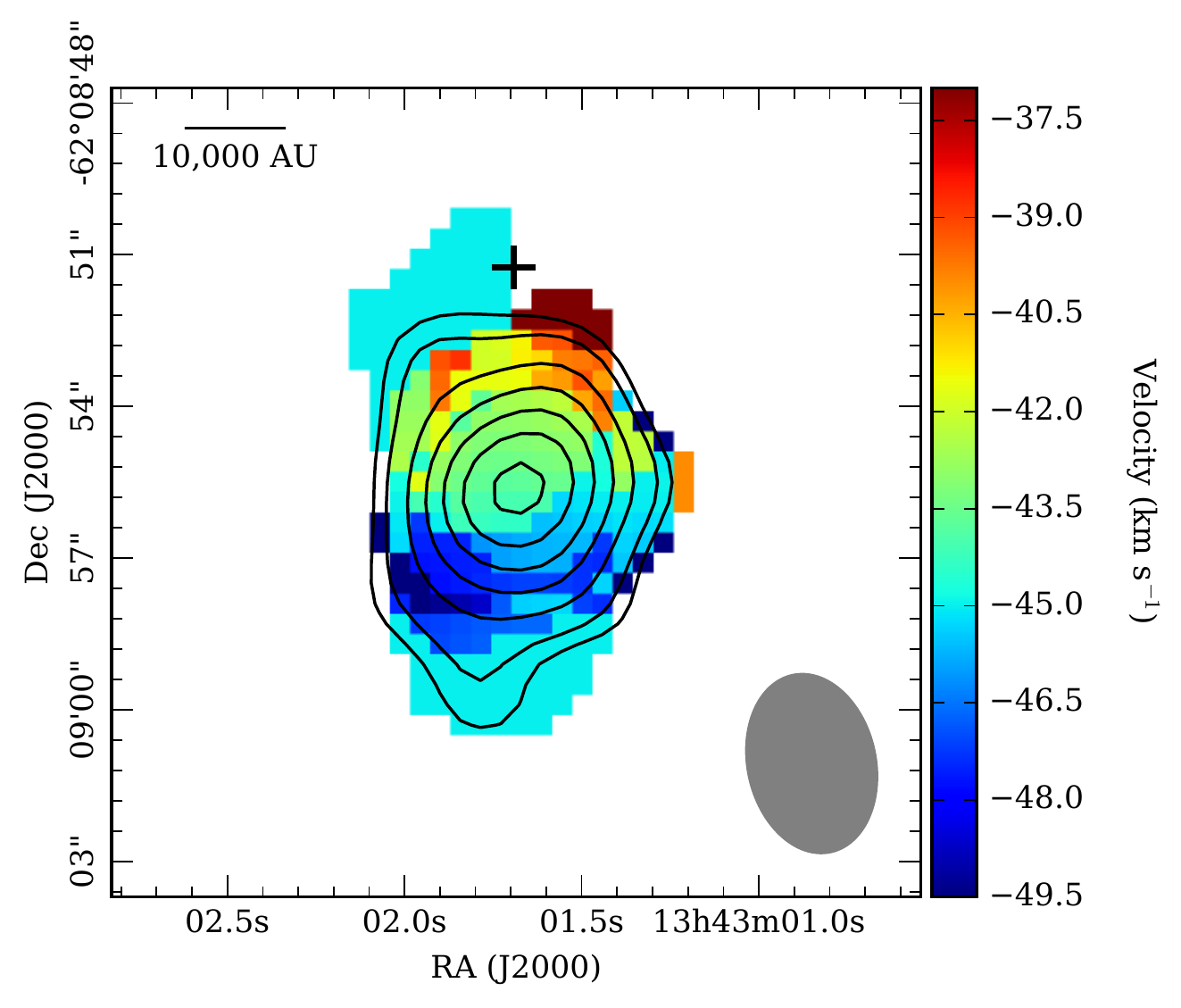}
\end{center}
\caption{First-moment map of the ATCA H68$\alpha$ emission shown in colorscale. Black contours show integrated H68$\alpha$ emission (local $\sigma$ = 30\,mJy\,beam$^{-1}$\,km\,s$^{-1} \times$ -5, 5, 6, 8, 10, 12, 14, 16). The black plus sign marks the position of AFGL\,4176 mm1. The beam is shown in the bottom right corner; its properties are given in Table~\ref{ATCAspws}. \label{ATCA_H68alpha}}
\end{figure}

Hydrogen radio recombination line (RRL) emission was observed and detected with both  ATCA and ALMA. The zeroth- and first-moment map of the H68$\alpha$ line observed with ATCA in April 2012 is shown in Fig.~\ref{ATCA_H68alpha}. We display this line as it was detected with the highest signal-to-noise of all the observed ATCA RRLs. The HII region traced by the H68$\alpha$ emission, like most of the 1.23\,cm continuum, lies to the south of AFGL\,4176 mm1. The first-moment map shows that there is an NNW-SSE velocity gradient (red to blueshifted, respectively) in the ionized gas on scales of 10,000\,au. This does not agree with the velocity gradient seen in the NH$_3$ emission, which agrees more closely with that of the disk rotation seen in tracers like CH$_3$CN. It is also interesting to note that the central velocity of the H68$\alpha$ line ($\sim-44$\,km\,s$^{-1}$) is different to that of the molecular tracers, which lie around $-52$ to $-53$\,km\,s$^{-1}$ (e.g., Fig.\,\ref{vel_vfwhm_trends}). We fitted the mean H68$\alpha$ spectrum measured within a 1$''$ radius circular aperture centered on 13$^h$43$^m$01$^s.$672 -62$^{\circ}$08$'$55.5$''$ (J2000). The resulting flux, central velocity, and linewidth are 18.0$\pm$1.1\,mJy\,beam$^{-1}$, -44.4$\pm$0.8\,km\,s$^{-1}$, and 29.0$\pm$2.0\,km\,s$^{-1}$, respectively, with the stated uncertainties determined only from the scatter in the data. Interestingly, the emission becomes more redshifted closer to AFGL\,4176 mm1. If the ionized gas is flowing from AFGL\,4176 mm1, this would indicate that the redshifted side of the ionized flow has a higher redshifted velocity closer to the source compared to the systemic velocity of mm1 ($v_{LSR}$=52-53\,km\,s$^{-1}$). This velocity structure might be explained by a turbulent jet model \citep{arce07a} and the fact that the gas is ionized by radiative ionization or strong shocks in the outflow. Alternatively, the velocity gradient could be tracing dynamics driven by a separate source within this HII region, and/or by champagne flows due to varying density in the surrounding cloud. This seems the most likely explanation, as the bright bar of 1.23\,cm continuum emission in the south of Fig.~\ref{ATCA_contin} is coincident with the peak of the integrated H68$\alpha$ emission seen in Fig.~\ref{ATCA_H68alpha} and is centered on the velocity gradient within it, indicating that an independent source associated with this bar of continuum emission is probably producing the H68$\alpha$ emission.
 
We were only able to detect the remaining RRLs observed in December 2012 (H64$\alpha$, H65$\alpha$, H67$\alpha$, and H68$\alpha$) by finding their mean spectra over the area of sky covering the April H68$\alpha$ emission. We also fit these lines with Gaussians, and determine their peak flux densities to be 13.3$\pm$1.4, 12.9$\pm$1.9, 14.6$\pm$1.9, and 14.9$\pm$1.5\,mJy\,beam$^{-1}$, respectively, which are consistent to within 1$\sigma$. Similarly, the velocities of the lines are $-43.9\pm$1.3, $-47.4\pm$2.4, $-41.9\pm$1.9, and $-44.5\pm$1.2\,km\,s$^{-1}$, and the linewidths are 25.5$\pm$3.2, 33.7$\pm$5.7, 30.5$\pm$4.5, and 24.4$\pm$2.9\,km\,s$^{-1}$, which are consistent to within 2$\sigma$. As no trends can be deduced from these, the line properties are thus consistent with a single flux, velocity, and linewidth and therefore likely trace a similar range of densities within the HII region.

We also detect H29$\alpha$ with ALMA. This H29$\alpha$ line emission is extended and faint, so that the resulting images are not useful. However, we measured a mean spectrum in an 5$''$(R.A.) $\times$ 3.5$''$(Dec.) ellipse centered on the same position as the circular aperture used for the ATCA RRLs. This resulted in a 9.3$\sigma$ detection; the peak flux density, central velocity and linewidth from a gaussian fit are 0.58$\pm$0.1\,mJy\,beam$^{-1}$, -47.8$\pm$2.1\,km\,s$^{-1}$, and 25.7$\pm$5.0\,km\,s$^{-1}$, respectively. Although, due to large uncertainties, the linewidths of the H29$\alpha$ and H64, 65, 67, and 68$\alpha$ lines are consistent within errors, we calculated the probability distribution of electron density $n_e$ for each pair of lines (i.e. H29$\alpha$ paired with one of the remaining lines) using Monte Carlo statistics and the method outlined in \citet{keto08}. We then multiplied these distributions to find the combined probability distribution for $n_e$ and found the 99.7th\,percentile provided an upper limit of $n_e < 6.6 \times 10^5$\,cm$^{-3}$, which given the size of the emission would be consistent with a compact HII region.

\section{Discussion} \label{sec:discussion}
\subsection{Comparison to Other Massive YSOs}  \label{sec:comparison}

\begin{deluxetable*}{llcccccccc}
\tablecaption{Source properties for new and updated sources displayed in Fig.~\ref{Beltranplot} \label{Beltrantable}}
\tablehead{\colhead{Name} & \colhead{d} & \colhead{$\mathrm{L_{\rm bol}}$} & \colhead{$\mathrm{M_{\rm gas}}$} & \colhead{$\mathrm{R_{\rm cont}}$} & \colhead{$\mathrm{R_{\rm vel}}$} & \colhead{$\mathrm{v_{\rm rot}}$} & \colhead{$\mathrm{M_{\star}}$} & \colhead{$\mathrm{t_{\rm ff}/t_{\rm rot}}$} & \colhead{References}\\ 
\colhead{ } & \colhead{(kpc)} & \colhead{($\mathrm{L_{\odot}}$)} & \colhead{($\mathrm{M_{\odot}}$)} & \colhead{(au)} & \colhead{(au)} & \colhead{$\mathrm{(km\,s^{-1})}$} & \colhead{($\mathrm{M_{\odot}}$)} & \colhead{}}
\startdata
AFGL\,4176 & 4.2 & 1 $\times$ 10$^5$ & 0.91 & 870 & 1700 & 4.0 & 25 & 1.03 & 1 \\
W3(H$_2$O)\,E & 2.0 & 2 $\times$ 10$^4$ & 8.8\tablenotemark{a} & 1150 & 1000 & 4.0 & 15 & 0.31 & 2\\
W3(H$_2$O)\,W & 2.0 & 2 $\times$ 10$^4$ & 6.5\tablenotemark{a} & 1150 & 1000 & 3.0 & 15 & 0.27 & 2\\
G328.2551$-$0.5321 & 2.5 & 1.3 $\times$ 10$^4$ & 0.15 & 250 & 800 & 9.0 & 13.5 & 3.90 & 3\\ 
G328.2551$-$0.5321 inner env. & 2.5 & 1.3 $\times$ 10$^4$ & 4.7 &1500 & 2000 & 4.5 & 13.5 & 0.55 & 3\\
G31.41$+$0.31 & 3.7 & 4.4 $\times$ 10$^4$ & 21\tablenotemark{a} & 1200 & 2400 & 7.5 & 38\tablenotemark{b} & 0.48 & 4,5\\
G23.01$-$0.41 & 4.6 & 4 $\times$ 10$^4$ & 1.5 & 2000 & 2000 & 3.5 & 20 & 0.76 & 6\\
IRAS\,18162$-$2048\,MM1 & 1.7 & 2 $\times$ 10$^4$ & 1.3 & 291 & 1700 & 3.4 & 14 & 0.73 & 7,8 \\
G11.92$-$0.61\,MM1 & 3.37 & 1 $\times$ 10$^4$ & 0.84 & 480 & 850 & 6.5 & 34  & 1.23 & 9\\
Orion Src I & 0.415 & 1 $\times$ 10$^5$ & 0.1 & 50 & 80 & 13.0 & 15 & 2.18 & 10,11\\ 
IRAS\,16547$-$4247 & 2.9 & 1 $\times$ 10$^5$ & 1.8 & $\sim$1000 & 1190 & 4.6\tablenotemark{c} & 20 & 0.70 &12\\ 
G17.64$+$0.16 & 2.2 & 1 $\times$ 10$^5$ & 1.0 & 120 & 120 & 17.9 & 45 & 1.16 &13\\
G339.88$-$1.26 & 2.1 & 4 $\times$ 10$^4$ & 2.8 & 530 & 530 & 6.0 & 12 & 0.49 &14 \\
\enddata
\tablecomments{References: 1: \citet{johnston15a}, 2: \citet{ahmadi18a}, 3: \citet{csengeri18a}, 4: \citet{beltran18a}, 5: \citet{beltran19a}, 6: \citet{sanna18a}, 7: \citet{girart17a} , 8: \citet{girart18a}, 9: \citet{ilee18a}, 10: \citet{ginsburg18a}, 11: \citet{plambeck16a}, 12: \citet{zapata19a}, 13:  \citet{maud19a}, 14: \citet{zhang19a}.
\tablenotetext{a}{These masses have not been recalculated and instead have been scaled from those given in the respective papers, as a varying temperature was used in their original calculation.}\tablenotetext{b}{Due to the updated distance to the source, this mass is an upper limit.}\tablenotetext{c}{This velocity was scaled to R$_{\rm vel}$ using the model in reference 11.}}
\end{deluxetable*}

\begin{figure}
\begin{center}
\includegraphics[width=9cm]{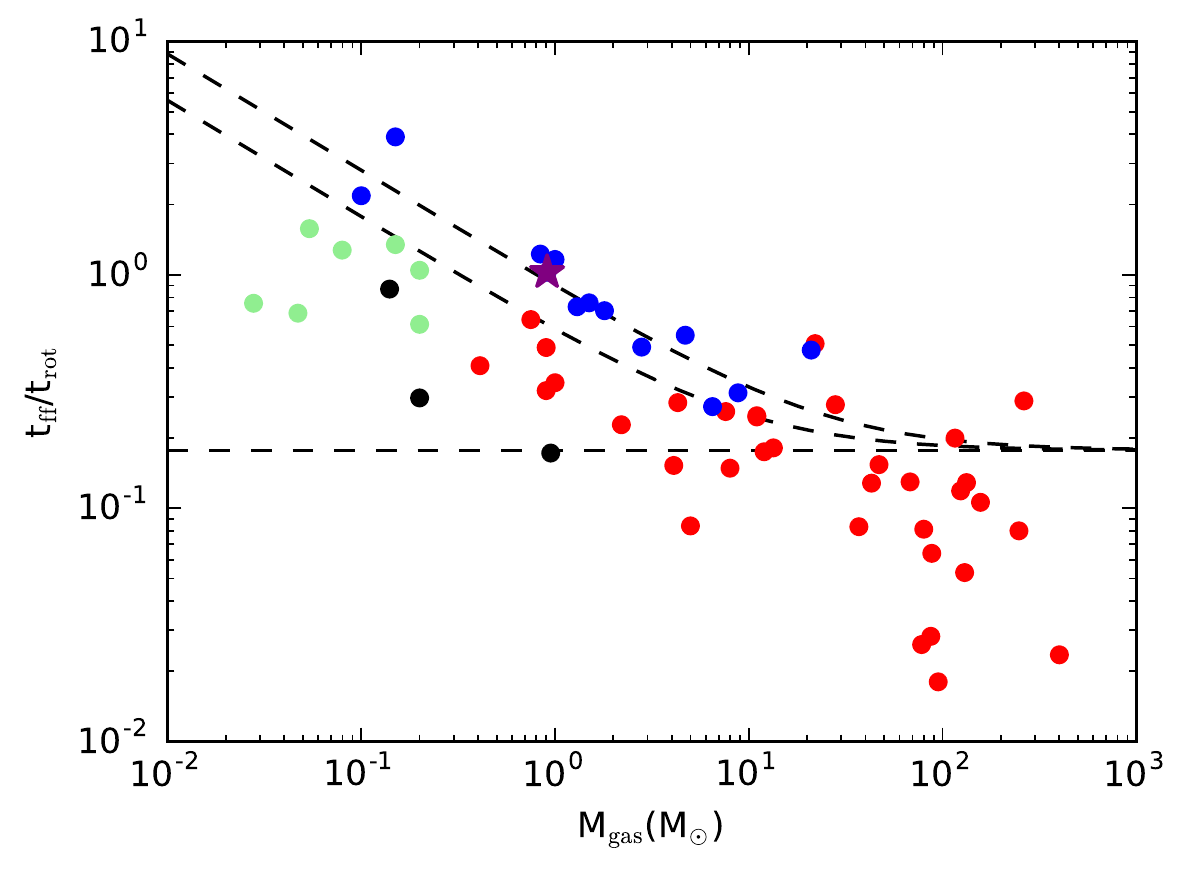}
\end{center}
\caption{Ratio of the freefall to rotation timescales against M$_{\rm gas}$ for rotating structures around forming stars. The purple star represents AFGL\,4176. The red, green, and black points represent the high-mass, intermediate-mass, and Herbig Ae sources presented in \citet{beltran16a}. The blue points represent new or updated sources, which are listed along with AFGL\,4176 in Table~\ref{Beltrantable}. From bottom to top, black dashed lines indicate theoretical values of $t_{\rm ff} / t_{\rm rot}$ for M$_{\star}$=0, 10, and 25\,M$_{\odot}$ (see Section \ref{sec:comparison} for further details). \label{Beltranplot} }
\end{figure}

We begin our comparison to other forming massive stars with rotating disk-like structures by presenting an updated version of the lower panel of Figure 14 shown by \citet{beltran16a} in our Figure~\ref{Beltranplot}. This figure plots the ratio of the freefall timescale ($t_{\rm ff}$) to the rotational period ($t_{\rm rot}$) against the gas mass of the structure ($M_{\rm gas}$). The $t_{\rm ff}/t_{\rm rot}$ ratio compares the timescale for the gas to collapse under self-gravity in absence of rotation ($t_{\rm ff}$) to the time for the structure to complete one rotation ($t_{\rm rot}$). Thus, low values of $t_{\rm ff}/t_{\rm rot}$ indicate that the structure will only complete a small number of rotations before it is accreted on to the central star, so it is unlikely to be able to reach a stable configuration within this time. 

Figure~\ref{Beltranplot} has previously been used to separate the two main types of rotating structures observed around forming massive stars: (1) toroids, rotating flattened envelopes, or pseudo-disks, which are massive ($>$100$M_{\rm \odot}$) large ($\sim$10000\,au) rotating structures that are not in centrifugal equilibrium and lie at the bottom right of the diagram, and (2) disks in Keplerian rotation that exist over many rotations and lie toward the top left. 

The properties of new or updated sources are given in Table~\ref{Beltrantable} and are shown as blue points in Figure~\ref{Beltranplot} (apart from AFGL\,4176, which is shown as a purple star). The updated values have been taken from the papers listed in Table~\ref{Beltrantable}; however, some values remain as stated in \citet{beltran16a}. 

As assumed in \citet{beltran16a}, we calculate the freefall timescale $t_{\rm ff}$ as

\begin{equation} t_{\rm ff}= \frac{\pi}{2} \sqrt{\frac{R_{\rm vel}^3}{2 G M_{\rm gas}}} \label{tff}, \end{equation}

\noindent and $t_{\rm rot}$ the rotation timescale as $t_{\rm rot} = {2 \pi R_{\rm vel}}/{v_{\rm rot}}$. We use the radius measured from the line emission ($R_{\rm vel}$) for the radius in all cases except for the case $R_{\rm vel} < R_{\rm cont}$ where we used the radius measured from the continuum ($R_{\rm cont}$) for the calculation of $t_{\rm ff}$. This was to ensure that $ t_{\rm ff}$ was evaluated at a radius that contained all of the mass of the structure. Apart from Src I, which has optically thick continuum emission, the masses determined from the dust-continuum emission $M_{\rm gas}$ for all of the sources in Table~\ref{Beltrantable} and Fig.~\ref{Beltranplot} (including AFGL\,4176 mm1) have been recalculated to be consistent with the set of assumptions stated in \citet{beltran16a}, namely that the gas-to-dust ratio is 100, the opacity at 1.4\,mm is 1\,cm$^{2}$\,g$^{-1}$ \citep{ossenkopf94}, and the dust opacity index $\beta$ is 2. The temperature used to calculate the mass for each source was taken from the papers in Table~\ref{Beltrantable}.

Figure~\ref{Beltranplot} also shows several dashed lines, which show the value of $t_{\rm ff}/t_{\rm rot}$ for structures in Keplerian rotation around stars with masses of $M_{\star} = 0, 10$ and $25\,M_{\odot}$, taking into account both the mass of the rotating structure and the star such that

\begin{equation} t_{\rm ff}/t_{\rm rot} = \sqrt{ \frac{M_{\rm gas} + M_{\star}}{32 M_{\rm gas}} }. \label{tratio} \end{equation}

Equation \ref{tff} assumes spherical symmetry, which given the flattened nature of most of these structures, leads to an overestimate of $t_{\rm ff}/t_{\rm rot}$ \citep{beltran14a}. In addition, Equation \ref{tff} does not include the mass of the central star $M_{\star}$, which could change the value of $t_{\rm ff}$ significantly in cases where $M_{\rm gas} \lesssim M_{\star}$. In fact, if $M_{\star}$ is included in the calculation of $t_{\rm ff}$ in both of these equations, this has the effect of ``flattening" Figure\,\ref{Beltranplot}, so that the y-axis then becomes a measure of the ``Keplerian-ness" of the structure; the dashed lines also become one horizontal line that shows the positions of structures in Keplerian rotation in the Figure. We will further discuss such theoretically motivated variants of Figure~\ref{Beltranplot} in Kee et al. (submitted), but for now we show Figure\,\ref{Beltranplot} as presented in previous studies for a consistent comparison. 

As can be seen in Figure\,\ref{Beltranplot}, the new or updated sources (blue points) lie mostly above the previously known sources in the ``disk" area of the diagram. Furthermore, several lie close to or above the upper dashed line for Keplerian structures around a 25\,$M_{\odot}$ star, indicating that recent studies have started to push the discovery of Keplerian-like structures around massive forming stars to even higher stellar masses, into the regime of forming O-type stars. AFGL\,4176 falls close to the dashed line representing Keplerian disks around a 25\,$M_{\odot}$ star, in agreement with our previous modeling (J15). 

Below, we expand on our brief discussion that was given in Section~\ref{sec:intro} on the variety of properties seen in disks around forming massive stars. While several sources typify the picture of a scaled-up version of a disk around a low-mass star \citep[e.g. AFGL\,4176, G11.92-0.61\,MM1, IRAS\,18162-2048 or GGD27\,MM1, G023.01-00.41, and G339.88-1.26; J15,][]{ilee16a, girart18a, sanna19a, zhang19b}, there are in fact a range of disk morphologies found toward high-mass stars. These can include multiple-disk systems \citep[e.g. NGC\,7538\,IRS1, W3(H$_2$O), and W33A\,MM1;][in the latter case fed by a large-scale accretion streamer]{beuther17b, ahmadi18a, izquierdo18a}, small disks \citep[e.g. G328.2551-0.5321 and G17.64+0.16;][]{csengeri18a, maud19a}, and no clear evidence of disks \citep[W51 and IRAS\,18566+0408;][]{ginsburg17a,silva17a}.

Within the cluster surrounding the forming high-mass star GGD 27 MM1, \citet{busquet19a} found a lack of disks close to MM1 ($<$0.02\,pc), and that larger and more massive disks only existed at distances greater than 0.04\,pc. They also found that the disks were on average smaller than more distributed environments such as Taurus. One explanation for this finding is that the rich cluster environment leads to truncation of these disks via more frequent interactions, which, along with photoevaporation, may also go toward the explanation of smaller and less massive disks in the Orion Nebula Cluster \citep{eisner18a}. As massive stars are almost always found in clustered environments \citep[e.g.][]{de-wit05}, it is expected that interactions and accretion streams in clusters are important mechanisms for shaping disks \citep{bonnell03b,bate18a}, whereas, despite their high EUV luminosity, disks around forming high-mass stars are found to be less affected by photoevaporation due to their high optical depth \citep{tanaka17a, kuiper18a}.

As further observations of forming OB stars at high resolutions are carried out, the distribution of their disk properties, for instance disk mass and radius, will act as an important discriminant between the importance of the cluster-scale environment in massive star formation. By definition, the cluster-scale environment does not play an important role for models of core accretion \citep[][]{mckee03,krumholz09}. If disk properties are found to be strongly affected by their cluster-scale environment, this suggests that a more holistic picture of star formation is required that takes into account not just the stellar core but the whole environment in which a massive star is formed.

\subsection{Chemistry in massive YSO disks \label{chemHMdisk}} 

High-mass star-formation regions are known for their chemical richness, especially in complex organic molecules (COMs), which are defined by astronomers as molecules with $\ge$6 carbon atoms \citep{herbst09a}, and this is indeed seen in our observations of AFGL\,4176. Observations of regions such as Orion KL and Sgr B2 have often been at the forefront in the detection of these molecules, due respectively to their proximity and luminosity \citep[e.g.][]{sutton85a, cummins86a}. 

Segregation between different groups of species such as that seen in our observations has also been observed in Orion KL and many other massive star-formation regions, most notably between N- and O-bearing species \citep{blake87b, feng15a}. Often, the N-bearing species are found to be more compact and closely associated with the hot cores, whereas the O-bearing species display more distributed emission \citep{blake87b, oberg13a, fayolle15a, feng15a}. There is a range of explanations put forward to account for this difference; a recent hypothesis by \citet{suzuki18a} is that the N-bearing molecules are enhanced by reactions in the hot gas ($>$100\,K) after the warm-up phase and evaporation of the icy dust mantles, as hydrogenation of these molecules into other species is not efficient at high temperatures. However, in contrast, \citet{quenard18a} find that formamide, one of the disk-tracing and N-bearing species we detect, requires radical-radical grain-surface reactions as well as gas-phase reactions to reproduce the observed abundances \citep[for further discussion, see][]{bogelund19a}. Thus, to-date there is no definitive explanation of the chemical segregation of N-/O-bearing species.

As presented in Section \ref{sec:morphologies}, the observed species in AFGL\,4176 (including 12 COMs) fall into four distinct morphological groups: disk-tracing, blue-dominant, red-dominant, and outflow-tracing lines. Except for HCOOH, all disk-tracing species contain nitrogen. Conversely, all molecules that we detect that contain nitrogen trace the disk  (excluding H$_2$CCN, which is too faint). As discussed in Section \ref{sec:disk-tracing} and can be seen from Fig.~\ref{momentsfig}, these species only trace the inner several hundred to thousand astronomical units of AFGL\,4176, similar to the results of previous studies mentioned above. In this way, AFGL\,4176 follows the template of N-bearing species being associated with the hottest and densest regions of hot cores.

Given previous difficulties in unambiguously identifying disks around high-mass stars, only a few studies have been carried out to-date that begin to examine their chemistry. \citet{isokoski13a} previously studied three high-mass YSOs with rotating disk-like structures and compared their chemistry to YSOs at the time thought to not contain disks, and found no significant differences. However, several of the forming stars that were chosen as part of the comparison group that did not have disks have since been shown to contain (multiple-)disk systems \citep[e.g.,][]{maud17a, ahmadi18a}.

Specific high-mass YSOs with disks that have been previously studied in relation to chemistry include:

\textit{AFGL\,2591 VLA3}. \citet{jimenez-serra12a} found chemical segregation toward an MYSO known to harbor a disk, with the emission from three groups of molecules falling into three morphological types: (1) those peaking on AFGL\,2591 VLA3 (H$_2$S and $^{13}$CS), (2) those that avoided the continuum position and were double-peaked (HC$_3$N, OCS, SO, and SO$_2$), and (3) CH$_3$OH, which presented a ring-like morphology. Their observations only covered one N-bearing molecule (HC$_3$N), which fell into the double-peaked morphological group. They explained the ring-like structure in CH$_3$OH emission by FUV photodissociation of this molecule within the inner regions of this MYSO. Whereas molecules such as H$_2$S and CS that peak at the central position can be reformed by gas-phase reactions, once destroyed methanol cannot be efficiently reformed in the gas phase \citep{garrod08a}. This picture assumes spherical symmetry, but in reality the disk will provide some degree of shielding from dissociation for CH$_3$OH. Therefore, it is unclear whether CH$_3$OH would be fully destroyed within several thousand astronomical units of the star.

\textit{NGC 6334 I(N) SMA 1b}. \citet{hunter14a} observed the emission from several different molecules toward this source, whose kinematics were found to be consistent with a rotating and infalling (sub-Keplerian) disk, including the N-bearing species HC$_3$N, CH$_3$CN, CH$_3$CH$_2$CN, and HNCO. However, in this source many of the other observed species (such as CH$_3$OH) also trace the disk, showing in this case that there does not appear to be any strong chemical segregation of N-bearing species. Instead, the high-temperature transitions are seen to peak toward the source position, whereas the lower-temperature transitions had a double-peaked morphology, indicating a temperature gradient.  However, one molecule that did not fit with this picture was HNCO, which had a compact morphology. \citet{hunter14a} suggested this was the result of destruction of larger parent molecules to form HNCO in the gas phase in the high-temperature inner regions of the MYSO. They also suggested that the disk morphology could provide shielding from dissociation for molecules such as HC$_3$N, explaining its emission at high velocities close to the source.

\textit{G11.92-0.61 MM1}. \citet{ilee16a,ilee18a} observed a range of species toward this MYSO, finding a Keplerian disk with an enclosed dynamical mass of 40$\pm$5\,M$_{\odot}$. Species that traced the disk include CH$_3$CN, CH$_3$CH$_2$CN, HNCO, DCN, OCS, H$_2$CO, CH$_3$OH, and CH$_3$OCHO. However, although they displayed the same velocity gradient as other tracers, the emission from the observed methanol lines was offset to the southeast compared to the dust continuum. Many of the observed molecules displayed double-peaked and/or asymmetric emission, except for OCS, which was coincident with the central source. In their modeling of the CH$_3$CN emission, they find evidence for two temperature components, which they suggest may originate from the hot ($>$150\,K) and dense gas near the midplane, where CH$_3$CN has thermally desorbed from ice mantles, and from the atmosphere of the disk where gas-phase formation of CH$_3$CN is more important. 

\textit{G345.4938+01.4677 / IRAS\,16562-3959}. \citet{guzman18a} analyzed emission from 22 different species toward the MYSO G345.4938+01.4677, which has previously been found to contain a compact rotating core seen in SO and SO$_2$ \citep{guzman14a}. They categorized their observed molecules into two groups. The first group (the ``Shock group'') is comprised of species whose emission was similar to SiO, and the second group (the ``Continuum group''), containing a broad range of species including COMs such as CH$_3$C$_3$N, is associated with the 3\,mm dust-continuum emission. However, they note that except for SO and SO$_2$, no other molecules display the previously seen velocity gradient that indicated a disk may be embedded in the rotating core. 

\textit{G35.20-0.74 N}. \citet{allen17a} found that N-bearing species, specifically cyanides, were found to be more abundant in part B3 of the source G35.20-0.74\,N\,B, which has previously been found to be a Keplerian disk by \citet{sanchez-monge14a}. They suggested that this chemical segregation could be from fragments or different sources within a disk or rotating torus, where hot gas-phase chemistry is more active in source B3, producing the larger abundance of N-bearing species.

\textit{GGD27\,MM1 (HH80-81)}. \citet{girart17a} found that sulfurated molecules such as SO$_2$ and SO, as well as H$_2$CO traced a rotating disk. They attributed the production of the sulfurated molecules to shocks occurring at the radius of the centrifugal barrier, also seen toward low-mass YSOs \citep{sakai14a, oya16a}. Another explanation they put forward was UV photodissociation of water in the disk, which allows the reactants to form molecules such as SO$_2$. In contrast, they found that methanol departs from the emission expected for a rotating disk, and that it instead traces the outflow cavity walls. Detected isotopologues of HCN and HC$_3$N also did not show a clear velocity gradient; however, this may have been due to the fact these detections had a low signal-to-noise.
 
\textit{Orion SrcI}. \citet{ginsburg18a,ginsburg19a} investigated the dust continuum and molecular lines detected at three frequencies between 0.87 and 3.0\,mm toward the disk and outflow in this source, which at a distance of 415\,pc constitutes the nearest example of a massive star with a disk. They detected SiO, water, and a host of unidentified lines, which they went on to identify as isotopologues of NaCl, KCl, and possibly AlO. By fitting the position-velocity diagram of the H$_2$O 5$_{5,0}$ - 6$_{4,3}$ line, which they find is tracing the upper envelope of the disk and the lower section of the outflow, they determine a central mass of 15$\pm$2\,M$_{\odot}$. They find that the salt lines also trace the base of the outflow or an upper layer of the disk, but one that lies closer to the disk midplane than water and non-vibrationally excited transitions of SiO. This interpretation is complicated by the fact SrcI is nearly edge-on and the dust emission is optically thick \citep{plambeck16a}, masking any line emission close to the midplane. Given the rarity of salt lines in the ISM, \citet{ginsburg19a} suggest that they may be a unique tracer of disks around massive stars, yet, at least in SrcI, they do not trace the midplane but instead an upper layer in the disk or the base of the outflow. 

\textit{G17.64+0.16}. This MYSO was one of six sources studied by \citet{cesaroni17a}, who found it was the most evolved in their sample, but also the most chemically rich. However, their analysis found clear evidence for a disk. Later, G17 was revisited by \citet{maud18a} and at higher resolution in \citet{maud19a}. \citet{maud18a} found a small ($\sim$200\,au radius) rotating disk in SiO, with a similar position angle to the continuum emission. Comparable to the results for SrcI \citep{ginsburg18a}, the presence of SiO in the disk around this source indicates the presence of possibly ionized and turbulent hot shocked gas. Indeed, the detection of compact H30$\alpha$ emission toward the source supports this. Both CH$_3$CN and CH$_3$OH are not found to trace the disk, but instead the cavity working surfaces at the point the wide-angle wind is interacting with the dense envelope. At a higher resolution of 20\,mas, or 44\,au at the distance of G17, the continuum observations of \citet{maud19a} uncovered a ring-like structure in a 120\,au radius disk. They also found Keplerian kinematics seen in highly excited vibrational lines of water, tracing the hot upper layers of the disk.
 
 \textit{G339.88-1.26}. In their observations of the MYSO G339.88-1.26, \citet{zhang19b} found that within the midplane CH$_3$OH and H$_2$CO trace the infalling, rotating envelope as well as the centrifugal barrier at 530\,au, whereas SO$_2$ and H$_2$S trace the centrifugal barrier and outer disk. The radial difference between these two groups of molecules can be attributed to a radial temperature dependence and higher upper energy levels for the molecular transitions observed for the second group. In comparison, they found SiO traces the Keplerian disk as well as the envelope and jet. They suggest that the enhancements in emission close to the centrifugal barrier are either explained by an accretion shock or the irradiated inner edge of the envelope. Outside of the midplane, CH$_3$OH emission is also found to be coincident with methanol masers. The thermal and maser methanol emission both extend perpendicular to the disk midplane in the outflow direction, lying to the south of the source on the blueshifted side of the disk (see their Figure 12a). \citet{zhang19b} explain this morphology, which is very similar to that seen for AFGL\,4176, by shocks along the cavity walls of the outflow. 
 
The blue-dominant lines in AFGL\,4176, which include CH$_3$OH and H$_2$CO and have a ring-like morphology that is brighter on the blueshifted side of the disk, may be explained by similar processes to those seen in G339.88-1.26. \citet{csengeri18a} also found that CH$_3$OH is tracing the centrifugal barrier in the MYSO G328.2551-0.5321. Thus, the enhancement of CH$_3$OH (and other blue-dominant molecules) at a specific radius from AFGL\,4176 may be due to accretion shocks in the centrifugal barrier. In addition, the blueshifted asymmetry in the blue-dominant molecules may be either due to self-absorption of the redshifted emission from the centrifugal barrier by the envelope or by an actual asymmetry in the accreting material. Given that no blueshifted asymmetries are seen in our model for the lower K transitions of CH$_3$CN (J15), which would include the effect of self-absorption by the envelope, and that maser emission is seen coincident with the blueshifted CH$_3$OH emission (as noted also by \citealt{zhang19b} for G339.88-1.26), we suggest that there is an intrinsic asymmetry in the AFGL\,4176 envelope and disk structure. This is also corroborated by the asymmetric NH$_3$(5,5) emission, which also peaks on the eastern side of the disk.
 
Looking back on the findings to-date for MYSOs with disks summarized above, it is clear that patterns are beginning to emerge, including the ring-like emission such as SO$_2$ and CH$_3$OH tracing shocks, possibly at the disk's centrifugal barrier and that N-bearing species often trace the disk. Nevertheless, there is still a large degree of inhomogeneity in the chemistry of these disks and their inner envelopes, and thus observations of more sources at 100-1000\,au scales are necessary to determine if any patterns persist and whether others emerge. 

\section{Conclusions} \label{sec:conclusions}

Using observations from ALMA, ATCA, and APEX, we present a detailed view of the circumstellar environment and disk of the MYSO AFGL\,4176. Our main results are as follows:

\begin{enumerate}
\item At millimeter wavelengths, we detect 17 continuum sources within 5$''$ of AFGL\,4176 mm1, which is the brightest source in the field and traces the dust emission from the disk found in J15. Their masses range from 0.2 to 13.3\,M$_{\odot}$ and their H$_2$ column densities range between 2.2 and 8.7 $\times$ 10$^{23}$\,cm$^{-2}$. The spectral index of mm1 between the two wide spectral windows at 1.2\,mm is 3.4$\pm$0.2, compatible within errors with ISM dust. 
\item We detect a compact continuum source associated with mm1 at 1.2\,cm with ATCA. Its deconvolved size is $<$2000 $\times$ 760\,au. The spectral index between the two continuum bands at 1.21\,mm and 1.23\,cm is 1.56$\pm$0.15. The dust contributes $>$87\% of the emission at 1.2\,mm and ionized gas contributes $>$96\% of the emission at 1.2\,cm. The spectral index of the ionized gas component is therefore $<$0.7.
\item We present ALMA spectra of mm1 at 1.2\,mm (in four spectral windows within 238.8376-256.5834\,GHz) and identify lines with fluxes $>$5$\sigma$. We detect lines from 25 different molecules, which we can separate into four different morphological types: disk-tracing, blue-dominant, red-dominant, and outflow-tracing.
\item In addition to HCOOH, and apart from H$_2$CCN, which is too faint, all detected lines that contain nitrogen trace the AFGL\,4176 mm1 disk. In particular, vibrationally excited CH$_3$CN and formamide or HC(O)NH$_2$ appear to be excellent disk tracers.
\item The line morphological types cluster in peak velocity and linewidth.  Blue-dominant molecules are predominantly oxygen-bearing. The brightest emission from these lines exhibits a bar-like structure perpendicular to the disk and is associated in position and velocity with a group of Class~II methanol masers. We suggest that this emission is due to shocks at the centrifugal barrier in the blueshifted part of the disk. Red-dominant molecules are comprised of SO and SO$_2$ and their isotopologues that are bright and extended in the redshifted side of the disk. 
\item The outflow-tracing molecules are C$^{34}$S, H$_2$CS, and CH$_3$CCN, which specifically trace emission from a slow, wide-angle wind or dense structures in the outflow cavity walls. We determine that the half-opening angle of the wide-angle outflow is $\sim$19$^{\circ}$ at 150,000\,au, which being parabolic is much wider at smaller distances from the star.
\item The lack of complex molecules toward mm2, combined with the fact its spectrum is rich in S-bearing molecular lines and that it lies along the disk axis, suggest that mm2 may instead be a knot in the blueshifted part of a jet emanating from mm1, instead of another protostar.
\item The NH$_3$(1,1) and (2,2) emission from the region traces a large-scale (r$\sim$0.5\,pc) rotating clump or toroid with a mass of several thousands of solar masses, which rotates in the same sense as the CH$_3$CN disk. AFGL\,4176 mm1 lies at the NW blueshifted end of the toroid, offset from its center. We determine that the temperature in the clump derived from NH$_3$(1,1) and (2,2) peaks toward the north of the cloud, suggesting it is heated by mm1 and/or the HII region to the northeast. NH$_3$(4,4) and (5,5) are detected close to mm1, and NH$_3$(5,5) likely traces hot gas in the blueshifted part of the disk, similar to the blue-dominant lines detected with ALMA such as methanol.
\item We detect the hydrogen recombination lines H29$\alpha$, H64$\alpha$, H25$\alpha$, H67$\alpha$, and H68$\alpha$. H68$\alpha$ traces a north-south velocity gradient in the extended HII region associated with AFGL\,4176. This velocity structure may be due to dynamics driven by a separate source.
\item In comparison to other MYSOs that have associated rotating structures, AFGL\,4176 mm1 lies, along with several other newly detected disks around massive YSOs, within the ``disks" area of the t$_{\rm ff}$/t$_{\rm rot}$ against M$_{\rm gas}$ figure often used to separate disks from toroids (e.g. Fig.~\ref{Beltranplot}). These new sources lie above the previously known disk sources in this Figure, confirming that these structures are more likely to be stable and are associated with more massive MYSOs.
\end{enumerate}
Recent studies have provided several results on the chemistry of disks around forming massive stars, which have begun to show patterns, such as that N-bearing species often trace the disk, specific molecules trace the centrifugal barrier, or methanol lines show unusual morphologies. However, we are still far from constructing a complete picture of the dynamical and chemical processes that govern disks around massive stars, an aim requiring a homogeneous survey of MYSOs to search for disks and study their structure and chemistry.

\acknowledgments

We thank the anonymous referee for their insightful comments, which helped improve the manuscript. K.G.J. and M.G.H. acknowledge support from the Science and Technology Facilities Council via grant number ST/P00041X/1. R.K. acknowledges financial support via the Emmy Noether Research Group on Accretion Flows and Feedback in Realistic Models of Massive Star Formation funded by the German Research Foundation (DFG) under grant No. KU 2849/3- 1 and KU 2849/3- 2. H.B. acknowledges support from the European Research Council under the Horizon 2020 Framework Program via the ERC Consolidator Grant CSF-648505, as well as support from the Deutsche Forschungsgemeinschaft in the Collaborative Research Center (SFB 881) ``The Milky Way System'' (subproject B1). P.B. was supported by the Russian Science Foundation under grant 18-72-10132. N.D.K. acknowledges support from the KU Leuven C1 grant MAESTRO C16/17/007. The Australia Telescope is funded by the Commonwealth of Australia for operation as a National Facility managed by CSIRO. This paper makes use of the following ALMA data: ADS/JAO.ALMA\#2012.1.00469.S. ALMA is a partnership of ESO (representing its member states), NSF (USA) and NINS (Japan), together with NRC (Canada), MOST and ASIAA (Taiwan), and KASI (Republic of Korea), in cooperation with the Republic of Chile. The Joint ALMA Observatory is operated by ESO, AUI/NRAO, and NAOJ. This publication is based on data acquired with the Atacama Pathfinder Experiment (APEX). APEX is a collaboration between the Max-Planck-Institut fur Radioastronomie, the European Southern Observatory, and the Onsala Space Observatory. This research has made use of NASA's Astrophysics Data System and the software listed below.


\vspace{5mm}
\facilities{ALMA, APEX, ATCA.}


\software{
  IPython \citep{ipython},
  SciPy,
  Matplotlib \citep{matplotlib},
  APLpy \citep{aplpy},
  DS9 \citep{ds9},
  spectral-cube \citep{spectralcube},
  Astropy \citep{astropy-collaboration13a, astropy-collaboration18a},
  CASA \citep{mcmullin07a},
  astrodendro.
}



\clearpage

\appendix
\renewcommand\thefigure{\thesection.\arabic{figure}} 
\renewcommand\theHfigure{\thesection.\arabic{figure}}       
\setcounter{figure}{0}

\renewcommand\thetable{\thesection.\arabic{table}}    
\renewcommand\theHtable{\thesection.\arabic{table}}   
\setcounter{table}{0}

\section{Spectra and Properties of Detected ALMA Lines}

\begin{figure*}[h]
\begin{center}
\includegraphics[width=18.cm,angle=270]{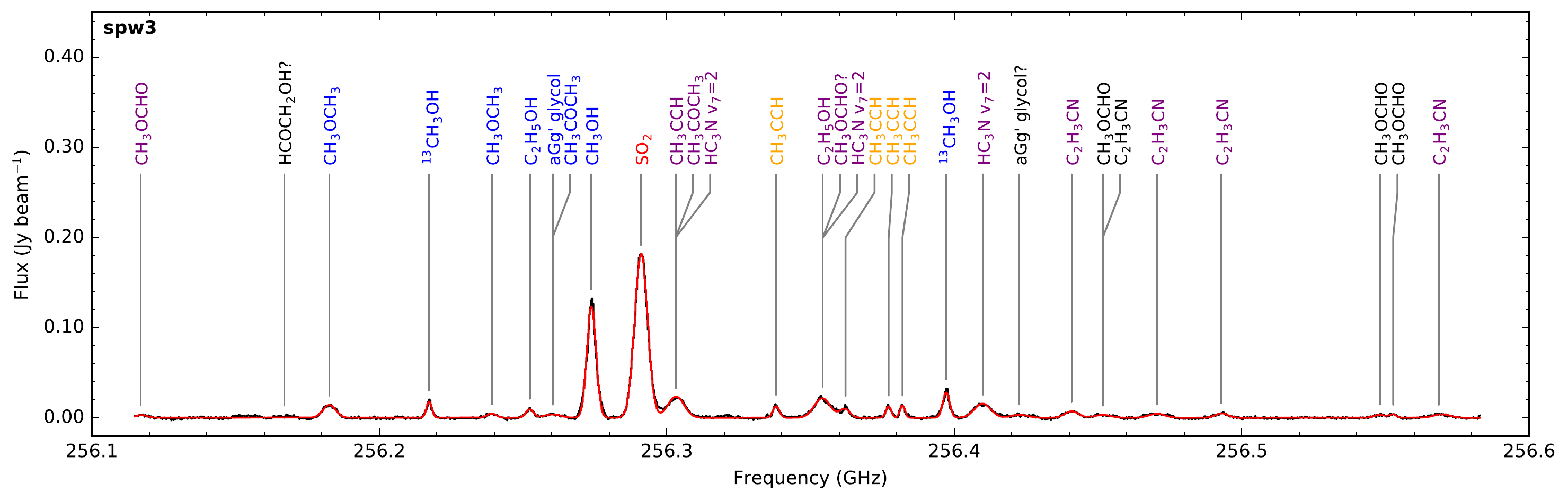}
\includegraphics[width=17.7cm,angle=270]{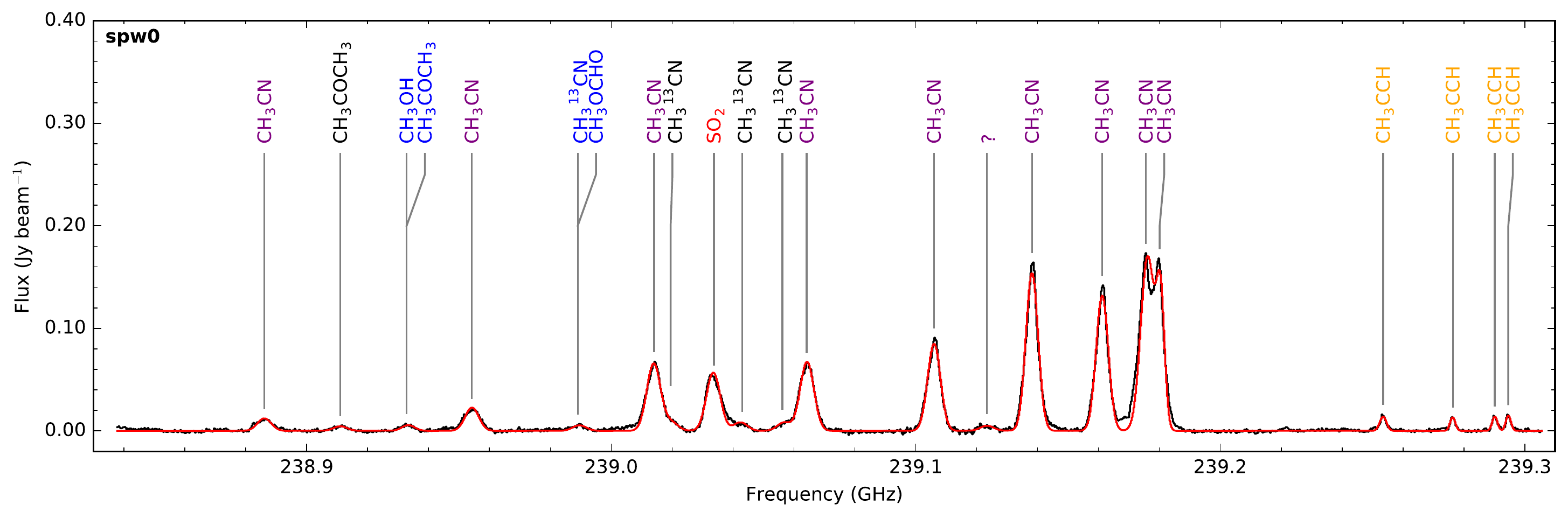}
\caption{Spectrum of the two ALMA narrow spectral windows covering 238.8376 -- 239.3064 and 256.1146 -- 256.5834\,GHz as well as the combined fit of all Gaussians to the lines shown in red. The different line label colors denote different types of line morphology: purple is disk-tracing, blue is blue-dominant, red is red-dominant, and orange is outflow-tracing. Line labels are black when no classification was possible.\label{figspectrumnarrow_wfit}}
\end{center}
\end{figure*}

\begin{figure*}[h]
\begin{center}
\includegraphics[width=22.0cm, angle=270]{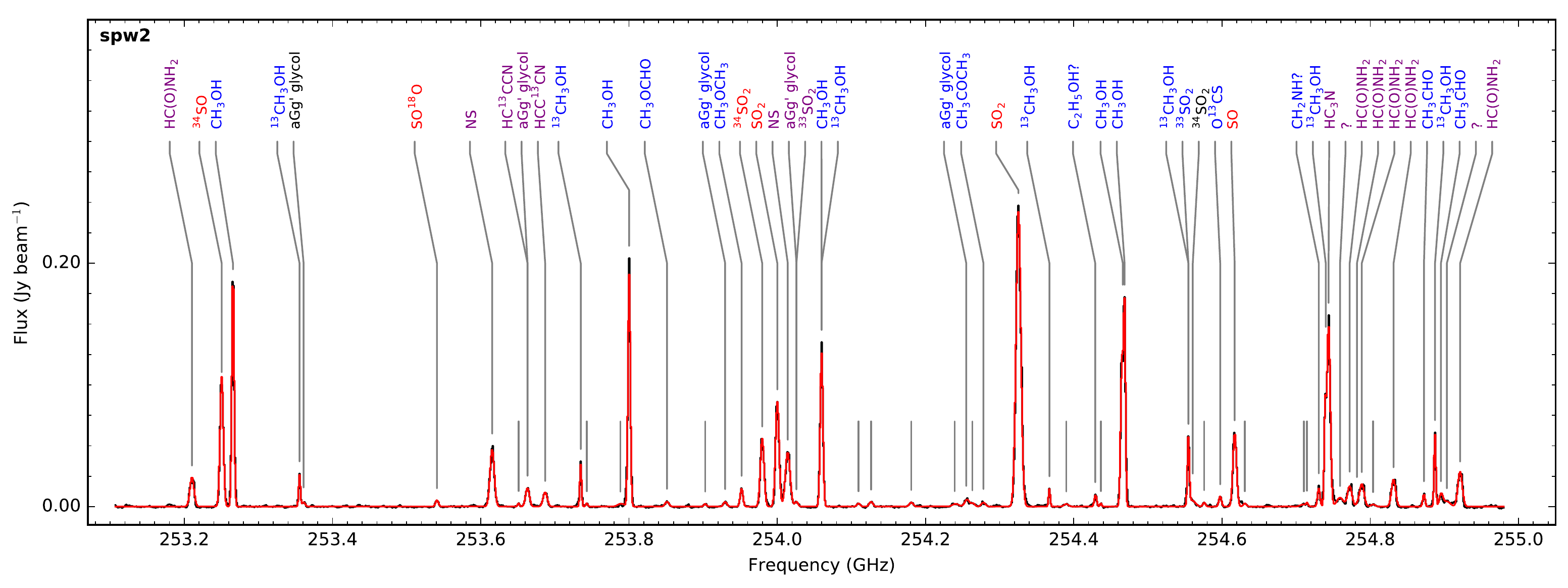}
\includegraphics[width=22.0cm, angle=270]{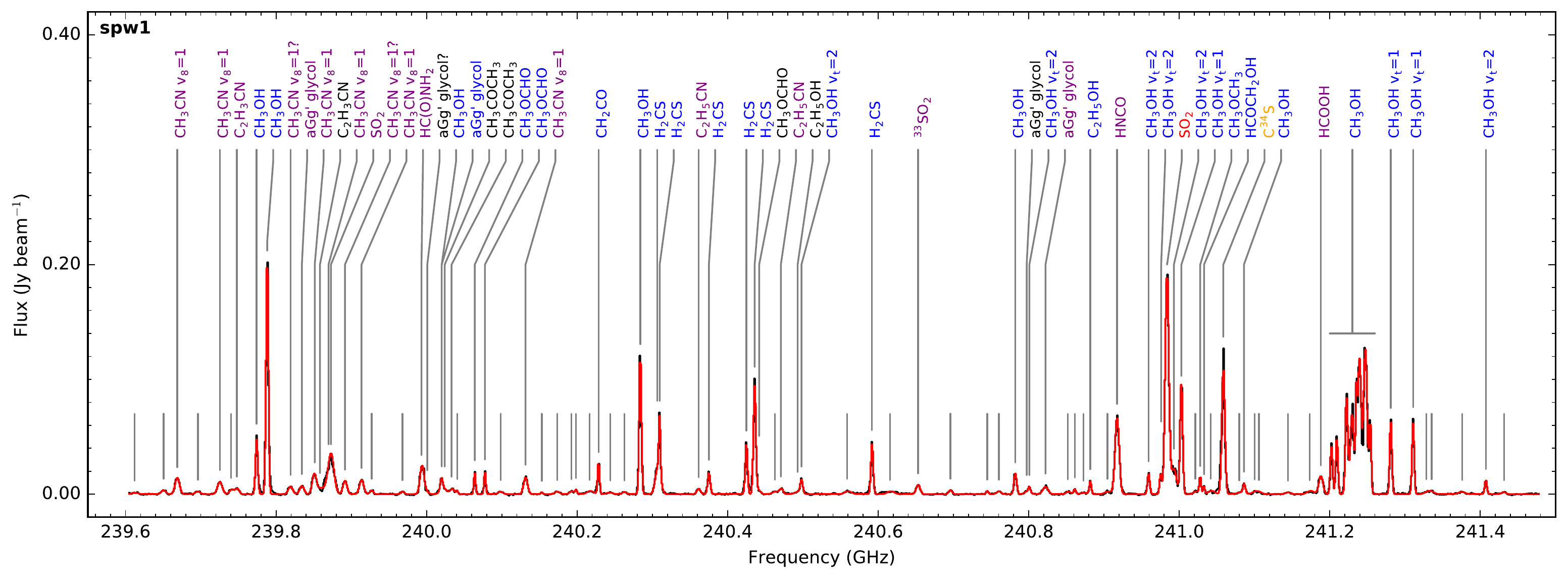}
\end{center}
\caption{Spectrum of the two ALMA wide spectral windows covering 239.6035 -- 241.4785 and 253.1055 -- 254.9805\,GHz as well as the combined fit of all Gaussians to the lines shown in red. The different line label colors denote different types of line morphology: purple is disk-tracing, blue is blue-dominant, red is red-dominant, and orange is outflow-tracing. Line labels are black when no classification was possible. \label{figspectrumwide_wfit}}
\end{figure*}

\begin{figure*}
\begin{center}
\includegraphics[width=15.8cm,angle=0]{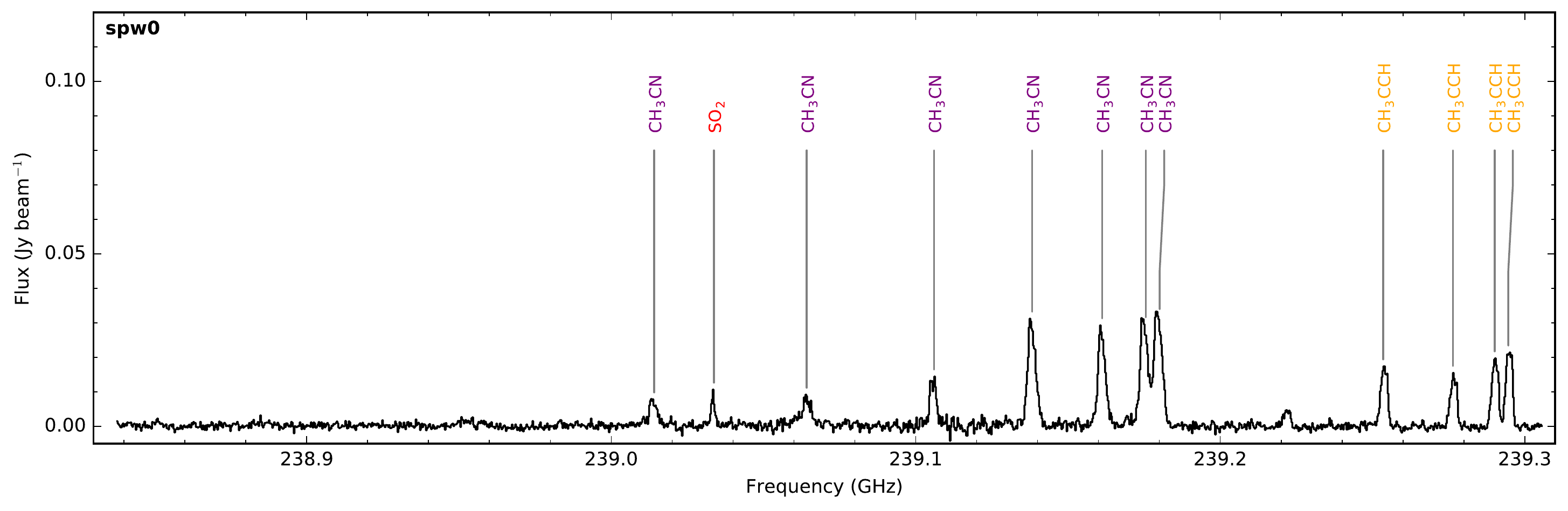}
\includegraphics[width=15.8cm,angle=0]{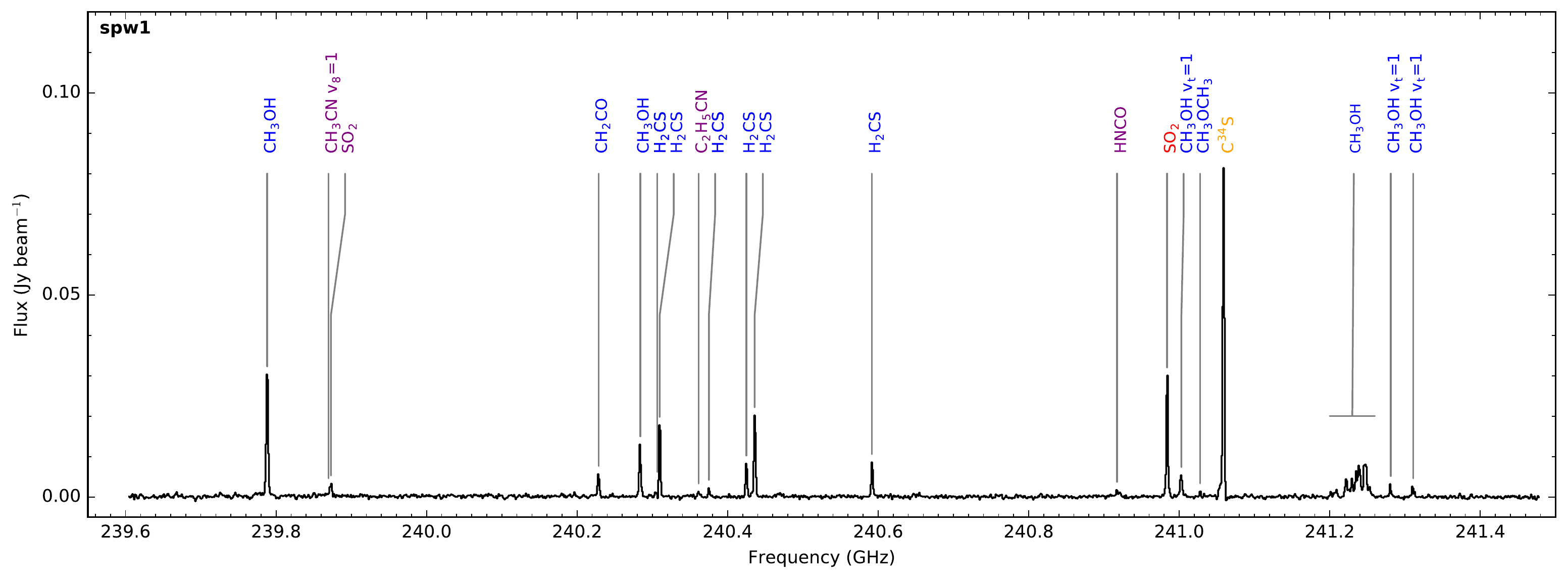}
\includegraphics[width=15.8cm,angle=0]{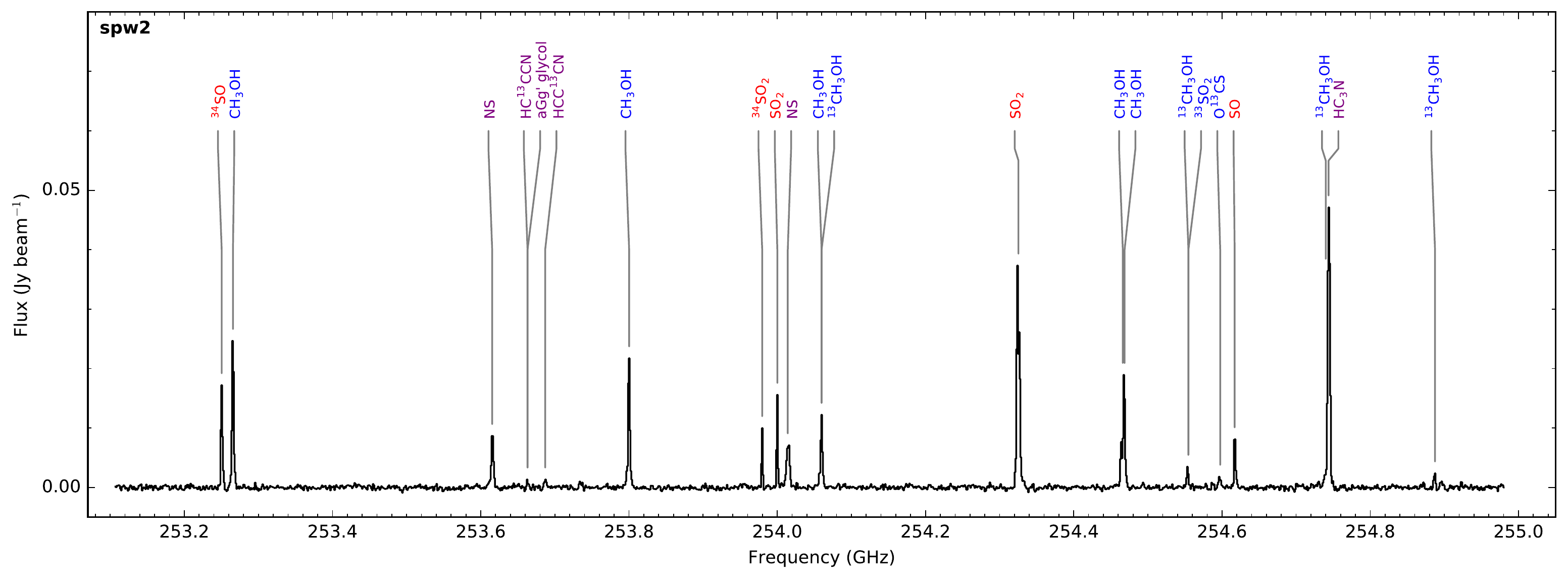}
\includegraphics[width=16.5cm,angle=0]{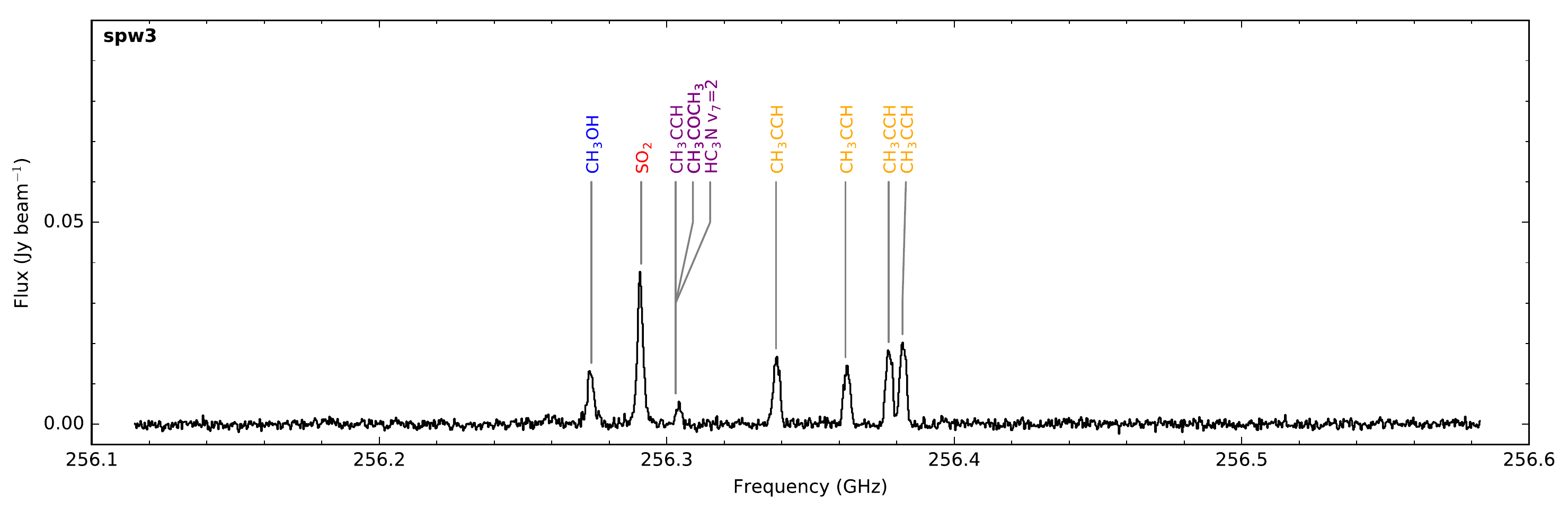}
\caption{Spectrum of the four spectral windows toward mm2. The different line label colors denote different types of line morphology toward mm1: purple is disk-tracing, blue is blue-dominant, red is red-dominant, and orange is outflow-tracing. \label{figmm2}}
\end{center}
\end{figure*}

In this Appendix, we include the spectra of the narrow and wide ALMA spws overlaid with the combined fit of all Gaussians to the detected lines in Figures~\ref{figspectrumnarrow_wfit} and \ref{figspectrumwide_wfit}, respectively. We also include the observed spectra in spws0-3 for mm2 in Figure~\ref{figmm2}. Tables \ref{spw37} through \ref{spw15} (also available online\footnote{http://doi.org/10.5281/zenodo.3369188}) provide the measured properties of the detected lines, and Table~\ref{linedats} provides a summary of the detected species, including the line databases used for each species, their molecular tag identifiers in each catalog, and the number of lines identified for each morphological type.

\clearpage
\begin{deluxetable*}{llcccccc}
\tablecaption{Lines detected in the first narrow spectral window (spw0) covering 238.8376 to 239.3064\,GHz\label{spw37}}
\tablehead{\colhead{Species} & \colhead{Transition} & \colhead{Rest Freq.} & \colhead{E$_{\rm up}$} & \colhead{Flux} & \colhead{v} & \colhead{v$_{\rm _{FWHM}}$} & \colhead{Line}\\ 
\colhead{ } & \colhead{ J(K$_a$,K$_c$)} & \colhead{$\mathrm{(GHz)}$} & \colhead{$\mathrm{(K)}$} & \colhead{$\mathrm{(mJy\,beam^{-1})}$} & \colhead{$\mathrm{(km\,s^{-1})}$} & \colhead{$\mathrm{(km\,s^{-1})}$} & \colhead{Morph.}\tablenotemark{a}}
\startdata
CH$_3$CN & 13(8) $\rightarrow$ 12(8) & 238.843926 & 537.04 & 12.28 & -52.9 & 6.7 & D \\
CH$_3$COCH$_3$ & 23(1,22) $\rightarrow$ 22(2,21)  EE & 238.868953 & 144.37 & 4.50 & -52.8 & 6.7 & U \\
CH$_3$COCH$_3$ & 23(2,22) $\rightarrow$ 22(2,21)  EE & 238.868953 & 144.37 & -- & -- & -- & U \\
CH$_3$COCH$_3$ & 23(1,22) $\rightarrow$ 22(1,21)  EE & 238.868953 & 144.37 & -- & -- & -- & U \\
CH$_3$COCH$_3$ & 23(2,22) $\rightarrow$ 22(1,21)  EE & 238.868953 & 144.37 & -- & -- & -- & U \\
CH$_3$OH & 29(3) $\rightarrow$ 29(2) A$^{-+}$ & 238.890424 & 1058.72 & 5.63 & -53.9 & 6.3 & B \\
CH$_3$COCH$_3$ & 23(2,22) $\rightarrow$ 22(2,21)  AA & 238.890805 & 144.29 & -- & -53.4 & -- & B \\
CH$_3$COCH$_3$ & 23(1,22) $\rightarrow$ 22(1,21)  AA & 238.890805 & 144.29 & -- & -53.4 & -- & B \\
CH$_3$CN & 13(7) $\rightarrow$ 12(7) & 238.912715 & 430.10 & 22.72 & -52.1 & 6.7 & D \\
CH$_3$$^{13}$CN & 13(4) $\rightarrow$ 12(4) & 238.946305 & 194.58 & 5.45 & -54.3 & 6.7 & B \\
CH$_3$OCHO & 19(3,16)  $\rightarrow$ 18(3,15)  E v$_t$=1 & 238.947229 & 308.63 & -- & -53.1 & -- & B \\
CH$_3$CN & 13(6) $\rightarrow$ 12(6) & 238.972389 & 337.37 & 65.57 & -52.1 & 6.7 & D \\
CH$_3$$^{13}$CN & 13(3) $\rightarrow$ 12(3) & 238.978371 & 144.59 & 8.34 & -51.9 & 6.7 & U \\
SO$_2$ & 21(7,15) $\rightarrow$ 22(6,16) & 238.992534 & 332.51 & 56.85 & -51.4 & 6.7 & R \\
CH$_3$$^{13}$CN & 13(2) $\rightarrow$ 12(2) & 239.001283 & 108.88 & 8.46 & -51.6 & 6.7 & U \\
CH$_3$$^{13}$CN & 13(1) $\rightarrow$ 12(1) & 239.015035 & 87.45 & 8.27 & -52.3 & 6.7 & U \\
CH$_3$$^{13}$CN & 13(0) $\rightarrow$ 12(0) & 239.019619 & 80.30 & -- & -46.6 & -- & U \\
CH$_3$CN & 13(5) $\rightarrow$ 12(5) & 239.022924 & 258.87 & 67.35 & -51.9 & 6.7 & D \\
CH$_3$CN & 13(4) $\rightarrow$ 12(4) & 239.064299 & 194.62 & 84.97 & -52.3 & 6.3 & D \\
? & -- & 239.123400 & -- & 5.01 & -- & 6.7 & D \\
\textbf{CH$_3$CN\tablenotemark{b}} & \textbf{13(3) $\rightarrow$ 12(3)} & \textbf{239.096497} & \textbf{144.63} & \textbf{153.81} & \textbf{-52.3} & \textbf{5.9} & \textbf{D} \\
CH$_3$CN & 13(2) $\rightarrow$ 12(2) & 239.119504 & 108.92 & 131.93 & -52.3 & 5.9 & D \\
CH$_3$CN & 13(1) $\rightarrow$ 12(1) & 239.133313 & 87.49 & 169.74 & -53.7 & 6.7 & D \\
CH$_3$CN & 13(0) $\rightarrow$ 12(0) & 239.137916 & 80.34 & 122.78 & -53.3 & 3.9 & D \\
CH$_3$CCH & 14(3) $\rightarrow$ 13(3) & 239.211216 & 150.92 & 13.91 & -53.0 & 3.3 & O \\
CH$_3$CCH & 14(2) $\rightarrow$ 13(2) & 239.234011 & 114.92 & 13.21 & -53.0 & 2.6 & O \\
CH$_3$CCH & 14(1) $\rightarrow$ 13(1) & 239.247727 & 93.32 & 13.97 & -53.2 & 2.6 & O \\
CH$_3$CCH & 14(0) $\rightarrow$ 13(0) & 239.252297 & 86.12 & 14.92 & -53.0 & 2.5 & O \\
\enddata
\tablecomments{When a line is unidentified, indicated by a ? in the Species column, the measured line frequency is given instead of the rest frequency. An em-dash (--) shown in the flux, velocity and linewidth columns indicates that the line identification and measured properties are shared with the last line where the properties are stated. If the transition quantum numbers are not known, a reference is given instead.}
\tablenotetext{a}{The morphology column sorts the lines into five different morphologies. These are, B: blue-dominant, D: disk-tracing, O: Outflow-tracing, R: red-dominant, and U: unknown.}
\tablenotetext{b}{Lines whose moment maps are shown in Fig. \ref{momentsfig} are shown in bold}
\end{deluxetable*}

\clearpage
\startlongtable
\begin{deluxetable*}{llcccccc}
\tablecaption{Lines detected in the second ALMA narrow spectral window (spw3) covering 256.1146 to 256.5834\,GHz\label{spw04}}
\tablehead{\colhead{Species} & \colhead{Transition} & \colhead{Rest Freq.} & \colhead{E$_{\rm up}$} & \colhead{Flux} & \colhead{v} & \colhead{v$_{\rm _{FWHM}}$} & \colhead{Line}\\ 
\colhead{ } & \colhead{ J(K$_a$,K$_c$)} & \colhead{$\mathrm{(GHz)}$} & \colhead{$\mathrm{(K)}$} & \colhead{$\mathrm{(mJy\,beam^{-1})}$} & \colhead{$\mathrm{(km\,s^{-1})}$} & \colhead{$\mathrm{(km\,s^{-1})}$} & \colhead{Morph.\tablenotemark{a}}}
\startdata
CH$_3$OCHO & 21(15,7)  $\rightarrow$ 20(15,6)  E v$_t$=1 & 256.071907 & 473.45 & 3.06 & -53.2 & 5.4 & D \\
HCOCH$_2$OH? & 33(3,30) $\rightarrow$ 33(3,31) v=2 & 256.123418 & 686.70 & 0.37 & -52.3 & 7.8 & U \\
CH$_3$OCH$_3$ & 19(5,14)  $\rightarrow$ 19(4,15)  EE & 256.137423 & 208.33 & 14.24 & -52.9 & 6.2 & B \\
$^{13}$CH$_3$OH & 6(1,5) $\rightarrow$ 5(2,3) E & 256.171587 & 68.63 & 17.95 & -53.6 & 2.6 & B \\
CH$_3$OCH$_3$ & 29(4,25)  $\rightarrow$ 29(3,26)  EE & 256.193781 & 419.59 & 4.55 & -53.1 & 4.6 & B \\
C$_2$H$_5$OH & 15(2,14)  $\rightarrow$ 14(2,13)  g$^+$ & 256.206340 & 160.74 & 9.40 & -53.8 & 3.7 & B \\
aGg'-(CH$_2$OH)$_2$ & 25(6,19) v=0 $\rightarrow$ 24(6,18) v=1 & 256.216551 & 178.92 & 3.83 & -51.2 & 7.7 & B \\
CH$_3$COCH$_3$ & 23(4,20) $\rightarrow$ 22(4,19)  EA & 256.213507 & 161.92 & -- & -54.7 & -- & B \\
CH$_3$COCH$_3$ & 23(3,20) $\rightarrow$ 22(3,19)  EA & 256.213507 & 161.92 & -- & -54.7 & -- & B \\
CH$_3$OH & 17(3,15) $\rightarrow$ 17(2,16) A$^{+-}$ & 256.228714 & 404.80 & 123.62 & -52.8 & 4.4 & B \\
SO$_2$ & 5(3,3) $\rightarrow$ 5(2,4) & 256.246945 & 35.89 & 182.00 & -51.7 & 6.2 & R \\
CH$_3$CCH & 15(4) $\rightarrow$ 14(4) & 256.258435 & 213.61 & 23.53 & -52.4 & 7.8 & D \\
CH$_3$COCH$_3$ & 23(3,20) $\rightarrow$ 22(4,19)  EE & 256.259007 & 161.87 & -- & -51.7 & -- & D \\
CH$_3$COCH$_3$ & 23(3,20) $\rightarrow$ 22(3,19)  EE & 256.259007 & 161.87 & -- & -51.7 & -- & D \\
CH$_3$COCH$_3$ & 23(4,20) $\rightarrow$ 22(4,19)  EE & 256.259007 & 161.87 & -- & -51.7 & -- & D \\
CH$_3$COCH$_3$ & 23(4,20) $\rightarrow$ 22(3,19)  EE & 256.259007 & 161.87 & -- & -51.7 & -- & D \\
HC$_3$N & 28 $\rightarrow$ 27 l=0 v$_7$=2 & 256.259616 & 820.06 & -- & -51.0 & -- & D \\
CH$_3$CCH & 15(3) $\rightarrow$ 14(3) & 256.292638 & 163.22 & 12.80 & -53.1 & 3.1 & O \\
C$_2$H$_5$OH & 15(2,14)  $\rightarrow$ 14(2,13)  g$^-$ & 256.307331 & 165.41 & 21.71 & -54.7 & 7.7 & D \\
CH$_3$OCHO? & \citet{mcmillan16a} & 256.309575 & -- & -- & -52.1 & -- & D \\
HC$_3$N & 28 $\rightarrow$ 27 l=2e v$_7$=2 & 256.311440 & 823.35 & -- & -49.9 & -- & D \\
CH$_3$CCH & 15(2) $\rightarrow$ 14(2) & 256.317078 & 127.22 & 9.92 & -52.5 & 4.7 & O \\
CH$_3$CCH & 15(1) $\rightarrow$ 14(1) & 256.331746 & 105.62 & 12.75 & -53.1 & 2.6 & O \\
CH$_3$CCH & 15(0) $\rightarrow$ 14(0) & 256.336636 & 105.62 & 13.76 & -53.0 & 2.4 & O \\
$^{13}$CH$_3$OH & 13(3,11) $\rightarrow$ 13(2,12) A$^{+-}$ & 256.351482 & 256.14 & 28.11 & -53.5 & 3.3 & B \\
HC$_3$N & 28 $\rightarrow$ 27 l=2f v$_7$=2 & 256.365922 & 823.37 & 15.81 & -51.4 & 7.8 & D \\
aGg'-(CH$_2$OH)$_2$? & 26(1,25) v=1 $\rightarrow$ 25(1,24) v=0 & 256.379254 & 167.28 & 3.41 & -51.0 & 7.8 & U \\
\textbf{C$_2$H$_3$CN\tablenotemark{b}} & \textbf{27(7,20) $\rightarrow$ 26(7,19)} & \textbf{256.397408} & \textbf{278.02} & \textbf{7.17} & \textbf{-50.8} & \textbf{6.7} & \textbf{D} \\
CH$_3$OCHO & 21(13,8)  $\rightarrow$ 20(13,7)  A v$_t$=1 & 256.403576 & 435.82 & 3.22 & -56.3 & 7.8 & U \\
CH$_3$OCHO & 21(13,9)  $\rightarrow$ 20(13,8)  A v$_t$=1 & 256.403576 & 435.82 & -- & -- & -- & U \\
C$_2$H$_3$CN & 27(8,19) $\rightarrow$ 26(8,18) & 256.409287 & 310.30 & -- & -49.6 & -- & U \\
C$_2$H$_3$CN & 27(8,20) $\rightarrow$ 26(8,19) & 256.409287 & 310.30 & -- & -49.6 & -- & U \\
C$_2$H$_3$CN & 27(6,21) $\rightarrow$ 26(6,20) & 256.426112 & 250.01 & 4.40 & -51.9 & 7.8 & D \\
C$_2$H$_3$CN & 27(9,18) $\rightarrow$ 26(9,17) & 256.448018 & 326.82 & 5.02 & -52.6 & 6.0 & D \\
C$_2$H$_3$CN & 27(9,19) $\rightarrow$ 26(9,18) & 256.448018 & 326.82 & -- & -- & -- & D \\
CH$_3$OCHO & 21(10,11)  $\rightarrow$ 20(10,10)  E v$_t$=1 & 256.501348 & 390.07 & 3.24 & -54.6 & 5.7 & U \\
CH$_3$OCHO & 21(12,9)  $\rightarrow$ 20(12,8)  A v$_t$=1 & 256.506751 & 419.12 & 3.72 & -53.7 & 3.0 & U \\
CH$_3$OCHO & 21(12,10)  $\rightarrow$ 20(12,9)  A v$_t$=1 & 256.506751 & 419.12 & -- & -- & -- & U \\
C$_2$H$_3$CN & 27(5,23) $\rightarrow$ 26(5,22) & 256.523007 & 226.30 & 3.99 & -54.5 & 7.7 & D \\
C$_2$H$_3$CN & 27(5,22) $\rightarrow$ 26(5,21) & 256.527505 & 226.30 & -- & -49.2 & -- & D
\enddata
\tablecomments{When a line is unidentified, indicated by a ? in the Species column, the measured line frequency is given instead of the rest frequency. An em-dash (--) shown in the flux, velocity and linewidth columns indicates that the line identification and measured properties are shared with the last line where the properties are stated. If the transition quantum numbers are not known, a reference is given instead.}
\tablenotetext{a}{The morphology column sorts the lines into five different morphologies. These are, B: blue-dominant, D: disk-tracing, O: Outflow-tracing, R: red-dominant, and U: unknown.}
\tablenotetext{b}{Lines whose moment maps are shown in Fig. \ref{momentsfig} are shown in bold}
\end{deluxetable*}

\clearpage
\startlongtable
\begin{deluxetable*}{llcccccc}
\tablecaption{Lines detected in the first wide spectral window (spw1) covering 239.6035 to 241.4785\,GHz\label{spw26}}
\tablehead{\colhead{Species} & \colhead{Transition} & \colhead{Rest Freq.} & \colhead{E$_{\rm up}$} & \colhead{Flux} & \colhead{v} & \colhead{v$_{\rm _{FWHM}}$} & \colhead{Line}\\ 
\colhead{ } & \colhead{ J(K$_a$,K$_c$)} & \colhead{$\mathrm{(GHz)}$} & \colhead{$\mathrm{(K)}$} & \colhead{$\mathrm{(mJy\,beam^{-1})}$} & \colhead{$\mathrm{(km\,s^{-1})}$} & \colhead{$\mathrm{(km\,s^{-1})}$} & \colhead{Morph.\tablenotemark{a}}}
\startdata
H$^{13}$CCCN & 27 $\rightarrow$ 26 l=2f v$_7$=2 & 239.571308 & 800.34 & 1.40 & -51.6 & 12.5 & U \\
aGg'-(CH$_2$OH)$_2$ & 24(11,13) v=0 $\rightarrow$ 23(11,12) v=1 & 239.605334 & 206.93 & 3.56 & -56.5 & 8.7 & B \\
CH$_3$OCHO & 19(5,14)  $\rightarrow$ 18(5,13)  A v$_t$=1 & 239.610154 & 317.09 & -- & -50.5 & -- & B \\
CH$_3$CN & 13(1)  $\rightarrow$ 12(1)  l=+1 v$_8$=1 & 239.627369 & 599.52 & 14.38 & -51.8 & 8.7 & D \\
HC(O)NH$_2$ & 11(1,10) $\rightarrow$ 10(1,9) v$_{12}$=1 & 239.653472 & 487.98 & 2.57 & -53.2 & 9.1 & D \\
CH$_3$CN & 13(5)  $\rightarrow$ 12(5)  l=-1 v$_8$=1 & 239.684649 & 850.61 & 10.87 & -50.9 & 9.3 & D \\
CH$_3$CN & 13(7)  $\rightarrow$ 12(7)  l=+1 v$_8$=1 & 239.699313 & 862.28 & 4.27 & -52.7 & 9.3 & D \\
C$_2$H$_3$CN & 26(0,26) $\rightarrow$ 25(0,25) & 239.708394 & 157.04 & 4.78 & -50.3 & 7.4 & D \\
CH$_3$OH & 16(7,10) $\rightarrow$ 17(6,12) A$^{-}$ & 239.731363 & 560.07 & 47.59 & -53.6 & 4.5 & B \\
CH$_3$OH & 5(1,5) $\rightarrow$ 4(1,4) A$^{+}$ & 239.746219 & 49.06 & 208.53 & -52.5 & 4.5 & B \\
CH$_3$CN & 13(3)  $\rightarrow$ 12(3)  l=-1 v$_8$=1 & 239.777130 & 709.90 & 6.81 & -52.6 & 8.8 & D \\
aGg'-(CH$_2$OH)$_2$ & 25(3,23) v=0 $\rightarrow$ 24(3,22) v=1 & 239.792798 & 161.62 & 7.20 & -51.8 & 8.6 & D \\
CH$_3$CN & 13(2)  $\rightarrow$ 12(2)  l=-1 v$_8$=1 & 239.808835 & 660.92 & 17.25 & -52.4 & 8.3 & D \\
C$_2$H$_3$CN & 25(1,24) $\rightarrow$ 24(1,23) & 239.816142 & 152.76 & 5.07 & -54.1 & 13.3 & U \\
CH$_3$CN & 13(4)  $\rightarrow$ 12(4)  l=+1 v$_8$=1 & 239.824766 & 666.76 & 15.34 & -57.6 & 11.6 & D \\
SO$_2$ & 15(3,13) $\rightarrow$ 16(0,16) & 239.832754 & 132.54 & 23.40 & -51.6 & 10.6 & D \\
CH$_3$CN & 13(3)  $\rightarrow$ 12(3)  l=+1 v$_8$=1 & 239.850003 & 630.10 & 11.71 & -51.9 & 8.0 & D \\
\textbf{CH$_3$CN\tablenotemark{b}} & \textbf{13(2)  $\rightarrow$ 12(2)  l=+1 v$_8$=1} & \textbf{239.871650} & \textbf{607.71} & \textbf{12.57} & \textbf{-52.4} & \textbf{8.3} & \textbf{D} \\
aGg'-(CH$_2$OH)$_2$ & 24(10,15) v=0 $\rightarrow$ 23(10,14) v=1 & 239.883541 & 196.68 & 3.40 & -54.3 & 6.6 & B \\
aGg'-(CH$_2$OH)$_2$ & 24(10,14) v=0 $\rightarrow$ 23(10,13) v=1 & 239.883585 & 196.68 & -- & -54.3 & -- & B \\
CH$_3$OCHO? & 44(8,36)  $\rightarrow$ 44(7,37)  E & 239.927523 & 644.58 & 2.41 & -50.5 & 6.3 & U \\
HC(O)NH$_2$ & 11(1,10) $\rightarrow$ 10(1,9) & 239.951800 & 72.33 & 24.63 & -52.7 & 8.0 & D \\
aGg'-(CH$_2$OH)$_2$? & 25(2,23) v=0 $\rightarrow$ 24(2,22) v=1 & 239.957158 & 161.61 & 1.99 & -53.2 & 8.7 & U \\
CH$_3$OH & 27(3,24) $\rightarrow$ 27(2,25) A$^{-+}$ & 239.977050 & 926.58 & 10.56 & -53.3 & 5.0 & B \\
aGg'-(CH$_2$OH)$_2$ & 23(4,20) v=1 $\rightarrow$ 22(4,19) v=0 & 239.980039 & 144.08 & -- & -49.6 & -- & B \\
CH$_3$COCH$_3$ & 24(1,24) $\rightarrow$ 23(1,23)  EA & 239.984779 & 146.60 & 4.78 & -47.7 & 13.3 & U \\
CH$_3$COCH$_3$ & 24(0,24) $\rightarrow$ 23(0,23)  EA & 239.984779 & 146.60 & -- & -- & -- & U \\
CH$_3$COCH$_3$ & 24(0,24) $\rightarrow$ 23(1,23)  EE & 239.991110 & 146.50 & 4.75 & -54.0 & 7.5 & U \\
CH$_3$COCH$_3$ & 24(1,24) $\rightarrow$ 23(1,23)  EE & 239.991110 & 146.50 & -- & -- & -- & U \\
CH$_3$COCH$_3$ & 24(0,24) $\rightarrow$ 23(0,23)  EE & 239.991110 & 146.50 & -- & -- & -- & U \\
CH$_3$COCH$_3$ & 24(1,24) $\rightarrow$ 23(0,23)  EE & 239.991110 & 146.50 & -- & -- & -- & U \\
CH$_3$COCH$_3$ & 24(0,24) $\rightarrow$ 23(1,23)  AA & 239.997383 & 146.40 & 3.10 & -53.8 & 3.7 & U \\
CH$_3$COCH$_3$ & 24(1,24) $\rightarrow$ 23(0,23)  AA & 239.997383 & 146.40 & -- & -- & -- & U \\
CH$_3$OCHO & 19(3,16)  $\rightarrow$ 18(3,15)  E & 240.021140 & 122.26 & 18.89 & -53.5 & 3.3 & B \\
CH$_3$OCHO & 19(3,16)  $\rightarrow$ 18(3,15)  A & 240.034673 & 122.25 & 19.63 & -53.5 & 3.4 & B \\
SO$_2$ & 11(5,7) $\rightarrow$ 12(4,8) v$_2$=1 & 240.057521 & 868.48 & 2.23 & -50.8 & 9.3 & U \\
CH$_3$CHO & 23(8,16)  $\rightarrow$ 24(7,18)  E & 240.057567 & 399.46 & -- & -50.7 & -- & U \\
CH$_3$CN & 13(1)  $\rightarrow$ 12(1)  l=-1 v$_8$=1 & 240.089530 & 599.67 & 14.75 & -52.5 & 7.8 & D \\
C$_2$H$_5$OH & 27(2,25)  $\rightarrow$ 27(1,26)  a & 240.110242 & 327.24 & 1.78 & -53.4 & 3.9 & U \\
NH$_2$CN & 12(3,10) $\rightarrow$ 11(3,9) & 240.130645 & 205.25 & 2.33 & -53.5 & 9.5 & D \\
\textbf{NH$_2$CN\tablenotemark{b}} & \textbf{12(3,9) $\rightarrow$ 11(3,8)} & \textbf{240.132690} & \textbf{205.25} & \textbf{--} & \textbf{-51.0} & \textbf{--} & \textbf{D} \\
aGg'-(CH$_2$OH)$_2$ & 24(3,21) v=0 $\rightarrow$ 23(3,20) v=1 & 240.147963 & 155.15 & 2.58 & -56.9 & 10.0 & U \\
OC$^{33}$S & 20 $\rightarrow$ 19 & 240.155961 & 121.03 & 3.31 & -53.5 & 3.3 & B \\
H$_2$CS? & 7(6,1) $\rightarrow$ 6(6,0) & 240.179077 & 518.82 & 2.84 & -48.2 & 13.3 & U \\
H$_2$CS? & 7(6,2) $\rightarrow$ 6(6,1) & 240.179077 & 518.82 & -- & -- & -- & U \\
CH$_2$CO & 12(1,12) $\rightarrow$ 11(1,11) & 240.187257 & 88.01 & 28.21 & -51.5 & 3.6 & B \\
HNCO? & 48(0,48) $\rightarrow$ 47(1,47) & 240.201778 & 1239.46 & 1.45 & -51.6 & 12.2 & U \\
HCOCH$_2$OH & 14(10,5) $\rightarrow$ 14(9,6) & 240.219091 & 119.02 & 2.01 & -54.4 & 7.7 & U \\
CH$_3$OH & 5(3,2) $\rightarrow$ 6(2,4) E & 240.241490 & 82.53 & 122.85 & -52.9 & 4.3 & B \\
H$_2$CS & 7(5,2) $\rightarrow$ 6(5,1) & 240.261988 & 374.70 & 23.57 & -56.8 & 10.2 & B \\
H$_2$CS & 7(5,3) $\rightarrow$ 6(5,2) & 240.261988 & 374.70 & -- & -- & -- & B \\
H$_2$CS & 7(0,7) $\rightarrow$ 6(0,6) & 240.266872 & 46.14 & 49.90 & -53.4 & 3.3 & B \\
\textbf{C$_2$H$_5$CN\tablenotemark{b}} & \textbf{28(1,28) $\rightarrow$ 27(1,27)} & \textbf{240.319337} & \textbf{169.27} & \textbf{4.80} & \textbf{-52.5} & \textbf{7.2} & \textbf{D} \\
H$_2$CS & 7(4,4) $\rightarrow$ 6(4,3) & 240.332190 & 256.59 & 18.70 & -53.5 & 5.1 & B \\
H$_2$CS & 7(4,3) $\rightarrow$ 6(4,2) & 240.332190 & 256.59 & -- & -- & -- & B \\
H$_2$CS & 7(2,6) $\rightarrow$ 6(2,5) & 240.382051 & 98.82 & 44.19 & -53.4 & 4.5 & B \\
H$_2$CS & 7(3,4) $\rightarrow$ 6(3,3) & 240.393762 & 164.60 & 93.89 & -52.5 & 4.7 & B \\
CH$_3$OCHO & 41(9,33)  $\rightarrow$ 41(8,34)  A & 240.400935 & 568.58 & 3.75 & -49.7 & 8.4 & U \\
CH$_3$OCHO? & 36(10,27)  $\rightarrow$ 36(9,28)  E & 240.417922 & 462.48 & 2.39 & -57.2 & 12.3 & D \\
C$_2$H$_5$CN & 28(0,28) $\rightarrow$ 27(0,27) & 240.429184 & 169.24 & 4.53 & -52.7 & 6.9 & D \\
C$_2$H$_5$OH & 26(3,24)  $\rightarrow$ 26(2,25)  a & 240.450946 & 305.41 & 2.34 & -53.8 & 13.3 & U \\
CH$_3$OH & 5(1,5) $\rightarrow$ 4(1,4) A$^{+}$ v$_t$ = 2 & 240.454848 & 717.41 & 11.31 & -54.0 & 4.7 & B \\
H$_2$CS? & 21(1,20) $\rightarrow$ 21(1,21) & 240.518081 & 399.25 & 2.65 & -52.3 & 13.3 & U \\
H$_2$CS & 7(2,5) $\rightarrow$ 6(2,4) & 240.549066 & 98.83 & 44.16 & -53.3 & 4.3 & B \\
HCOCH$_2$OH? & 11(10,1) $\rightarrow$ 11(9,2) & 240.572978 & 97.38 & 2.65 & -54.7 & 13.2 & U \\
$^{33}$SO$_2$ & 10(3,7) $\rightarrow$ 10(2,8) & 240.611132 & 72.27 & 8.12 & -52.2 & 7.9 & D \\
CH$_3$CHO & 12(2,11)  $\rightarrow$ 11(2,10)  E v$_t$=2 & 240.652038 & 462.31 & 3.65 & -54.8 & 6.6 & B \\
C$_2$H$_5$OH & 4(2,2)  g$^-$ $\rightarrow$ 3(1,2)  g$^+$ & 240.653983 & 74.79 & -- & -52.4 & -- & B \\
? & -- & 240.744900 & -- & 2.88 & -- & 4.5 & B \\
HC(O)NH$_2$ & 9(3,6) $\rightarrow$ 10(1,9) v$_{12}$=1 & 240.718643 & 488.03 & 2.58 & -51.9 & 8.7 & U \\
CH$_3$OH & 26(3,23) $\rightarrow$ 26(2,24) A$^{-+}$ & 240.738926 & 863.98 & 18.13 & -53.9 & 5.2 & B \\
aGg'-(CH$_2$OH)$_2$ & 50(13,38) v=1 $\rightarrow$ 50(12,38) v=0 & 240.753385 & 711.82 & 2.60 & -56.1 & 11.1 & U \\
CH$_3$OH & 5(2,3) $\rightarrow$ 4(2,2) A$^{-}$ v$_t$ = 2 & 240.757920 & 911.34 & 4.66 & -53.5 & 4.0 & B \\
aGg'-(CH$_2$OH)$_2$ & 25(1,25) v=1 $\rightarrow$ 24(1,24) v=0 & 240.778125 & 148.00 & 6.51 & -54.9 & 10.3 & D \\
aGg'-(CH$_2$OH)$_2$ & 24(8,17) v=0 $\rightarrow$ 23(8,16) v=1 & 240.807880 & 179.25 & 2.24 & -56.1 & 9.2 & B \\
CH$_3$OH & 5(1,4) $\rightarrow$ 4(1,3) E v$_t$ = 2 & 240.817972 & 833.64 & 4.41 & -53.8 & 4.1 & B \\
aGg'-(CH$_2$OH)$_2$ & 24(8,16) v=0 $\rightarrow$ 23(8,15) v=1 & 240.828886 & 179.25 & 1.29 & -55.8 & 13.2 & B \\
C$_2$H$_5$OH & 14(1,13)  g$^+$ $\rightarrow$ 13(0,13)  g$^-$ & 240.838747 & 147.10 & 11.00 & -53.7 & 3.8 & B \\
CH$_3$OH & 5(-4,2) $\rightarrow$ 4(-4,1) E v$_t$ = 2 & 240.861406 & 779.22 & 2.39 & -55.2 & 13.2 & B \\
\textbf{HNCO\tablenotemark{b}} & \textbf{11(1,11) $\rightarrow$ 10(1,10)} & \textbf{240.875777} & \textbf{112.64} & \textbf{66.54} & \textbf{-51.9} & \textbf{8.0} & \textbf{D} \\
CH$_3$OH & 5(3,3) $\rightarrow$ 4(3,2) A$^{-}$ v$_t$ = 2 & 240.916172 & 692.78 & 17.85 & -53.7 & 4.5 & B \\
CH$_3$OH & 5(3,2) $\rightarrow$ 4(3,1) A$^{+}$ v$_t$ = 2 & 240.916173 & 692.78 & -- & -53.7 & -- & B \\
CH$_3$OH & 5(4,1) $\rightarrow$ 4(4,0) A$^{+}$ v$_t$ = 2 & 240.932051 & 649.19 & 17.10 & -53.8 & 4.2 & B \\
CH$_3$OH & 5(4,2) $\rightarrow$ 4(4,1) A$^{-}$ v$_t$ = 2 & 240.932051 & 649.19 & -- & -- & -- & B \\
SO$_2$ & 18(1,17) $\rightarrow$ 18(0,18) & 240.942792 & 163.07 & 191.50 & -51.0 & 7.0 & R \\
CH$_3$OH & 5(2,4) $\rightarrow$ 4(2,3) E v$_t$ = 2 & 240.952056 & 620.93 & 22.12 & -50.9 & 10.3 & B \\
CH$_3$OH & 5(1,5) $\rightarrow$ 4(1,4) A$^{+}$ v$_t$ = 1 & 240.960557 & 359.95 & 97.83 & -52.7 & 4.8 & B \\
C$_2$H$_3$CN & 35(7,29) $\rightarrow$ 36(6,30) & 240.978552 & 392.88 & 2.55 & -54.8 & 10.2 & U \\
CH$_3$OCH$_3$ & 5(3,3)  $\rightarrow$ 4(2,2)  EE & 240.985078 & 26.31 & 15.03 & -53.4 & 3.6 & B \\
HCOCH$_2$OH & 28(6,22) $\rightarrow$ 27(7,21) v=2 & 240.989945 & 626.66 & 6.48 & -53.4 & 3.3 & B \\
CH$_3$COCH$_3$ & 12(12,1) $\rightarrow$ 11(11,1)  EE & 240.998755 & 74.49 & 2.94 & -55.1 & 13.2 & B \\
C$^{34}$S & 5 $\rightarrow$ 4 & 241.016194 & 34.70 & 109.64 & -52.7 & 5.8 & O \\
CH$_3$OCHO & 33(10,23)  $\rightarrow$ 33(9,24)  E & 241.036087 & 399.46 & 1.49 & -55.8 & 13.2 & U \\
CH$_3$OH & 22(-6,16) $\rightarrow$ 23(-5,18) E & 241.042589 & 775.57 & 8.41 & -54.2 & 5.1 & B \\
CH$_3$OH & 9(3,7) $\rightarrow$ 10(0,10) E & 241.057109 & 152.17 & 1.96 & -54.9 & 13.2 & B \\
CH$_3$COCH$_3$ & 12(12,0) $\rightarrow$ 11(11,0)  EE & 241.062732 & 74.61 & 1.33 & -54.0 & 13.2 & B \\
? & -- & 241.144500 & -- & 1.69 & -- & 9.9 & B \\
HCOCH$_2$OH & 22(2,20) $\rightarrow$ 21(3,19) & 241.131839 & 142.80 & 2.98 & -53.2 & 13.2 & U \\
\textbf{HCOOH\tablenotemark{b}} &  \textbf{11(0,11) $\rightarrow$ 10(0,10)} &  \textbf{241.146330} &  \textbf{70.19} &  \textbf{15.88} &  \textbf{-52.1} &  \textbf{8.2} &  \textbf{D} \\
CH$_3$OH & 5(4,2) $\rightarrow$ 4(4,1) E v$_t$ = 1 & 241.159199 & 398.11 & 41.13 & -53.5 & 4.5 & B \\
CH$_3$OH & 5(3,2) $\rightarrow$ 4(3,1) E v$_t$ = 1 & 241.166580 & 452.14 & 48.24 & -53.3 & 4.2 & B \\
CH$_3$OH & 5(-3,3) $\rightarrow$ 4(-3,2) E v$_t$ = 1 & 241.179886 & 357.36 & 84.30 & -52.7 & 4.8 & B \\
CH$_3$OH & 5(-2,4) $\rightarrow$ 4(-2,3) E v$_t$ = 1 & 241.187428 & 399.34 & 69.80 & -52.7 & 7.5 & B \\
CH$_3$OH & 5(2,4) $\rightarrow$ 4(2,3) A$^{+}$ v$_t$ = 1 & 241.192856 & 333.40 & 77.20 & -52.9 & 2.9 & B \\
CH$_3$OH & 5(2,3) $\rightarrow$ 4(2,2) A$^{-}$ v$_t$ = 1 & 241.196430 & 333.40 & 117.60 & -53.5 & 6.5 & B \\
CH$_3$OH & 5(1,5) $\rightarrow$ 4(1,4) E v$_t$ = 1 & 241.203706 & 326.20 & 126.20 & -54.2 & 6.9 & B \\
CH$_3$OH & 5(0,5) $\rightarrow$ 4(0,4) E v$_t$ = 1 & 241.206035 & 335.31 & -- & -51.3 & -- & B \\
CH$_3$OH & 5(2,3) $\rightarrow$ 4(2,2) E v$_t$ = 1 & 241.210764 & 434.63 & 63.20 & -53.5 & 3.7 & B \\
CH$_3$OH & 5(-1,4) $\rightarrow$ 4(-1,3) E v$_t$ = 1 & 241.238144 & 448.11 & 63.72 & -53.4 & 4.2 & B \\
CH$_3$OH & 5(0,5) $\rightarrow$ 4(0,4) A$^{+}$ v$_t$ = 1 & 241.267862 & 458.39 & 61.57 & -53.3 & 4.1 & B \\
CH$_3$COCH$_3$ & 12(12,1) $\rightarrow$ 11(11,0)  AA & 241.286250 & 74.53 & 1.19 & -53.5 & 13.2 & U \\
aGg'-(CH$_2$OH)$_2$ & 24(5,20) v=0 $\rightarrow$ 23(5,19) v=1 & 241.291269 & 160.66 & 2.22 & -53.6 & 13.2 & U \\
H$_2$CCN & 12(0,12) $\rightarrow$ 11(0,11) J=25/2 $\rightarrow$ 23/2 & 241.333542 & 75.30 & 1.98 & -51.8 & 13.2 & U \\
CH$_3$OH & 5(1,4) $\rightarrow$ 4(1,3) A$^{-}$ v$_t$ = 2 & 241.364143 & 717.54 & 12.09 & -53.7 & 4.3 & B \\
H$_2$CCN & 12(2,11) $\rightarrow$ 11(2,10) J=25/2 $\rightarrow$ 23/2 & 241.391395 & 128.09 & 1.91 & -49.5 & 8.2 & U
\enddata
\tablecomments{When a line is unidentified, indicated by a ? in the Species column, the measured line frequency is given instead of the rest frequency. An em-dash (--) shown in the flux, velocity and linewidth columns indicates that the line identification and measured properties are shared with the last line where the properties are stated. If the transition quantum numbers are not known, a reference is given instead.}
\tablenotetext{a}{The morphology column sorts the lines into five different morphologies. These are, B: blue-dominant, D: disk-tracing, O: Outflow-tracing, R: red-dominant, and U: unknown.}
\tablenotetext{b}{Lines whose moment maps are shown in Fig. \ref{momentsfig} are shown in bold}
%\tablenotetext{c}{May instead be NCC(O)NH$_2$ 69(16,54)-69(15,55) at 241.146439\,GHz}
\end{deluxetable*}

\clearpage
\startlongtable
\begin{deluxetable*}{llcccccc}
\tablecaption{Lines detected in the second wide spectral window (spw2) covering 253.1055 to 254.9805\,GHz\label{spw15}}
\tablehead{\colhead{Species} & \colhead{Transition} & \colhead{Rest Freq.} & \colhead{E$_{\rm up}$} & \colhead{Flux} & \colhead{v} & \colhead{v$_{\rm _{FWHM}}$} & \colhead{Line}\\ 
\colhead{ } & \colhead{ J(K$_a$,K$_c$)} & \colhead{$\mathrm{(GHz)}$} & \colhead{$\mathrm{(K)}$} & \colhead{$\mathrm{(mJy\,beam^{-1})}$} & \colhead{$\mathrm{(km\,s^{-1})}$} & \colhead{$\mathrm{(km\,s^{-1})}$} & \colhead{Morph.\tablenotemark{a}}}
\startdata
HC(O)NH$_2$ & 12(2,11) $\rightarrow$ 11(2,10) & 253.165793 & 91.10 & 24.18 & -52.8 & 7.7 & D \\
$^{34}$SO & $^{3}\Sigma$ N,J = 6,6 $\rightarrow$ 5,5 & 253.207017 & 55.69 & 106.61 & -51.5 & 6.2 & R \\
CH$_3$OH & 13(3,11) $\rightarrow$ 13(2,12) A$^{+-}$ & 253.221376 & 260.99 & 192.46 & -52.5 & 4.3 & B \\
$^{13}$CH$_3$OH & 14(3,11) $\rightarrow$ 14(2,12) A$^{-+}$ & 253.310162 & 287.84 & 25.93 & -53.6 & 3.3 & B \\
aGg'-(CH$_2$OH)$_2$ & 24(9,16) v=1 $\rightarrow$ 23(9,15) v=0 & 253.314615 & 187.76 & 3.61 & -54.9 & 6.0 & U \\
aGg'-(CH$_2$OH)$_2$ & 24(9,15) v=1 $\rightarrow$ 23(9,14) v=0 & 253.315874 & 187.76 & -- & -53.4 & -- & U \\
SO$^{18}$O & 14(1,14) $\rightarrow$ 13(0,13) & 253.497415 & 89.36 & 5.25 & -51.4 & 5.5 & R \\
\textbf{NS\tablenotemark{b}} & $^{2}\Pi_{1/2}$ N,J,F,P = 6,5.5,6.5,-1 $\rightarrow$ 5,4.5,5.5,1 & \textbf{253.570476} & \textbf{38.70} & \textbf{46.77} & \textbf{-53.1} & \textbf{8.0} & \textbf{D} \\
\textbf{NS\tablenotemark{b}} &  $^{2}\Pi_{1/2}$ N,J,F,P = 6,5.5,5.5,-1 $\rightarrow$ 5,4.5,4.5,1 &  \textbf{253.570476} &  \textbf{38.70} &  \textbf{--} &  \textbf{--} &  \textbf{--} &  \textbf{D} \\
CH$_3$OCHO & 17(4,14)  $\rightarrow$ 16(3,13)  E & 253.606166 & 101.43 & 2.71 & -53.1 & 3.6 & B \\
aGg'-(CH$_2$OH)$_2$ & 25(3,23) v=1 $\rightarrow$ 24(3,22) v=0 & 253.616767 & 161.96 & 15.22 & -52.0 & 7.2 & D \\
HC$^{13}$CCN & 28 $\rightarrow$ 27 & 253.619052 & 176.51 & -- & -49.3 & -- & D \\
HCC$^{13}$CN & 28 $\rightarrow$ 27 & 253.643519 & 176.52 & 11.75 & -51.2 & 7.8 & D \\
$^{13}$CH$_3$OH & 13(3,10) $\rightarrow$ 13(2,11) A$^{-+}$ & 253.689530 & 256.15 & 34.52 & -53.6 & 3.2 & B \\
C$_2$H$_5$OH & 15(2,13)  g$^-$ $\rightarrow$ 14(3,11)  g$^+$ & 253.697863 & 167.90 & 2.77 & -53.3 & 3.8 & B \\
CH$_3$OCHO & 20(5,15)  $\rightarrow$ 19(5,14)  A v$_t$=1 & 253.743207 & 329.27 & 2.49 & -54.7 & 12.6 & B \\
CH$_3$OH & 14(3,12) $\rightarrow$ 14(2,13) A$^{+-}$ & 253.755809 & 293.47 & 190.72 & -52.5 & 4.4 & B \\
CH$_3$OCHO & 21(4,18)  $\rightarrow$ 20(4,17)  A v$_t$=1 & 253.807849 & 333.90 & 4.00 & -51.3 & 7.3 & B \\
aGg'-(CH$_2$OH)$_2$ & 24(8,16) v=1 $\rightarrow$ 23(8,15) v=0 & 253.856295 & 179.56 & 2.26 & -54.9 & 5.1 & B \\
aGg'-(CH$_2$OH)$_2$ & 24(3,21) v=1 $\rightarrow$ 23(3,20) v=0 & 253.883009 & 155.53 & 3.85 & -55.2 & 6.5 & B \\
CH$_3$OCH$_3$ & 20(5,15)  $\rightarrow$ 20(4,16)  EE & 253.906820 & 226.60 & 14.81 & -53.2 & 5.3 & B \\
$^{34}$SO$_2$ & 11(3,9) $\rightarrow$ 11(2,10) & 253.936317 & 81.93 & 55.87 & -51.3 & 6.9 & R \\
SO$_2$ & 15(6,10) $\rightarrow$ 16(5,11) & 253.956570 & 198.60 & 86.01 & -51.4 & 6.6 & R \\
NS & $^{2}\Pi_{1/2}$ N,J,F,P = 6,5.5,5.5,1 $\rightarrow$ 5,4.5,4.5,-1 & 253.970581 & 38.82 & 44.72 & -51.5 & 8.7 & D \\
NS & $^{2}\Pi_{1/2}$ N,J,F,P = 6,5.5,4.5,1 $\rightarrow$ 5,4.5,3.5,-1 & 253.970581 & 38.82 & -- & -- & -- & D \\
aGg'-(CH$_2$OH)$_2$ & 24(6,18) $\rightarrow$ 23(6,18) & 253.981324 & 166.26 & 4.00 & -52.6 & 6.0 & D \\
$^{33}$SO$_2$ & 30(4,26) $\rightarrow$ 30(3,27) & 253.982600 & 470.68 & -- & -51.1 & -- & D \\
CH$_3$OH & 2(0,2) $\rightarrow$ 1(-1,1) E & 254.015377 & 20.09 & 126.43 & -52.5 & 5.1 & B \\
$^{13}$CH$_3$OH & 12(3,9) $\rightarrow$ 12(2,10) A$^{-+}$ & 254.012582 & 226.71 & -- & -55.8 & -- & B \\
SO$^{18}$O & 10(3,8) $\rightarrow$ 10(2,9) & 254.066539 & 69.37 & 2.65 & -50.9 & 6.0 & U \\
CH$_3$COCH$_3$ & 21(6,16) $\rightarrow$ 20(5,15)  EE & 254.082229 & 151.84 & 3.99 & -52.3 & 6.9 & D \\
C$_2$H$_3$CN & 27(2,26) $\rightarrow$ 26(2,25) & 254.137460 & 180.05 & 3.41 & -51.2 & 7.3 & D \\
SO$_2$? & 41(11,31) $\rightarrow$ 42(10,32) & 254.194875 & 1087.68 & 2.26 & -54.0 & 12.1 & R \\
aGg'-(CH$_2$OH)$_2$ & 28(1,28) v=0 $\rightarrow$ 27(1,27) v=1 & 254.208718 & 183.96 & 5.83 & -55.9 & 8.9 & B \\
aGg'-(CH$_2$OH)$_2$? & 29(13,17) v=1 $\rightarrow$ 29(12,18) v=1 & 254.217765 & 297.15 & 2.58 & -54.8 & 6.9 & U \\
CH$_3$COCH$_3$ & 37(8,29) $\rightarrow$ 37(7,30)  AE & 254.233997 & 448.79 & 3.33 & -52.0 & 7.9 & B \\
CH$_3$COCH$_3$ & 37(9,29) $\rightarrow$ 37(8,30)  AE & 254.233997 & 448.79 & -- & -- & -- & B \\
SO$_2$ & 6(3,3) $\rightarrow$ 6(2,4) & 254.280536 & 41.40 & 242.40 & -52.9 & 9.0 & R \\
$^{13}$CH$_3$OH & 4(2,2) $\rightarrow$ 5(1,5) A$^{+}$ & 254.321721 & 60.46 & 14.57 & -53.6 & 3.1 & B \\
? & -- & 254.389800 & -- & 2.10 & -- & 8.5 & B \\
C$_2$H$_5$OH? & 7(3,4)  $\rightarrow$ 6(2,5)  a & 254.384085 & 34.84 & 9.37 & -53.4 & 4.6 & B \\
CH$_3$OCHO & 21(4,18)  $\rightarrow$ 20(4,17) E v$_t$=1 & 254.391288 & 333.63 & 2.82 & -53.4 & 3.1 & B \\
CH$_3$OH & 11(5,7) $\rightarrow$ 12(4,8) E & 254.419419 & 289.23 & 119.98 & -53.8 & 4.9 & B \\
CH$_3$OH & 15(3,13) $\rightarrow$ 15(2,14) A$^{+-}$ & 254.423520 & 328.26 & 156.62 & -53.1 & 3.6 & B \\
$^{13}$CH$_3$OH & 10(3,7) $\rightarrow$ 10(2,8) A$^{-+}$ & 254.509364 & 174.63 & 55.28 & -53.4 & 3.6 & B \\
$^{33}$SO$_2$ & 9(3,7) $\rightarrow$ 9(2,8) & 254.509400 & 63.04 & -- & -53.4 & -- & B \\
$^{34}$SO$_2$ & 14(6,8) $\rightarrow$ 15(5,11) & 254.516776 & 181.58 & 5.14 & -50.3 & 12.5 & U \\
CH$_3$OCHO & 20(5,15)  $\rightarrow$ 19(5,14) E v$_t$=1 & 254.530213 & 329.14 & 3.33 & -53.9 & 5.0 & B \\
O$^{13}$CS & 21 $\rightarrow$ 20 & 254.552731 & 134.39 & 8.35 & -52.8 & 5.0 & B \\
SO & $^{3}\Sigma$ N,J = 8,9 $\rightarrow$ 8,8 & 254.573628 & 99.70 & 60.39 & -51.3 & 6.7 & R \\
$^{13}$CH$_3$OH? & 19(3,17) $\rightarrow$ 18(4,14) A$^{+}$ & 254.586974 & 420.20 & 2.60 & -51.7 & 6.0 & U \\
? & -- & 254.710400 & -- & 1.82 & -- & 12.5 & U \\
CH$_3$CCNC? & 58(3) $\rightarrow$ 57(3) & 254.670887 & 427.53 & 2.18 & -51.7 & 3.1 & B \\
CH$_2$NH? & 4(0,4) $\rightarrow$ 3(0,3) & 254.685200 & 30.62 & 14.90 & -53.4 & 4.8 & B \\
$^{13}$CH$_3$OH & 9(3,6) $\rightarrow$ 9(2,7) A$^{-+}$ & 254.693481 & 151.98 & 69.60 & -53.7 & 3.5 & B \\
\textbf{HC$_3$N\tablenotemark{b}} & \textbf{28 $\rightarrow$ 27} & \textbf{254.699500} & \textbf{177.26} & \textbf{149.40} & \textbf{-52.1} &\textbf{6.3} & \textbf{D} \\
? & -- & 254.759300 & -- & 7.35 & -- & 12.5 & D \\
HC(O)NH$_2$ & 12(9,3) $\rightarrow$ 11(9,2) & 254.727017 & 320.46 & 16.68 & -53.4 & 8.2 & D \\
HC(O)NH$_2$ & 12(9,4) $\rightarrow$ 11(9,3) & 254.727017 & 320.46 & -- & -- & -- & D \\
HC(O)NH$_2$ & 12(10,2) $\rightarrow$ 11(10,1) & 254.737874 & 376.87 & 2.23 & -53.4 & 6.3 & D \\
HC(O)NH$_2$ & 12(10,3) $\rightarrow$ 11(10,2) & 254.737874 & 376.87 & -- & -- & -- & D \\
HC(O)NH$_2$ & 12(7,5) $\rightarrow$ 11(7,4) & 254.743840 & 225.35 & 18.48 & -53.0 & 8.3 & D \\
HC(O)NH$_2$ & 12(7,6) $\rightarrow$ 11(7,5) & 254.743840 & 225.35 & -- & -- & -- & D \\
HC(O)NH$_2$ & 12(11,1) $\rightarrow$ 11(11,0) & 254.757032 & 439.17 & 1.79 & -55.0 & 8.3 & U \\
HC(O)NH$_2$ & 12(11,2) $\rightarrow$ 11(11,1) & 254.757032 & 439.17 & -- & -- & -- & U \\
HC(O)NH$_2$ & 12(6,7) $\rightarrow$ 11(6,6) & 254.786445 & 186.68 & 22.28 & -53.0 & 7.8 & D \\
HC(O)NH$_2$ & 12(6,6) $\rightarrow$ 11(6,5) & 254.786445 & 186.68 & -- & -- & -- & D \\
CH$_3$CHO & 13(2,11)  $\rightarrow$ 12(2,10)  E & 254.827152 & 94.09 & 10.14 & -53.3 & 4.3 & B \\
$^{13}$CH$_3$OH & 8(3,5) $\rightarrow$ 8(2,6) A$^{-+}$ & 254.841818 & 131.59 & 59.55 & -53.6 & 3.2 & B \\
CH$_3$CHO & 13(2,11)  $\rightarrow$ 12(2,10)  A & 254.850487 & 94.07 & 10.10 & -53.2 & 6.7 & B \\
? & -- & 254.903400 & -- & 4.30 & -- & 12.5 & B \\
\textbf{HC(O)NH$_2$\tablenotemark{b}} & \textbf{12(5,8) $\rightarrow$ 11(5,7)} & \textbf{254.876330} & \textbf{153.95} & \textbf{28.62} & \textbf{-52.8} & \textbf{8.3} & \textbf{D} \\
HC(O)NH$_2$ & 12(5,7) $\rightarrow$ 11(5,6) & 254.876647 & 153.95 & -- & -52.4 & -- & D
\enddata
\tablecomments{When a line is unidentified, indicated by a ? in the Species column, the measured line frequency is given instead of the rest frequency. An em-dash (--) shown in the flux, velocity and linewidth columns indicates that the line identification and measured properties are shared with the last line where the properties are stated. If the transition quantum numbers are not known, a reference is given instead.}
\tablenotetext{a}{The morphology column sorts the lines into five different morphologies. These are, B: blue-dominant, D: disk-tracing, O: Outflow-tracing, R: red-dominant, and U: unknown.}
\tablenotetext{b}{Lines whose moment maps are shown in Fig. \ref{momentsfig} are shown in bold}
\end{deluxetable*}

\startlongtable
\begin{deluxetable*}{lllrrrrrr}
\tablecaption{Line databases used for line identification and summary of identified lines. \label{linedats}}
\tablehead{\colhead{Species} & \colhead{Line Database} & \colhead{Molecule Tag} & \multicolumn{6}{c}{Number of Lines Identified}\\
\colhead{ }  & \colhead{ } & \colhead{ } & \colhead{B} & \colhead{D} & \colhead{R} & \colhead{O} & \colhead{U} & \colhead{Total}} 
\startdata
HCOOH & JPL & 46005 & \nodata & 1(0) & \nodata & \nodata & \nodata & 1(0) \\ 
CH$_2$CO & JPL & 42002 & 1(0) & \nodata & \nodata & \nodata & \nodata & 1(0) \\
CH$_3$OH v$_t$ = 0,1,2 & JPL & 32003 & 27(6) & \nodata & \nodata & \nodata & \nodata & 27(6)\\
$^{13}$CH$_3$OH & CDMS & 33502 & 7(2) & \nodata & \nodata & \nodata & 1(0) & 8(2)\\
CH$_3$CHO & JPL & 44003 & 2(1) & \nodata & \nodata & \nodata & 0(1) & 2(2) \\
CH$_3$OCHO & JPL & 60003 & 7(2) & 2(1) & \nodata & \nodata & 4(2) & 13(5) \\
HCOCH$_2$OH & JPL & 60006 & 1(0) & \nodata & \nodata & \nodata & 4(0) & 5(0)\\
C$_2$H$_5$OH & CDMS & 46524 & 4(1) & 0(1) & \nodata & \nodata & 2(0) & 6(2) \\
CH$_3$OCH$_3$ & JPL & 46008 & 4(0) & \nodata & \nodata & \nodata & \nodata & 4(0) \\
CH$_3$COCH$_3$ & JPL & 58003 & 2(3) & 1(1) & \nodata & \nodata & 1(4) & 4(8) \\
aGg\'-(CH$_2$OH)$_2$ & CDMS & 62503 & 5(4) & 2(2) & \nodata & \nodata & 6(1) & 13(7)\\
HC$_3$N & CDMS & 51501 & \nodata & 1(0) & \nodata & \nodata & \nodata & 1(0)\\
HC$_3$N v$_7$ = 2 & CDMS & 51503 & \nodata & 1(2) & \nodata & \nodata & \nodata & 1(2)\\
H$^{13}$CCCN v$_7$ = 2 & CDMS &  52518 & \nodata & \nodata & \nodata & \nodata & 1(0) & 1(0) \\
HC$^{13}$CCN & CDMS & 52510 & \nodata & 0(1) & \nodata & \nodata & \nodata & 0(1) \\
HCC$^{13}$CN & CDMS & 52511 & \nodata & 1(0) & \nodata & \nodata & \nodata & 1(0) \\
H$_2$CCN & CDMS & 40505 & \nodata & \nodata & \nodata & \nodata & 2(0) & 2(0) \\
NH$_2$CN & JPL & 42003 & \nodata & 0(1) & \nodata & \nodata & \nodata & 0(1) \\
CH$_3$CN & JPL & 41001 & \nodata & 9(0) & \nodata & \nodata & \nodata & 9(0) \\
CH$_3^{13}$CN & JPL & 42007 & 0(1) & \nodata & \nodata & \nodata & 2(1) & 2(2) \\
CH$_3$CN v$_8$ = 1 & CDMS & 41509 & \nodata & 9(0) & \nodata & \nodata & \nodata & 9(0) \\
C$_2$H$_3$CN & CDMS & 53515 & \nodata & 4(2) & \nodata & \nodata & 2(1) & 6(3) \\
C$_2$H$_5$CN & CDMS & 55502 & \nodata & 2(0) & \nodata & \nodata & \nodata & 2(0) \\
CH$_3$CCNC\tablenotemark{a} & CDMS & 65505 & 1(0) & \nodata & \nodata & \nodata & \nodata & 1(0)\\
NS & CDMS & 46515 & \nodata & 0(2) & \nodata & \nodata & \nodata & 0(2) \\
HNCO & JPL & 43002 & \nodata & 1(0) & \nodata & \nodata & 1(0) & 2(0) \\
HC(O)NH$_2$ & CDMS & 45512 & \nodata & 2(5) & \nodata & \nodata & 0(1) & 2(6) \\
HC(O)NH$_2$ v$_{12}$ = 1 & CDMS & 45516 & \nodata & 1(0) & \nodata & \nodata & 1(0) & 2(0) \\
C$^{34}$S & JPL & 46001 & \nodata & \nodata & \nodata & 1(0) & \nodata & 1(0)\\
H$_2$CS & CDMS & 46509 & 4(2) & \nodata & \nodata & \nodata & 1(1) & 5(3) \\
SO & JPL & 48501 & \nodata & \nodata & 1(0) & \nodata & \nodata & 1(0)\\
$^{34}$SO & JPL & 50001 & \nodata & \nodata & 1(0) & \nodata & \nodata & 1(0) \\
SO$_2$ & CDMS & 64502 & \nodata & 1(0) & 6(0) & \nodata & \nodata & 7(0) \\
$^{33}$SO$_2$ & CDMS & 65501 & 0(1) & 1(1) & \nodata & \nodata & \nodata & 1(2) \\
$^{34}$SO$_2$ & JPL & 66002 & \nodata & \nodata & 1(0) & \nodata & 1(0) & 2(0) \\
SO$_2$ v$_2$ = 1 & CDMS & 64503 & \nodata & \nodata & \nodata & \nodata & 0(1) & 0(1) \\
SO$^{18}$O$_2$ & CDMS & 66502 & \nodata & \nodata & 1(0) & \nodata & 1(0) & 2(0) \\
O$^{13}$CS & CDMS & 61502 & 1(0) & \nodata & \nodata & \nodata & \nodata & 1(0) \\
OC$^{33}$S & CDMS & 61503 & 1(0) & \nodata & \nodata & \nodata & \nodata & 1(0) \\
CH$_3$CCH & JPL & 40001 & \nodata & 0(1) & \nodata & 8(0) & \nodata & 8(1) \\
CH$_2$NH\tablenotemark{a} & JPL & 29003 & 1(0) & \nodata & \nodata & \nodata & \nodata & 1(0)\\
? & \nodata & \nodata & 4(0) & 2(0) & \nodata & \nodata & 1(0) & 7(0)\\
\enddata
\tablecomments{The number of lines identified for each species and morphology type (Blue-dominant, Disk-tracing, Red-dominant, Outflow-tracing, and Unknown) are given in the penultimate five columns, along with the total number of lines for that species in the final column. The numbers in parentheses give the number of ambiguous line identifications. The final row lists unidentified lines, which are indicated by a question mark in the first column of Tables \ref{spw37} to \ref{spw15}.}
\tablenotetext{a}{These species have only tentative identifications.}
\end{deluxetable*}

\clearpage
%\bibliographystyle{aasjournal}
%\bibliography{paper_library}
\bibliography{}



\end{document}